\documentclass{aa}  

\usepackage{graphicx}
\usepackage{txfonts}
\usepackage[bookmarks=false,colorlinks=true,linkcolor=black,citecolor=cyan,filecolor=black,urlcolor=cyan]{hyperref}
\usepackage{natbib}

\usepackage{float}
\usepackage{supertabular}

\usepackage[symbol]{footmisc}
\usepackage[usenames,dvipsnames]{color}

\begin{document}

   \title{The LOFAR - eFEDS survey: The incidence of radio and X-ray AGN and the disk-jet connection\thanks{
   The source catalogue is available at the CDS via anonymous ftp to cdsarc.u-strasbg.fr (130.79.128.5) or via \url{http://cdsweb.u-strasbg.fr/cgi-bin/qcat?J/A+A/} or on the LOFAR Surveys DR website: \url{https://lofar-surveys.org/efeds.html}}}

   \author{Z. Igo
          \inst{1,2}\thanks{\email{zigo@mpe.mpg.de}},
          A. Merloni\inst{1},
          D. Hoang\inst{3},
          J. Buchner\inst{1,2},
          T. Liu\inst{1,4},
          M. Salvato\inst{1,2},
          R. Arcodia\inst{5},
          S. Bellstedt\inst{6},
          M. Br\"uggen\inst{3},\\
          J. H. Croston\inst{7},
          F. de Gasperin\inst{3,8}
          A. Georgakakis\inst{9},
          M. J. Hardcastle\inst{10},
          K. Nandra\inst{1},
          Q. Ni\inst{1},
          T. Pasini\inst{3,8},\\
          T. Shimwell\inst{11,12},
          J. Wolf\inst{13}
          }

   \authorrunning{Z. Igo et al.}
   
\institute{Max-Planck-Institut für Extraterrestrische Physik (MPE), Giessenbachstrasse 1, 85748 Garching bei München, Germany
\and
    Exzellenzcluster ORIGINS, Boltzmannstr. 2, 85748, Garching, Germany 
\and
    Hamburger Sternwarte, Gojenbergsweg 112, 21029 Hamburg, Germany
\and
University of Science and Technology of China,
No.96, JinZhai Road Baohe District, Hefei, Anhui, 230026, P.R.China
\and
    MIT Kavli Institute for Astrophysics and Space Research, Massachusetts Institute of Technology, Cambridge, MA 02139, USA
\and
ICRAR, The University of Western Australia, 35 Stirling Highway, Crawley, WA 6009, Australia
\and
School of Physical Sciences, The Open University,
Walton Hall, Milton Keynes, MK7 6AA, United Kingdom
\and
Istituto di Radioastronomia IRA-INAF, via Gobetti 101, 40129 Bologna, Italy
    \and 
    Institute for Astronomy and Astrophysics, National Observatory of Athens, V. Paulou and I. Metaxa, 11532, Greece
    \and
Centre for Astrophysics Research, University of Hertfordshire, College Lane, Hatfield AL10 9AB, UK
\and
ASTRON, the Netherlands Institute for Radio Astronomy, Oude Hoogeveensedijk 4, 7991 PD Dwingeloo, The Netherlands
\and
Leiden Observatory, Leiden University, PO Box 9513, NL-2300 RA, Leiden, The Netherlands
    \and
     Max-Planck Institut f{\"u}r Astronomie, Königstuhl 17, 69117 Heidelberg
}

   \date{}

 
  \abstract
   {Radio jets are present in a diverse sample of AGN. However, the mechanisms of jet powering are not fully understood, and it is yet unclear to what extent they obey mass-invariant scaling relations, similar to those found for the triggering and fuelling of X-ray selected AGN.}
   {This work uses the multi-wavelength data in the eFEDS field observed by eROSITA/\textit{Spectrum-Roentgen-Gamma} (SRG) and LOFAR to study the incidence of X-ray and radio AGN as a function of several stellar mass ($M_*$) normalised AGN power indicators.}
   {A new sample of radio AGN from the LOFAR - eFEDS survey, with host galaxy counterparts from Legacy Survey DR9, is defined via a radio-excess relative to the star formation rates in their hosts. We further subdivide the sample into compact and complex radio morphologies. The subset matching to the well-characterised, highly complete spectroscopic GAMA09 galaxies ($0<z<0.4$) is used in this work. We release this value-added LOFAR-eFEDS catalogue$^*$. The fraction of GAMA09 galaxies hosting radio, X-ray and both radio and X-ray AGN are calculated as a function of the specific black hole kinetic ($\lambda_{\rm Jet}$) and radiative ($\lambda_{\rm Edd}$) power.}
   {Despite the soft-X-ray eROSITA selected sample, the incidence of X-ray AGN as a function of $\lambda_{\rm Edd}$ shows the same mass-invariance and power-law slope of $-0.65$ as found in past studies, once corrected for completeness. Across the $M_*$ range probed, the incidence of compact radio AGN as a function of $\lambda_{\rm Jet}$ is described by a power-law with constant slope, showing that it is not only high mass galaxies hosting high power jets and vice versa. This slope is steeper than that of the X-ray incidence, with a value around $-1.5$. Furthermore, higher mass galaxies are more likely to host radio AGN across the $\lambda_{\rm Jet}$ range, indicating some residual mass dependence of jet powering. Upon adding complex radio morphologies, including 34 FRIIs, three of which are giant radio galaxies, the incidence not only shows a larger mass dependence but also a jet power dependence, being clearly boosted at high $\lambda_{\rm Jet}$ values. Importantly, the latter effect cannot be explained by such radio AGN residing in more dense environments (or more massive dark matter haloes). The similarity in the incidence of quiescent and star-forming radio AGN reveals that radio AGN are not only found in `red and dead' galaxies. Overall, our incidence analysis reveals some fundamental statistical properties of radio AGN samples, but highlights a number of open questions on the use of a single radio luminosity--jet power conversion. We explore how different mass and accretion rate dependencies of the incidence can explain the observed results for varying disk-jet coupling models.}
   {} 


   \maketitle

%

\section{Introduction}
\label{sec:intro}

It is now widely accepted that supermassive black holes (SMBHs) populate the centres of galaxies and co-evolve with their hosts, undergoing different stages of feeding and feedback. The sub-population of SMBHs which are actively accreting matter from the surrounding gas, usually in the form of an accretion disk, are called active galactic nuclei (AGN). Depending on their accretion rate, AGN exhibit different observational properties which we observe over more than ten decades of frequency from radio to gamma ray wavelengths \citep[e.g.][and references therein]{AlexanderANDHickox2012,HeckmanandBest2014, HardcastleANDCroston2020}. 

For highly accreting systems, with Eddington ratios $\gtrsim 0.01-1$,
the situation is often thought to be well described in terms of an optically thick, geometrically thin standard Shakura-Sunyaev disk \citep{SS1973} with most of the energy being released radiatively, or in the form of wide-angle winds \citep[e.g.][and references therein]{Fabian2012}. These `radiatively efficient' AGN are dominantly detected in the optical/UV and at X-ray wavelengths. A small fraction of radiatively efficient, luminous accretion disks have been associated with radio jets; historically, the first identified quasars were indeed discovered as powerful radio sources \citep[e.g. ][]{Schmidt1963}.  

On the other hand, for low accretion rates, the disk cannot efficiently radiate energy away, thus developing an advection dominated inner accretion flow (ADAF), as a `puffed up' hot disk \citep[e.g.][]{NarayanANDYi1994, NarayanANDYi1995}. In these systems the main route to energy release is kinetic, also in the form of a relativistic particle jet \citep{Begelman1984, Blandford2019}. These `radiatively inefficient' AGN are often detected at radio wavelengths due to synchrotron emission from collimated jets \citep{Condon1992}. The population of `radio AGN', whether clearly associated with spatially resolved relativistic jets or not, and launched from accretion disks in either radiatively efficient or inefficient states, are the focus of this paper. 

AGN jets emitting at radio frequencies involve processes acting at multiple scales (from sub-parsec to kilo-parsec), starting from jet-launching, propagation, collimation to the interaction of the jet with the interstellar medium \citep[see the recent reviews from][and references therein]{Saikia2022, HardcastleANDCroston2020}. Jets are able to deposit enough energy to alter the evolution of galaxies, groups and clusters, which is why radio AGN may be key to solving the `cooling flow problem' and delivering AGN feedback required to fix the over-prediction of over-massive, over-luminous galaxies in simulations \citep[e.g.][]{Fabian1984, Croton2006, Sijacki2007, Morganti2017,McNamara2012}.

One of the main open questions in the study of AGN of different classes is whether there exists some `unified scheme' able to explain, with limited numbers of fundamental parameters, the vast and varied observational phenomena from these extreme objects. Starting from the `unified model of AGN' \citep{Antonucci1993}, aiming to categorise AGN by orientation only, to the so-called `evolutionary scheme' of AGN \citep{Hopkins2006, Hopkins2008, Klindt2019}, there have been many attempts to get a holistic understanding of AGN feedback and evolution. \citet{Merloni2003}, underpinned by theory from \citet{HeinzandSunyaev2003}, discovered a `Fundamental Plane of black hole accretion' (henceforth, FP), which unifies stellar and supermassive black holes of different accretion rates within the radiatively inefficient regime \citep[in the `low kinetic', LK,  branch as defined in][]{MerloniHeinz2008} on a single relationship. Similarly, \citet{Falcke2004, Kording2005, Kording2006} find tight radio-to-X-ray correlations and expanded the FP to include different accretion mode branches for X-ray binaries to AGN, ultimately concluding that, even with the large range in black hole masses, jet formation may be a universal, mass-invariant process.

The advent of wide-area, large, multi-wavelength surveys allows us to study the detailed physics of black hole processes in various accretion states \citep[e.g.][]{AlexanderANDHickox2012,HeckmanandBest2014} and test unification schemes on statistical grounds. In particular, one can use the fraction of galaxies from a complete parent sample that host varying types of AGN detected in different wavelengths, namely the `incidence of AGN', as a powerful statistical tool to make inferences about the underlying physical models.

Past works in this field include those of 
\citet{Aird2012}, \citet{Bongiorno2012}, \citet{Georgakakis2017}, \citet{Birchall2022}, to name a few, who focused on X-ray selected AGN from deep extragalactic fields to measure the incidence of X-ray AGN as a function of host stellar mass (as a proxy of black hole mass) and accretion rate. They all found that the probability of a galaxy hosting an X-ray AGN is described, to first order, by a universal Eddington ratio distribution independent of the host galaxy stellar mass, leading to the conclusion that the same physical mechanisms are in charge of triggering and fuelling AGN activity in all moderately massive galaxies \citep{Aird2012}. 

However, it remains unknown whether a similar mass-invariant relation can be found for radio AGN as a function of mass-normalised jet power in a given accretion mode, especially for different radio morphologies or host galaxy types. The nature of the jet power distributions are also unclear: are jets ubiquitous in AGN, simply possessing different jet powering efficiencies \citep[e.g.][]{FalckeBiermann1995, Macfarlane2021} or is there a `radio-loud-radio-quiet' dichotomy? Robust statistical insight on corona-jet-disk coupling from combined AGN incidences in radio and X-ray regimes has been hindered by the lack of co-spatial, large volume, deep spectroscopic surveys.

Recently, thanks to the sensitivity of the Low Frequency Array (LOFAR) Two-metre Sky Survey (LoTSS) \citep{Shimwell2017}, radio AGN studies have been able to expand on past key results, one being that the incidence of radio AGN is a strongly increasing function with host galaxy stellar mass \citep{Best2005, Smolcic2009}, populating `red and dead' galaxies \citep{BestandHeckman2012}. For example, \citet{Sabater2019} used a large sample of LOFAR radio AGN in the local universe ($z<0.3$), as a function of stellar mass ($M_*$), black hole mass and radio luminosity, to find that the most massive galaxies ($>10^{11}~M_{\odot}$) are always `switched on' as radio AGN. \citet{Kondapally2022} went further by selecting a sample of $10,481$ low-excitation radio galaxies (LERGs), which are considered to be accreting in a radiatively inefficient mode. They investigated the differences between the incidence of star-forming and quiescent LERGs out to $z\sim2.5$, to find a significant population of LERGs in bluer star-forming galaxies whose incidence displays a flatter stellar mass dependence compared to quiescent-LERGs, implying different fuelling mechanisms at play.

To build on these results, we present here robust measurements of the incidence of X-ray and radio AGN as the main statistical probe to infer general properties of accreting SMBHs and the processes governing their observational appearance. 
In particular, we investigate if jet powering shows the same mass-invariance as is suggested by the X-ray AGN incidence, and we explore the constraints that can be placed on the disk-jet connection by the measured incidences of radio and X-ray AGN.

Section 2 describes the multiwavelength surveys used in this work. Section 3 describes the characterisation of the X-ray, radio and both X-ray and radio AGN sample, including the optical counterpart finding, radio morphology categorisation, host galaxy classification and details on the completeness considerations applied. The value-added LOFAR-eFEDS catalogue is released with this work and explained in Appendix \ref{appendix:catalogues}. Section 4 and 5 present the methods to calculate AGN incidences and the results obtained. Lastly, Section 6 and 7 discuss the results in the context of disk-jet coupling and highlight the uncertainty in the determination of jet power.

A standard flat cosmology with $H_0 = 70\rm{~km~s^{-1}~Mpc^{-1}}$, $\Omega_M=0.3$ and $\Omega_{\lambda}=0.7$ is used throughout and all magnitudes are AB magnitudes, corrected for galactic extinction.

\section{Data}
\label{sec:sample definition}

\begin{figure}[H]
\centering
\includegraphics[width=0.99\linewidth]{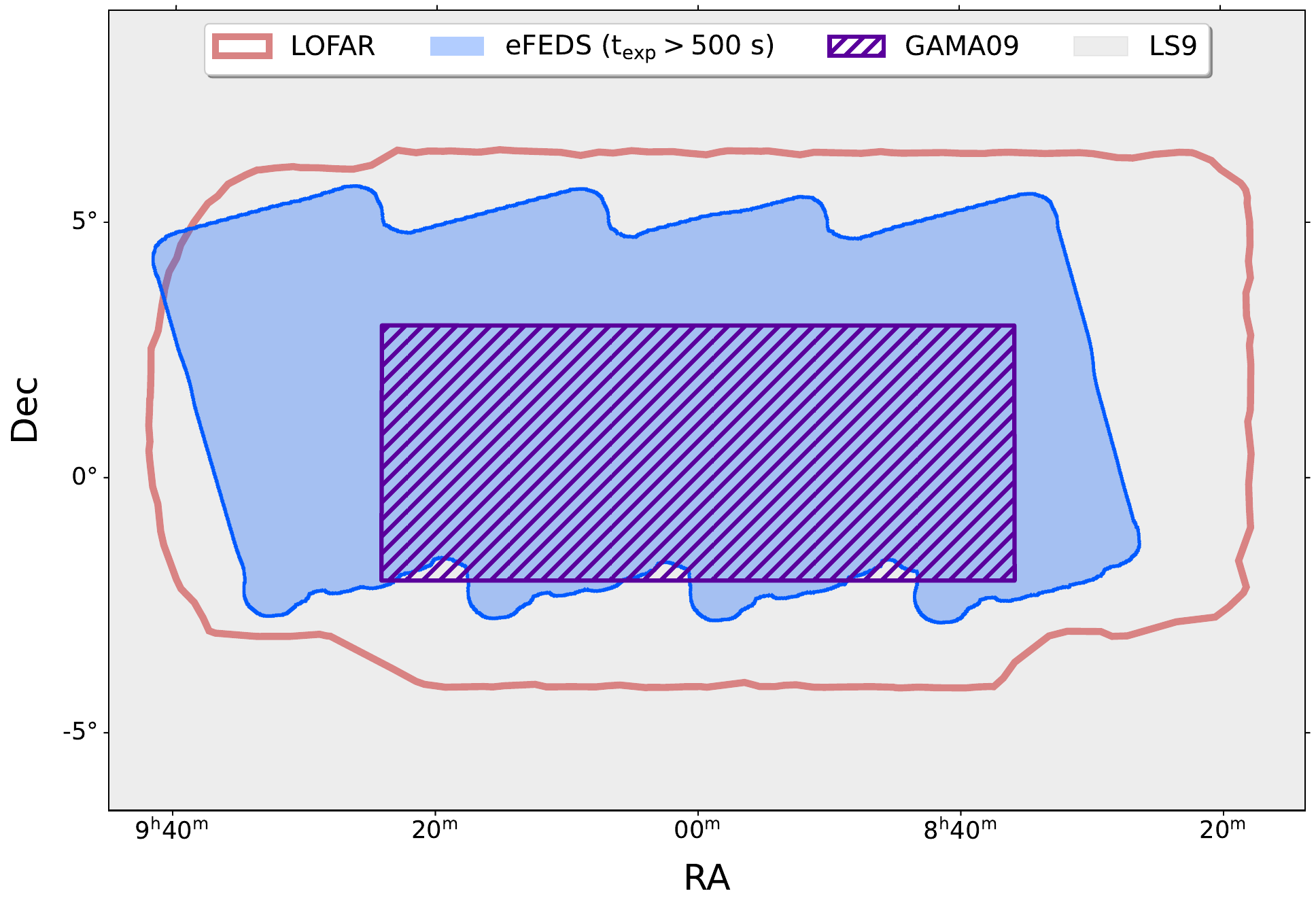}

\caption{Sky plot showing the distribution of radio (LOFAR; light red), X-ray (eROSITA eFEDS; blue) and optical/UV (GAMA09; purple, LS9; grey) sources in this equatorial field. Note that the eFEDS area has been cut to the region where the vignetted exposure time exceeds 500~s.}
\label{fig:catskyplot}
\end{figure}

For this study we combine multi-wavelength information from X-ray, radio and optical/UV catalogues. They are introduced below and visualised in sky coordinates in Fig.~\ref{fig:catskyplot}. Table \ref{tableOfsources} also presents the number of sources in each catalogue.

\subsection{eROSITA eFEDS X-ray catalogue}

The eROSITA Final Equatorial Depth Survey  \citep[eFEDS;][]{Brunner2022} is the $\sim$140~deg$^2$ pilot survey of the eROSITA All Sky Survey (eRASS), having a uniform exposure of $\sim$~1.2~ks (after correcting for vignetting) and observing most sensitively in the soft X-ray energies, with half-energy width of 30\arcsec\ in the $0.2-2.3$~keV band \citep{Merloni2024}. The point source catalogue, including 21,952 candidate AGN detected in the main $0.2-2.3$~keV band, is presented by \citet{TengeFEDS2022}, along with detailed X-ray spectral analysis results. The data were processed with eROSITA Standard Analysis Software System pipeline version c001 \citep[eSASS; version eSASS\_users201009,][]{Brunner2022}. In this work, only sources from the main eFEDS catalogue are used, as supplementing this with additional sources uniquely detected in the hard $2.3-5$~keV band \citep[][]{Brunner2022, Nandra2024}, would add a negligible amount of AGN ($<$~10). Furthermore, an eFEDS Multi-Order Coverage map (MOC), marking the area where the 0.2-2.3 keV vignetted exposure, $t_{\rm exp}$, exceeds 500s, selects the sources used for further analysis. This area cut was applied equally to all other catalogues described in this section. The nominal eFEDS 1$\sigma$ positional error is $\approx$ 4.5\arcsec\ \citep{MaraCTPeFEDS, Brunner2022}.

\subsection{LOFAR radio catalogue}
\label{sec:lofarintro}

The Low-Frequency Array (LOFAR) High Band Antenna (HBA) observations of the eFEDS field provide the 144~MHz radio data (rms noise level of $\sim$~135 $\mu$~Jy beam$^{-1}$) for this work \citep{Pasini2022}. LOFAR, with its excellent baseline coverage, is the ideal survey instrument for this study as it enables the detection of structures even in compact sources thanks to its high 8\arcsec\ $\times$ 9\arcsec\ angular resolution, as well as the identification of larger scale diffuse emission \citep{Shimwell2022}. We refer to \citet{Pasini2022} (Sect. 2.2) for details on the calibration, generation and validation of the radio source catalogue used in this study. For completeness, we briefly outline the procedure here. Standard calibration techniques, equivalent to those used for the LOFAR Two-metre Sky Survey (LoTSS) DR1, were applied, including direction-dependent calibration to account for varying ionosphere effects \citep{Shimwell2017, Shimwell2019, Tasse2021}. The `Python Blob Detector and Source Finder' (PyBDSF) \citep{PyBDSF2015} was run on the 8\arcsec\ $\times$ 9\arcsec\ high resolution mosaic images created for the full LOFAR-eFEDS field in order to model the radio emission with Gaussian components and produce the final source catalogue. A peak flux detection threshold of 5$\sigma$, with sigma the local rms noise, was imposed for a source to be detected, calculated via sliding a 150 $\times$ 150 pixel box in 15 pixel steps. The astrometric accuracy of LOFAR benefits from faced-based astrometric correction with PanSTARRS \citep{Shimwell2019}, leading to mean positional uncertainties around 1\arcsec\ (see Sect. \ref{sec:nwaylofarlegacy_radio} for more discussion).

\subsection{DESI Legacy Imaging Survey DR9}

The DESI Legacy Imaging Survey DR9 \citep[hereafter: LS9;][]{Dey2019} is used here in the counterpart identification process for both the radio and X-ray sources presented above. This is because of its large, homogeneous sky coverage, photometric depth and accurate astrometry. Along with optical photometry in the {\it g, r, z} bands (limiting AB magnitudes: 23.95, 23.54 and 22.50, respectively), the LS9 catalogue includes Wide-field Infrared Survey Explorer (WISE) forced photometry at the optical source coordinates, following \citet{Lang2014, Lang2016}, at 3.4~$\mu$m, 4.6~$\mu$m, 12~$\mu$m and 22~$\mu$m. This method ensures matched aperture photometry, making use of the higher optical angular resolution compared to that of WISE, and constructs reliable spectral energy distributions (SEDs), which are fundamental for robust counterpart identification (see Sect. \ref{sec:nwayefedslegacy} and \ref{sec:nwaylofarlegacy_radio}). The positional accuracy of these optical sources ($\sim$0.1\arcsec) exceeds that of radio and X-ray astrometry.

\begin{table*}[t!]
\caption{Summary table of the number of sources in each multiwavelength catalogue and the different subsets explained in Sect. \ref{sec:methods}.}
\label{tableOfsources}
\begin{center}
\begin{tabular}{lrl}
\hline
\hline
\textbf{Catalogue}             & \textbf{Number} & \textbf{Comments}                                                                            \\
\hline
eFEDS Main                     & 27,021*                    & * = cut to eFEDS MOC $t_{\rm exp}>500$s                                                   \\
LS DR9                         & 11,255,466*                  & duplicate (“DUP”) sources removed                                                            \\
LOFAR                          & 36,631*                    & 24,613 compact, 12,018 complex                                 \\
GAMA09                       & 48,190                      &  \texttt{SC} $\geq6$; $z<0.4$; 21,462 mass-complete                                                                           \\
\hline
eFEDS-LS9                      & 20,696                      & \citet{MaraCTPeFEDS}; \texttt{p\_any} \textgreater{} 0.035, \texttt{CTP\_quality} \textgreater{} 2 \\
eFEDS-LS9-GAMA09               & 584     &                                               \\

\hline
LOFAR-LS9                      & 22,759$^{a}$                     & \texttt{p\_any} \textgreater{} 0.06, SNR \textgreater{} 5; 16130 compact, 6629 complex.                    \\
LOFAR-LS9-GAMA09               & 2,619$^{b}$                        & 1901 compact, 718 complex      \\

\hline

GAMA09 (G9) X-ray AGN & 523 & 325 mass-complete \\
G9 X-ray AGN in Quiescent gal. & 147 & 124/147 mass-complete \\
G9 X-ray AGN in Star-forming gal. & 376 & 201/376 mass-complete \\
\hline
G9 Radio AGN                      & 764                       & 682 mass-complete (404/445 compact; 278/319 complex)  \\    
G9 Radio AGN in Quiescent gal.    & 646                        & 595/646 mass-complete (354/385 compact; 241/261 complex)     
\\
G9 Radio AGN in Star-forming gal. & 118                        & 87/118 mass-complete (50/60 compact; 37/58 complex)    \\
\hline
G9 Radio + X-ray sources & 121 &  74 (32) /92 mass-complete X-ray (radio) AGN; 24 Radio + X-ray AGN \\
\hline
\hline
\end{tabular}
\end{center}
\tablefoot{$^a$ The LOFAR-LS9 NWAY match yields 22,754 sources, to which 5/6 large FRII sources are added manually (the remaining large FRII source with LOFAR Source id 8153 was already present among the matched sources, albeit the optical CTP had to be corrected). $^b$ The mass-complete sources were visually inspected to confirm correct counterpart association and classify radio morphology. Two radio sources (LOFAR Source id: 10347, 27051) were found to be consistent with noise fluctuations of the background and are excluded from this point on (see more details in Appendix~\ref{appendix:nwaylofarlegacy_visinspect}).}
\end{table*}

\subsection{GAMA09 spectroscopic galaxy catalogue}
\label{sec:GAMA}

The Galaxy and Mass Assembly (GAMA) DR4 survey, specifically the 9hr field (hereafter: GAMA09) that is almost entirely contained within the eFEDS field, serves as the parent galaxy sample for this investigation due to its high spectroscopic completeness \citep{Driver2022}. Broadband coverage from the far-UV ($\sim$1500~$\AA$) to the far-infrared ($\sim$500~$\mu$m) allows for the creation of wide SEDs for individual galaxies and the determination of galaxy properties through SED fitting \citep{Robotham2020, Bellstedt2020, Bellstedt2021}. This includes stellar mass and star formation rate (SFR) estimates from stellar population synthesis modelling of SEDs, using \citet{BruzalandCharlot2003} stellar evolution models, taking a \citet{Chabrier2003} initial mass function (IMF) and \citet{Calzetti2000} dust curves (catalogue name: \texttt{gkvProSpectv02}). Considering only the subsample\footnote{\url{http://www.gama-survey.org/dr4/schema/table.php?id=684}} with quality flag \texttt{SC}~$\geq$~6 and $0<z<0.4$ results in a cleaned catalogue with 90\% completeness limit for $r < 19.77$ (more detailed discussion on completeness will follow in Sect. \ref{sec:mass_lumin_completeness}). By construction, spectroscopic redshifts exist for all 48,190 sources selected in this way. In the following, we focus our attention to the sources in the overlap region between eFEDS, LOFAR and GAMA09, as shown in Fig.~\ref{fig:catskyplot}.

\section{Characterization of X-ray and radio AGN samples}
\label{sec:methods}

\subsection{Characterization of the X-ray AGN sample}
\label{methods:xrays}

This section describes the X-ray AGN sample, their optical host galaxies and their properties, along with the considerations taken to control for stellar mass and X-ray luminosity completeness. For a summary of the number of sources at each step, see Table~\ref{tableOfsources}.

\subsubsection{Optical counterparts of the X-ray sources}
\label{sec:nwayefedslegacy}

The identification of the optical counterparts for the X-ray sources is crucial to later classify the host galaxy properties, including stellar mass and star-formation rate. This is done using a Bayesian cross-matching algorithm called NWAY \citep{maraNWAY2018}, which uses not only source sky density, distance priors and positional accuracy, but also additional priors based on observable characteristics (e.g. magnitudes, colours). Notably, it provides a best match flag (\texttt{match\_flag} = 1), a probability for the match being the correct one (\texttt{p\_i}) and a probability of the source in question having any counterpart at all in the search region (\texttt{p\_any}). 

The counterpart identification of the eFEDS X-ray sources has already been presented by \citet{MaraCTPeFEDS}\footnote{Available at: \url{ https://erosita.mpe.mpg.de/edr/eROSITAObservations/Catalogues/}}, who used external pre-constructed priors trained on X-ray sources with secure counterparts from Legacy Survey DR8 \citep{Dey2019} in 3XMM and Chandra catalogues (adjusted to have `eFEDS-like' source properties). An updated version of their catalog has been used in this work, where the eROSITA positional error ({\tt RADEC\_ERR}) was divided by $\sqrt{2}$ (1-dimensional positional error). This new version (V18) is consistent with the originally released V17 catalog, with minimal (5\%) change in the counterparts, of which only 0.3\% have \texttt{CTP\_quality}$>$2 (sources with secure counterparts). There is also improved counterpart association as most of the original sources with \texttt{CTP\_quality}$=$2 (i.e. with more than one possible counterpart) now have a secure and unique match (Saxena et al, in prep.).

\begin{figure}[!t]
    \centering
    \includegraphics[width=\linewidth]{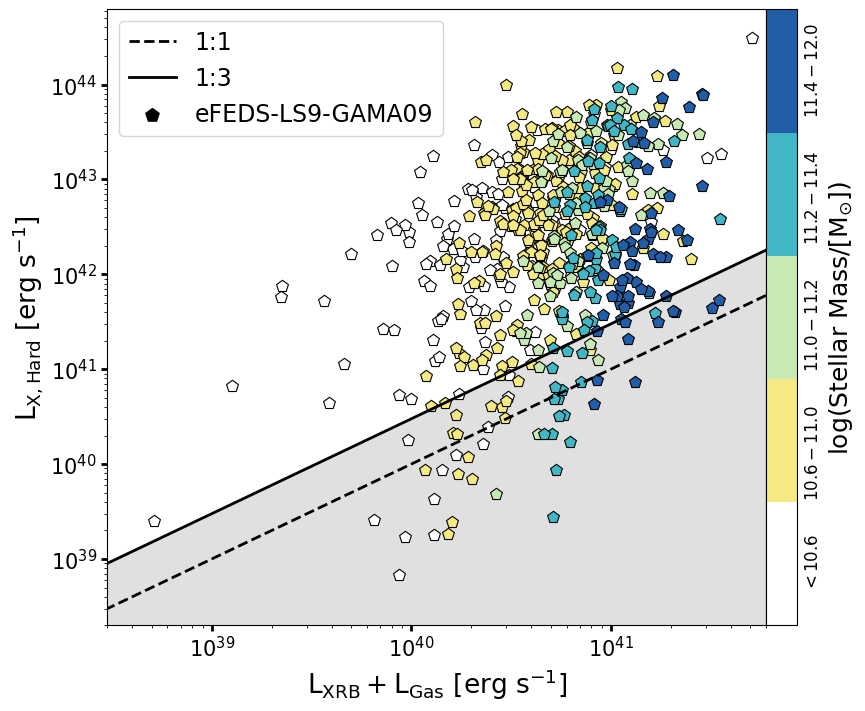}
    \caption{X-ray luminosity expected from X-ray binaries and hot gas of the GAMA09 matched galaxies versus the intrinsic $2-10$~keV luminosity of each eFEDS X-ray source, colour coded by the stellar mass (explained in Sect. \ref{sec:mass_lumin_completeness}). Sources lying above the 1:3 solid black line have X-ray emission securely dominated by AGN processes and constitute our X-ray AGN sample; sources in the shaded area are compatible with non-AGN emission processes, and excluded from the analysis.}
    \label{fig:xrbhotgas}
\end{figure}

\begin{figure*}[ht!]
\centering
\includegraphics[width=1.07\linewidth]{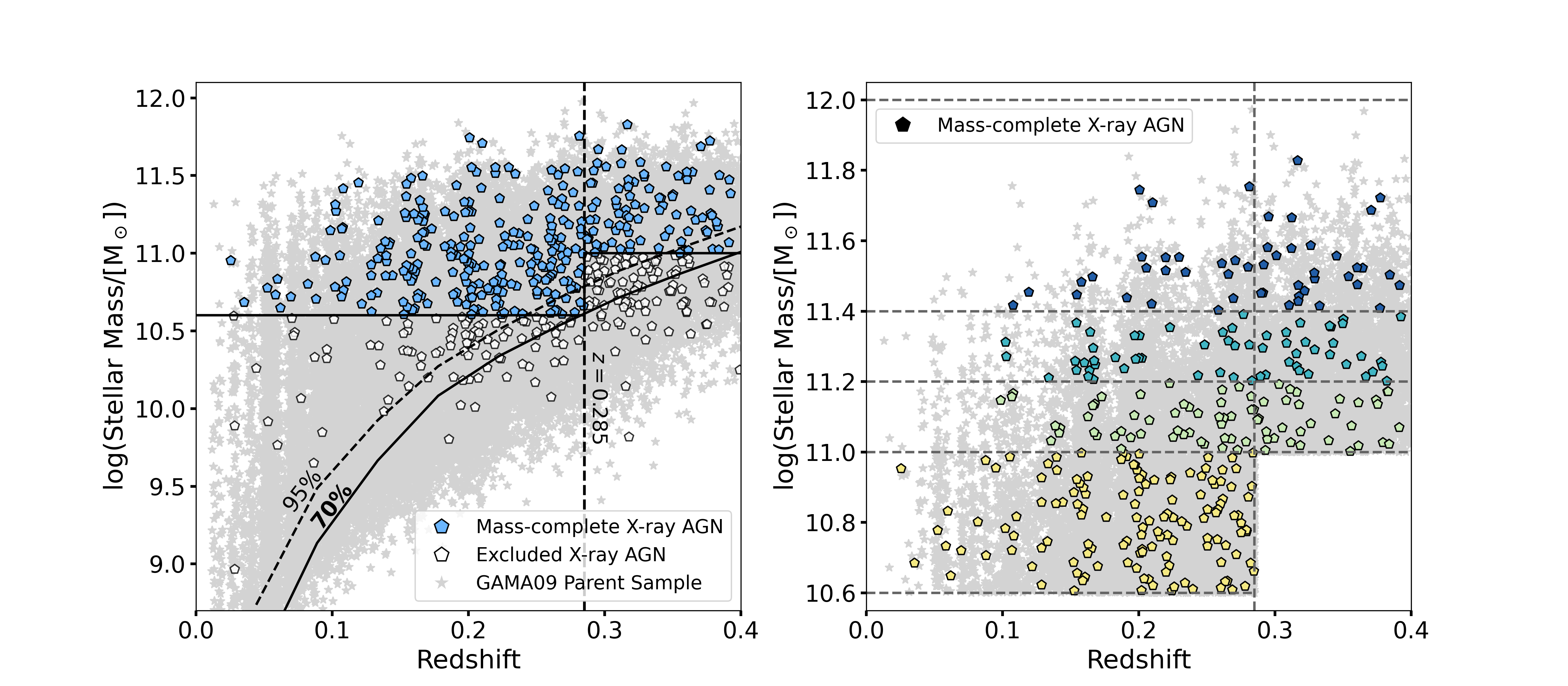}
\caption{Left: Stellar mass versus redshift distribution of the galaxies in GAMA09 (grey points) and of the X-ray AGN (blue filled pentagons). A vertical line divides the sample into two redshift bins. Completeness curves (70\%, 95\% with solid, dashed black lines, respectively) and horizontal thresholds are used to exclude sources incomplete in stellar mass (unfilled markers). Right: Zoom-in of the mass-complete sample, split into four stellar mass bins.}
\label{fig:xray_zvsMstar}
\end{figure*}

\begin{figure}
\centering
\includegraphics[width=0.9\linewidth]{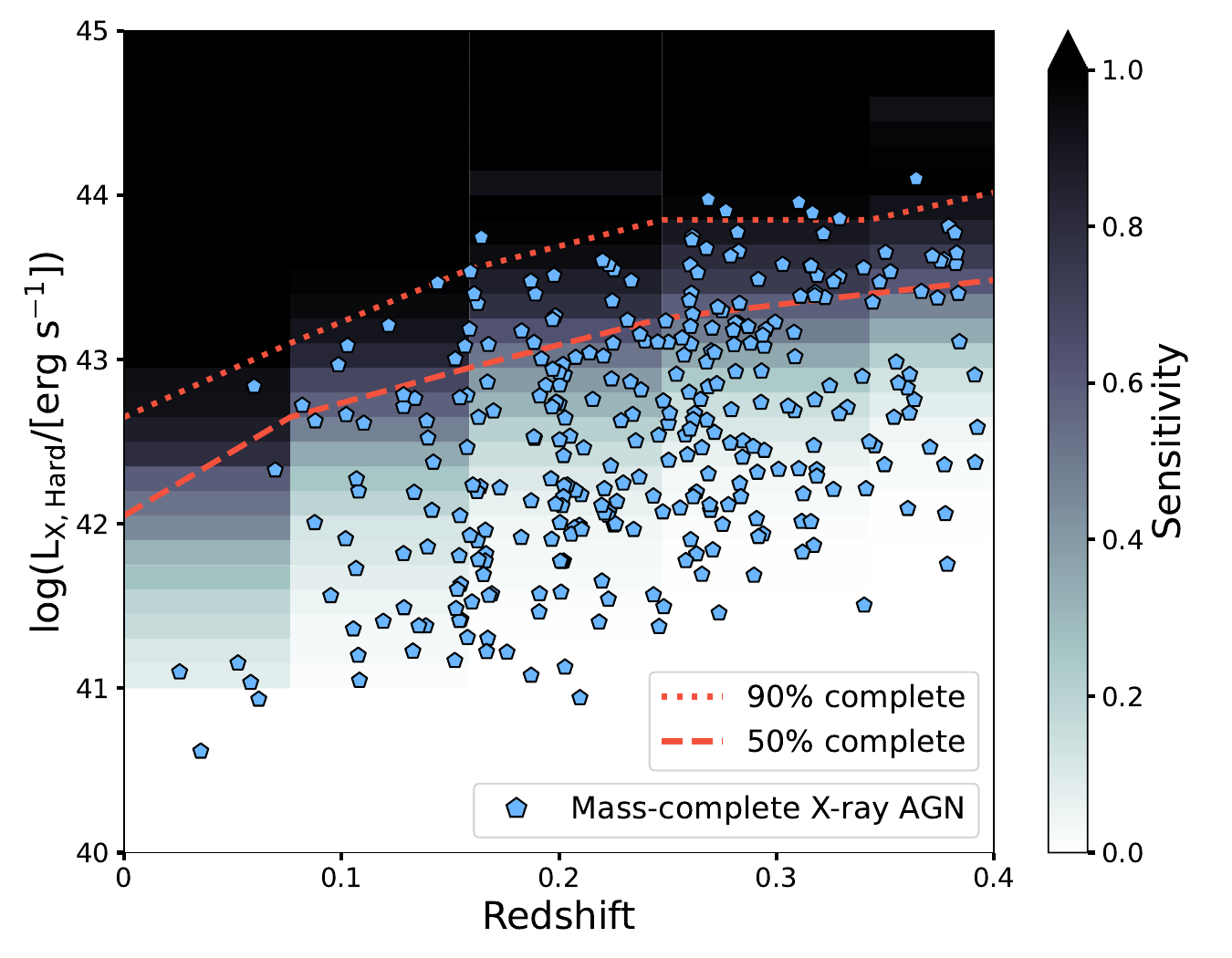}
\caption{Redshift versus intrinsic $2-10$~keV X-ray luminosity of the G9 X-ray AGN (blue pentagons). A sensitivity grid, for both obscured and unobscured sources, is plotted in the background from simulations done by \citet{TengSimulation2022} and used to compute the 50\% (orange, dashed) and 90\% (orange, dotted) X-ray luminosity completeness limits, respectively.}
\label{fig:xray_zvsLx}
\end{figure}

At the time of writing, a new data release of the DESI Legacy Survey (DR9), with improved flux calibration and source detection near bright sources, also became available and is thus adopted for this work. A simple 1\arcsec\ positional cross-match between Legacy Survey DR8 and DR9 provides the final X-ray catalogue. As in \citet{MaraCTPeFEDS}, thresholds of \texttt{p\_any}$>$0.035 and \texttt{CTP\_quality}$>$2 are applied.

In addition, \citet{TengeFEDS2022} provides X-ray spectroscopic results for all eFEDS sources. The absorbed power-law models are used to calculate intrinsic $2-10$~keV luminosities (see Sect. \ref{sec:mass_lumin_completeness}). An updated version, with improved spectroscopic redshifts from SDSS-V DR18 \citep{Almeida2023}, is used here and presented in Appendix~\ref{appendix:catalogues}.  

\begin{figure*}[ht!]
    \centering
    \includegraphics[width=0.85\linewidth]{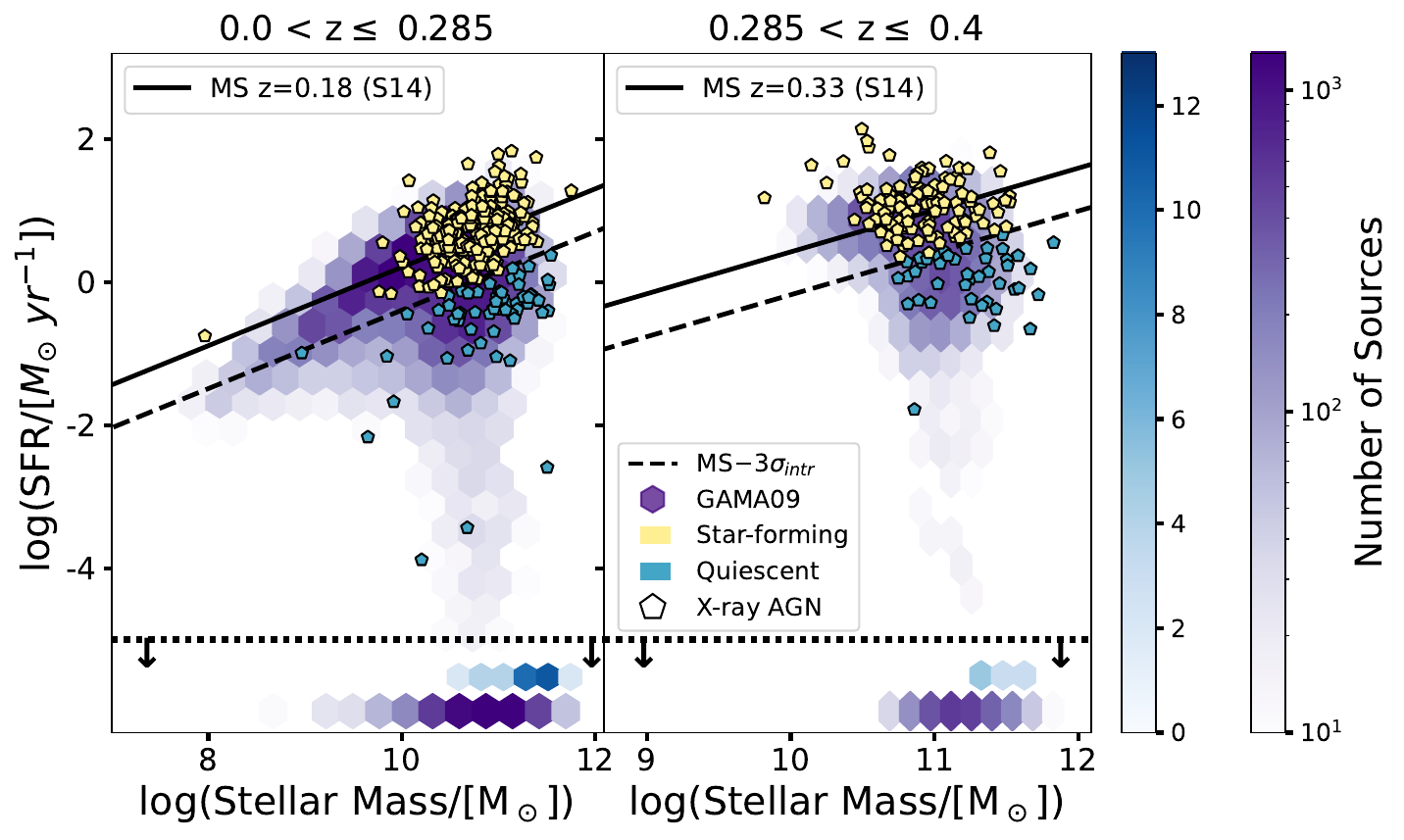}
    \caption{Stellar mass versus SFR of the GAMA09 galaxy sample (purple hexbins) and the X-ray AGN (pentagons). The solid black line in each redshift panel marks the star-forming galaxy main sequence from \citep[S14;][]{Speagle2014}. Sources $3\sigma$ below this line (black, dashed) are considered to be quiescent galaxies (blue), otherwise they are classified as star-forming (yellow). Quiescent sources below $\log \rm{(SFR)}=-5$ are marked as upper limits and their distributions are shown in the bottom of each panel.}
    \label{fig:qvsSF_xray}
\end{figure*}

\subsubsection{X-ray AGN among the GAMA09 galaxies} 

As the parent sample is the GAMA09 galaxies, the X-ray AGN catalogue is necessarily also limited to this region for the scope of this work. This is done using a simple 2\arcsec\ (to account for the fibre sizes) positional match between the LS9 optical coordinates and those of the GAMA09 galaxies. 

Then, five sources (eROIDs: 584, 1730, 6498, 9305, 14520) with discrepant redshifts between GAMA and SDSS, for which a visual inspection of the SDSS spectra indicate an AGN at $z>0.4$, have been removed from the sample, leaving a total of 584 X-ray sources. However, not all of these X-ray sources are AGN because at low luminosities the sample starts to be dominated by the collective (unresolved) X-ray emission from X-ray binaries (XRBs) and emission from hot diffuse gas within the hosts. The X-ray luminosity of the sources is calculated in the standard way using the absorption corrected flux in rest-frame $0.5-2$~keV from the work of \citet{TengeFEDS2022}. These were then converted to hard $2-10$~keV luminosities using the modelled photon index.
To select out the X-ray AGN, the relations from \citet{Lehmer2016}, Eq.~\ref{LehmerEq} below, and \citet{Mineo2012}, Eq.~\ref{MineoEq} below, are used to estimate the corresponding X-ray binary and hot gas emission (in erg~s$^{-1}$), respectively, for a given $M_*$ and SFR. 
\begin{equation}
    \label{LehmerEq}
    L_{\mathrm{XRB}}=\alpha_0 (1+z)^{\gamma}~M_*/[M_{\odot}]+\beta_0 (1+z)^\delta ~ {\rm SFR}/[M_{\odot}~yr^{-1}].
\end{equation}
The parameters have the following values for $2-10$~keV X-ray luminosity: $\rm{log_{10}}(\alpha_0)=29.37 \pm 0.15$, $\gamma=2.03 \pm 0.60$, $\rm{log_{10}}(\beta_0)=39.28 \pm 0.03$ and $\delta=1.31 \pm 0.13$ \citep[from Sect. 6.3.2. of][]{Lehmer2016}. 
\begin{equation}
    \label{MineoEq}
    L_{\mathrm{Gas}}=(8.3\pm0.1) \times 10^{38} ~ {\rm SFR}/[M_{\odot}~yr^{-1}].
\end{equation}

Fig.~\ref{fig:xrbhotgas} shows the $2-10$~keV luminosity expected from the X-ray binaries and hot gas versus the $2-10$~keV luminosity of each detected X-ray source, colour coded by their stellar mass (explained in the next section). Black dashed and solid lines mark the 1:1 and 1:3 levels, respectively. The 523/584 sources ($\sim$90\%) that lie above the grey-shaded region defined by the 1:3 line, are the sources in which the X-ray emission is dominated by AGN processes and will henceforth be referred to as the X-ray AGN among the GAMA09 galaxies, or `G9 X-ray AGN' (see Table \ref{tableOfsources}). This AGN sample is free of \texttt{CLUSTER\_CLASS}$=$5 sources, meaning that the AGN X-ray luminosities are not biased by additional emission potentially coming from hot cluster gas \citep[see][and details therein]{MaraCTPeFEDS}.

\subsubsection{Stellar Mass and X-ray luminosity complete X-ray AGN samples}
\label{sec:mass_lumin_completeness}

Fig.~\ref{fig:xray_zvsMstar} (left) shows the redshift vs. stellar mass distribution of the GAMA09 parent sample (grey points) and of the X-ray AGN among them (blue pentagons). The stellar masses are taken from the \texttt{StellarMass\_50} GAMA catalogue entry, the median of the posterior distribution from the Bayesian SED fitting (with Markov Chain Monte Carlo; MCMC). Although no AGN component was used in the SED fitting to determine the host galaxy stellar mass, we show in Section \ref{discussion} that these measurements are robust and in agreement with later works that do account for AGN \citep{Thorne2022,Aihara2018,Li2023}. A vertical line at $z=0.285$ divides the sample into low and high redshift bins (see Sect. \ref{sec:mstarzbinning}).

As this investigation deals with fractions of galaxies hosting AGN, it is vital to ensure that these samples are complete in both stellar mass and AGN luminosity. Firstly, the stellar mass completeness limits are calculated using the limiting stellar mass method, $M_{*, \rm{lim}}$ \citep[e.g.][]{Pozzetti2010, Moustakas2013, Mountrichas2022}, shown in Eq.~\ref{eq_mlim}, where $M_{*, {\rm lim}}$ is the stellar mass a given galaxy would have if its r-band magnitude ($r_{\rm mag}$) was equal to the limiting r-band magnitude of the survey ($r_{\rm lim}=19.8$):
\begin{equation}
\label{eq_mlim}
    \log_{10} M_{*, {\rm lim}} = \log_{10} M_* + 0.4~ (r_{\rm mag}-r_{\rm lim}).
\end{equation}
Then, for each redshift interval ($\Delta z=0.04$), the cumulative distribution of $M_{*, {\rm lim}}$ was used to calculate the 70\% (solid, black) and 95\% (dashed, black) completeness limits and plot this against the maximum redshift in each given interval. As shown in Fig.~\ref{fig:xray_zvsMstar} (left) the completeness function is rather steep compared to the change in stellar mass value, therefore, the 70\% limit is used for this work, in order to maximise source numbers.

Solid black horizontal lines at $\log(M_*/M_{\odot})=$ 10.6 and 11.0 mark the stellar mass completeness limits for the low and high redshift bins, respectively and the white-filled markers are the X-ray AGN which are excluded as a result of this cut. Overall, there are 325 X-ray AGN and 21,462 GAMA09 galaxies included in this `mass-complete' (to 70\%), volume-limited samples in the redshift range $0<z<0.4$. Thus, the total percentage of these galaxies hosting X-ray AGN detected by eROSITA is about 1.5\%. Fig.~\ref{fig:xray_zvsMstar} (right) shows a zoom-in of this `mass-complete' sample, splitting up the data into four mass bins (yellow, green, teal, blue), which will be elaborated upon in Sect. \ref{sec:calc_incidences}. 

Fig.~\ref{fig:xray_zvsLx} shows the distribution of intrinsic, hard $2-10$~keV X-ray luminosities ($L_{\rm X, Hard}$ in erg~s$^{-1}$) versus redshift for the G9 X-ray AGN (blue pentagons). An X-ray luminosity sensitivity grid from detailed simulations \citep[][their Fig.~8 and Section 4.1]{TengSimulation2022} is also plotted in the background. X-ray sensitivity functions are a complex combination of redshift, absorbing column density ($N_H$), spectral shapes and k-correction factor dependent parameters. This is rigorously taken into account, as the mock eFEDS AGN catalogue \citep{Comparat2019} used for these simulations is highly representative of the real eFEDS data. Thus, these parameters, along with any correlation of $N_H$ with $M_*$ \citep{Buchner2017} for example, have been folded into the X-ray luminosity completeness curves shown on Fig.~\ref{fig:xray_zvsLx}. They are valid for the entire $N_H$ distribution of the sample (both unobscured and obscured sources). Given the soft X-ray response of eROSITA \citep{Predehl2021}, most obscured AGN are missed, and the majority of sources lie below the orange dashed, 50\% and orange dotted, 95\%, limits.

In order to compare with past work dealing with X-ray AGN incidences using a hard X-ray band selection (e.g. {\it XMM-Newton}), these full X-ray completeness correction functions can be used to implement a weighting per bin in stellar mass, redshift, luminosity and $\lambda_{\rm{Edd}}$ when computing the incidences (see details in Sect. \ref{sec:calc_incidences}).

In Appendix \ref{appendix:softsel} the purely unobscured (soft X-ray) eFEDS sensitivity functions are presented and applied to the results, in order to show the impacts of obscuration on the measured X-ray AGN incidence.

\subsubsection{Host galaxy properties of X-ray AGN}

Having defined a clean sample of G9 X-ray AGN and shown the stellar mass and X-ray luminosity distributions as functions of redshift, this section discusses the properties of the host galaxies of these AGN.

AGN with star-forming host galaxies tend to scatter around the so-called main sequence (MS) of star forming galaxies, a well-studied relation in the literature \citep[e.g.][and references therein]{Speagle2014,PopessoMS2023}. There are variances in derived MS relations, especially at the high stellar mass end, which is why two well-founded, yet analytically different ones, namely those presented in \citet{Speagle2014} and \citet{PopessoMS2023}, were tested during the radio and X-ray incidence analysis. Ultimately, both produced similar results and thus the simplest of the two was chosen, for which the best fit MS relation is Eq.~28 from \citet{Speagle2014} with intrinsic scatter $\sigma_{\rm intr}=0.2$ (see also their Fig.~8 for a visual representation of the relation). 

Fig.~\ref{fig:qvsSF_xray} shows the stellar mass versus star-formation rate (\texttt{SFR\_50}) of the GAMA09 parent sample (purple hexbins), X-ray AGN (yellow, blue) and this MS relation in black for the two redshift bins introduced in Sect. \ref{sec:mass_lumin_completeness}, along with a dashed line marking SFRs $3\sigma_{\rm intr}$ below the MS. The MS is calculated using the mean redshifts of the GAMA09 sample within that redshift bin. The dashed line is used as a demarcation between the quiescent (blue) and star-forming (yellow) X-ray AGN. All sources with $\log \rm{(SFR)}\leq-5$ (below the black dotted line) are marked as upper limits to indicate their quenched nature \citep{Bellstedt2020}. Such low SFRs are possible because a skewed Normal parameterisation is used to fit the star formation history of each GAMA galaxy, from which the SFR is obtained by averaging over the past 100~Myr. Therefore, galaxies with SFRs peaking in the early universe could have $\log \rm{(SFR)}\leq-5$ at present day. Their stellar mass distributions are plotted as blue (X-ray AGN) and purple (GAMA09 galaxies) hexbins, shifted to an arbitrary low SFR value to avoid overlap. In the low (high) redshift bin, there are a total of 33 (11) and 5453 (2121) X-ray AGN and GAMA09 galaxies, respectively, which have $\log \rm{(SFR)}\leq-5$.

\subsection{Characterization of the radio AGN sample}
\label{methods:radio}

Having presented the X-ray AGN in the G9 field, we now move to the analysis of the radio AGN sample. We will first study the distinction between compact and complex radio morphologies among radio sources; then proceed with the identification of the optical counterparts to the radio sources using LS9 and GAMA09; then to the definition of radio AGN as `radio-excess' sources; and finally to the classification of the optical host galaxies into quiescent and star-forming. For a summary of the number of sources at each step, see Table \ref{tableOfsources}. The value-added LOFAR-eFEDS catalogue presented here is made publicly available and further details are given in Appendix \ref{appendix:catalogues}.

\begin{figure}[t!]
\centering
\includegraphics[width=\linewidth]{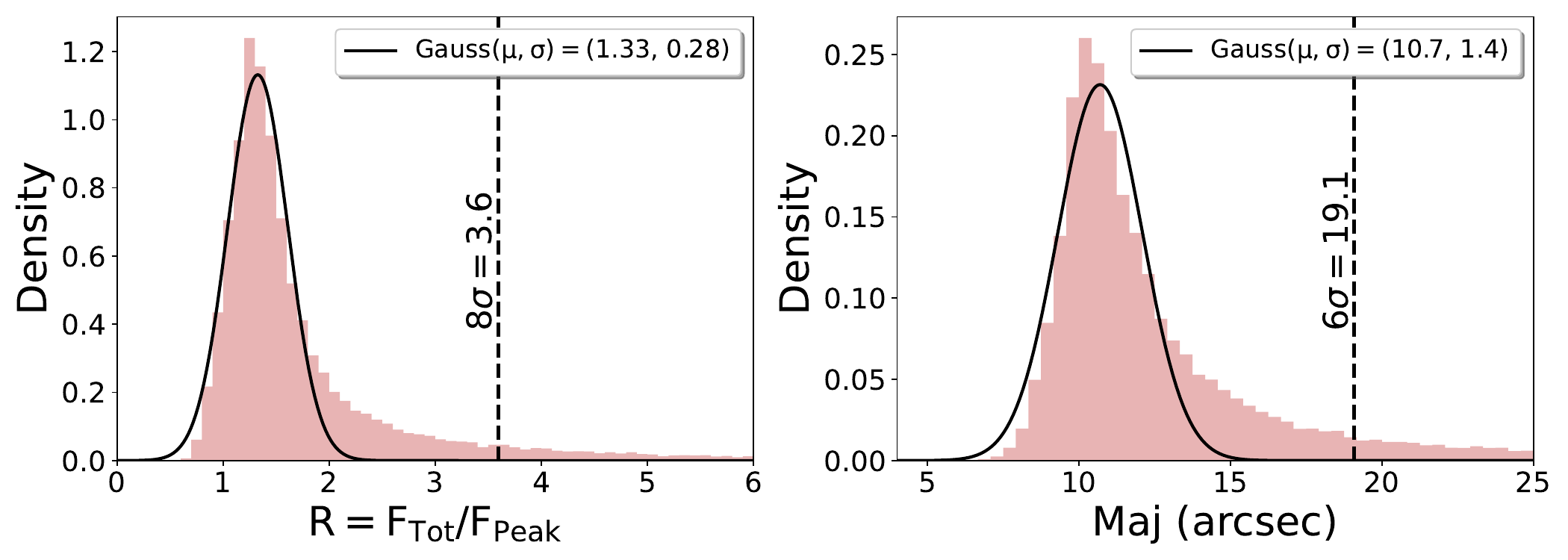}
\caption{Histograms showing the flux ratio (total to peak flux; left) and major axis (right) distributions for the LOFAR sample of 36,631 sources in the eFEDS field, each fit by a Gaussian (black curves) to determine the thresholds for being a compact radio emitter. These are shown as black dashed vertical lines (see text for more details).}
\label{fig:flux_maj_cuts}
\end{figure}

\begin{figure}[t!]
\centering
\includegraphics[width=0.95\linewidth]{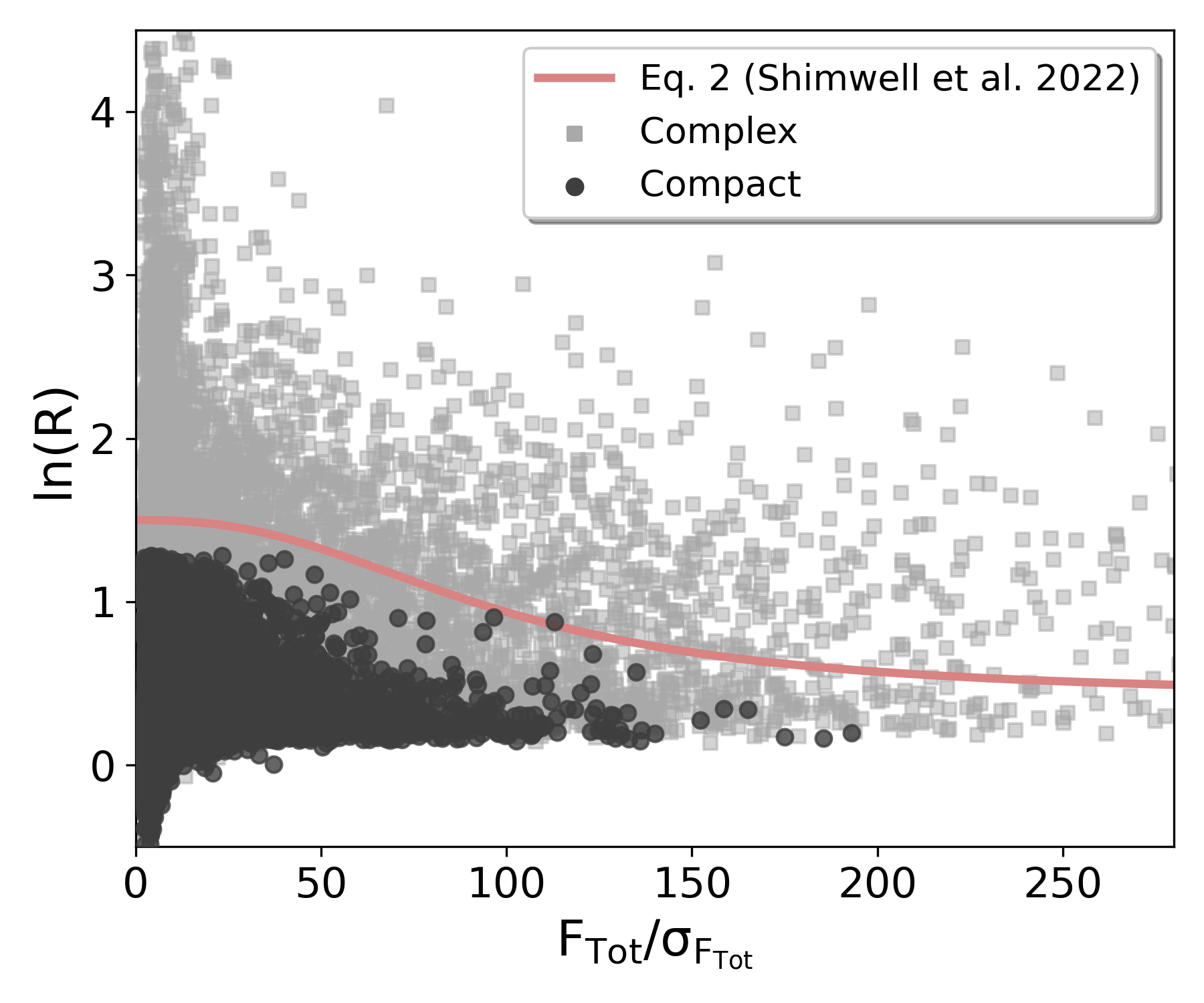}
\caption{Signal-to-Noise ratio, calculated by dividing the total radio flux by its associated error, versus the logarithm of the ratio of total to peak fluxes for the LOFAR sample of 36,631 sources. Sources classified as compact are shown in black, complex ones in light grey. The light red curve is taken from \citet{Shimwell2022} Eq.~2, below which 99.9\% of all compact or unresolved sources lie.}
\label{fig:shimwell_compact_cut}
\end{figure}

\begin{figure*}[ht!]
\centering
\includegraphics[width=0.49\linewidth]{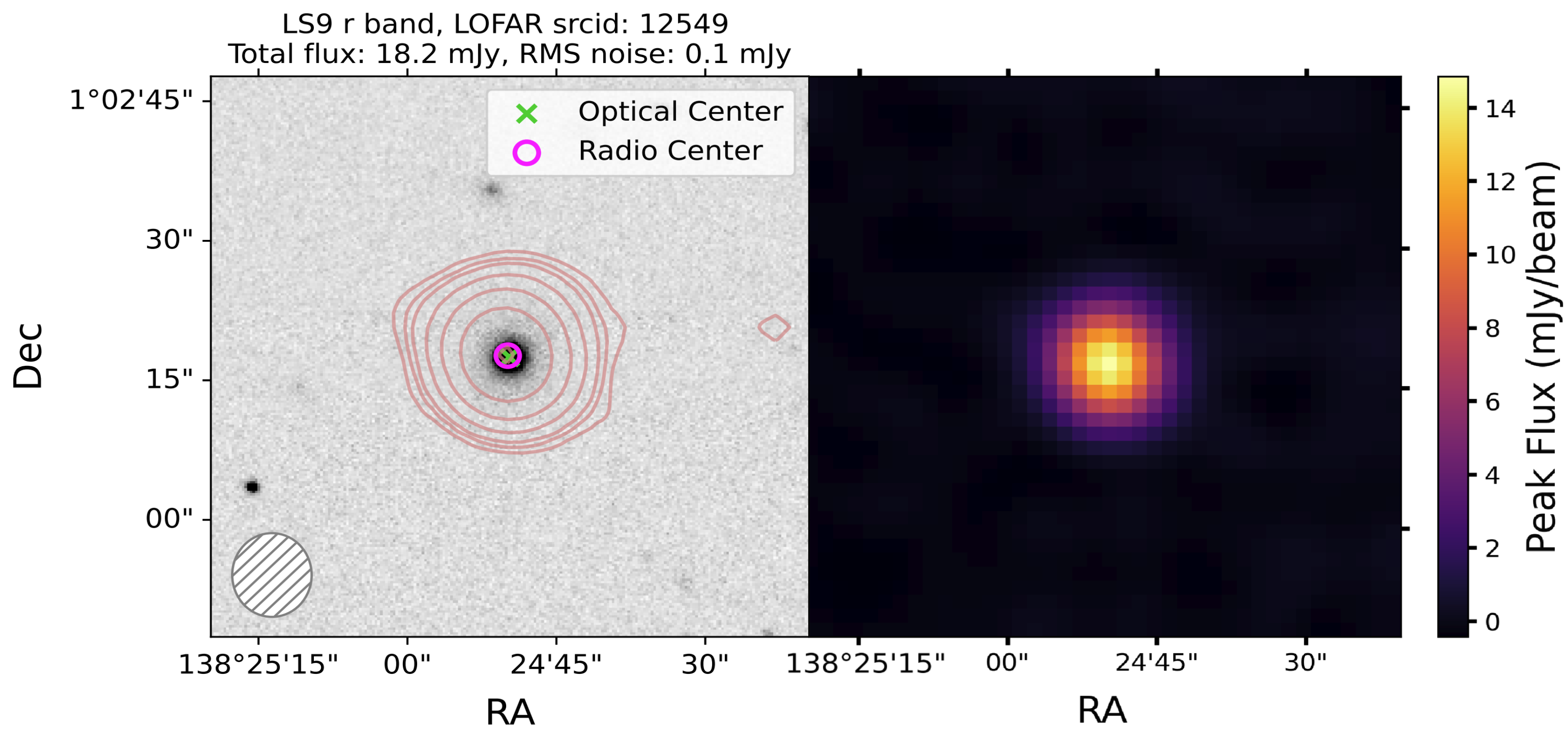}
\includegraphics[width=0.485\linewidth]{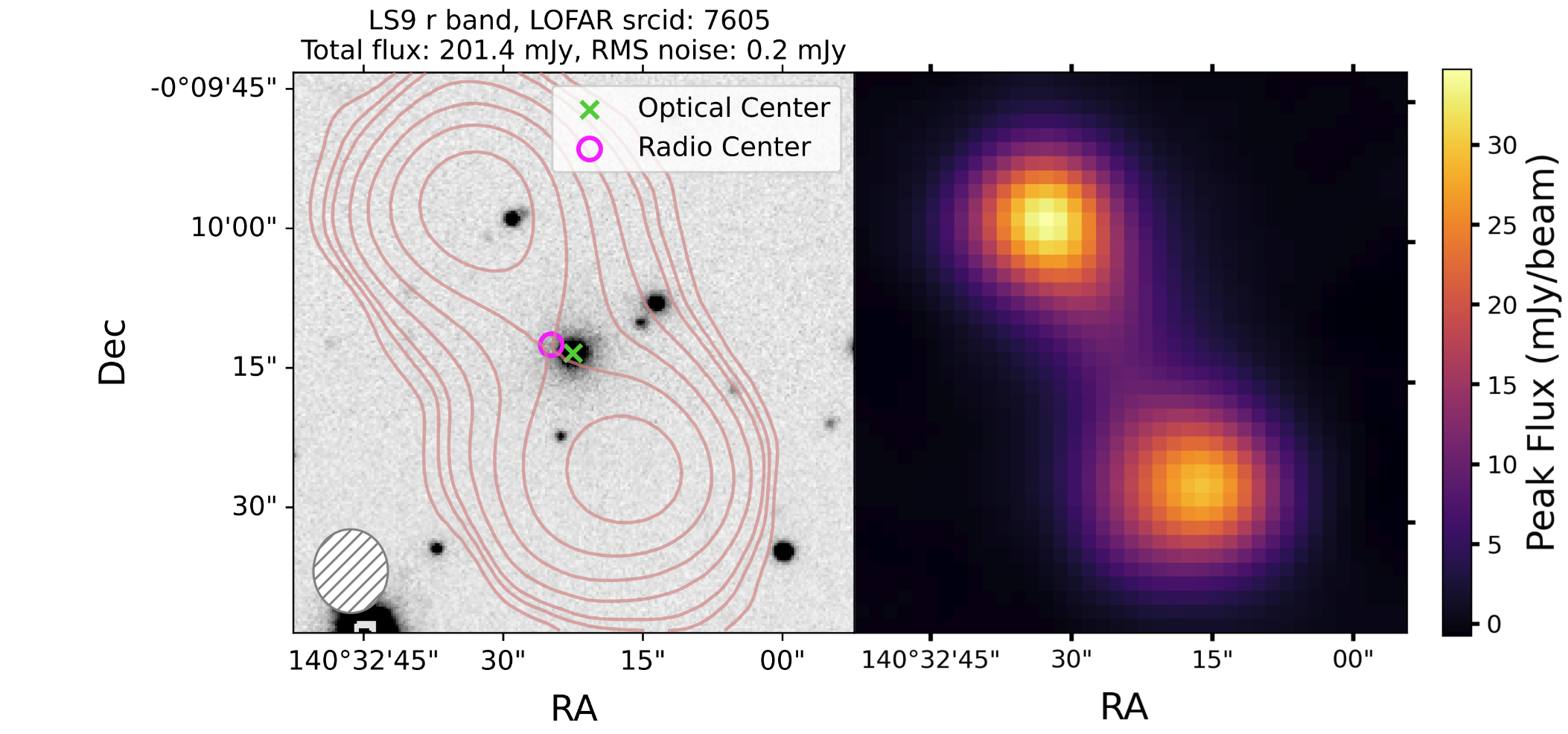}
\caption{Cutouts (60\arcsec\ $\times$ 60\arcsec) showing prototypical compact (left) and complex (right) radio morphologies with light red contours marking several factors of the local noise (at rms $\times$[2, 4, 6, 12, 24, 48]) and pink circle indicating the radio center. The beam size is depicted as a hatch-filled circle in the bottom left corner. In greyscale, the LS9 r-band image depicts the host galaxy, with its optical centre marked with a green cross. Radio intensity maps, with a colourbar indicating the peak flux, are also shown for these two examples.}
\label{fig:compact_complex_cutouts}
\end{figure*}

\subsubsection{Compact vs. complex radio morphology}
\label{sec:compactvscomplex}

Using the LOFAR catalogue described in Sect. \ref{sec:lofarintro}, consisting of 36,631 sources (in the eFEDS $t_{\exp}>500$s region), the first step is to distinguish between radio sources with different morphologies (compact versus complex/extended), as they may be governed by different physical process, either local to the source or on larger scales. Radio sources with markedly different morphologies may also require different cross-identification procedures, so they need to be classified first.

Following broadly \citet{Williams2019}, a set of four criteria must be fulfilled for a LOFAR source to be classified as `compact'. Firstly, we consider the fact that perfect `compact' (point-like, i.e. unresolved) sources have a ratio of the total integrated flux density to peak flux ($R=F_{\rm Tot}/F_{\rm Peak}$) equal to unity and reside completely within the size of the restoring beam \citep{Shimwell2022}. Considering that calibration is not perfect, we fit a Gaussian to the distribution of $F_{\rm Tot}/F_{\rm Peak}$ to determine the correct threshold to isolate compact galaxies; see Fig.~\ref{fig:flux_maj_cuts}, left. Compact sources are defined as those below the threshold marked by the vertical black dashed line at $R<3.6$ (8$\sigma$).

Secondly, compact emitters tend to have smaller sizes, measured for example by the full width half maximum (FWHM) of the major axis (\texttt{Maj}) of the source. The right panel of Fig.~\ref{fig:flux_maj_cuts} shows the distribution of major axes fit by a Gaussian for the LOFAR sample. Compact sources are defined as those below the threshold of \texttt{Maj}$<19.1\arcsec$ (6$\sigma$; vertical black dashed line).  

The third criterion is that a compact source must be fit with only a single Gaussian by PyBDSF, that is \texttt{S\_Code}$=$S. This will exclude those sources fit with multiple Gaussians, \texttt{S\_Code}$=$M, or sources fit with a single Gaussian but being located in the same island as other radio sources, \texttt{S\_Code}$=$C \citep{PyBDSF2015}.

Lastly, compact sources must be in an isolated region without any other catalogued LOFAR sources (no nearest neighbours) within 45\arcsec. This is to remove cases where far-away lobes/hotspots, associated to the same host galaxy, are catalogued as two different radio sources and are indeed not a compact emitter; or the case of dense cluster regions with multiple nearby radio emitters.

We consider that all four criteria have to be simultaneously fulfilled in order for a source to be considered compact, having \texttt{LOFAR\_compact\_flag} set to True in the catalogue (see Table \ref{table:catalogue_cols}). As a validation of this approach, Fig.~\ref{fig:shimwell_compact_cut} plots the signal to noise, defined in this case as total flux divided by the error on the total flux (note that in the rest of the work, $F_{\rm Peak}$ is used to calculate SNR), versus the natural logarithm of flux ratio, $\ln(R)$. Compact or unresolved sources are likely to lie under the black dashed line 99.9\% of the time \citep[see Eq. 2 and further discussion in][]{Shimwell2022}. This is indeed the case for the compact LOFAR-eFEDS sample we defined, but note that the inverse is not true and the curve cannot be used for selecting `complex' sources. Instead, we simply define as `complex' all those sources which do not satisfy at least one of our compactness criteria described above. 

\begin{figure*}[ht!]
\centering
\includegraphics[width=0.90\linewidth]{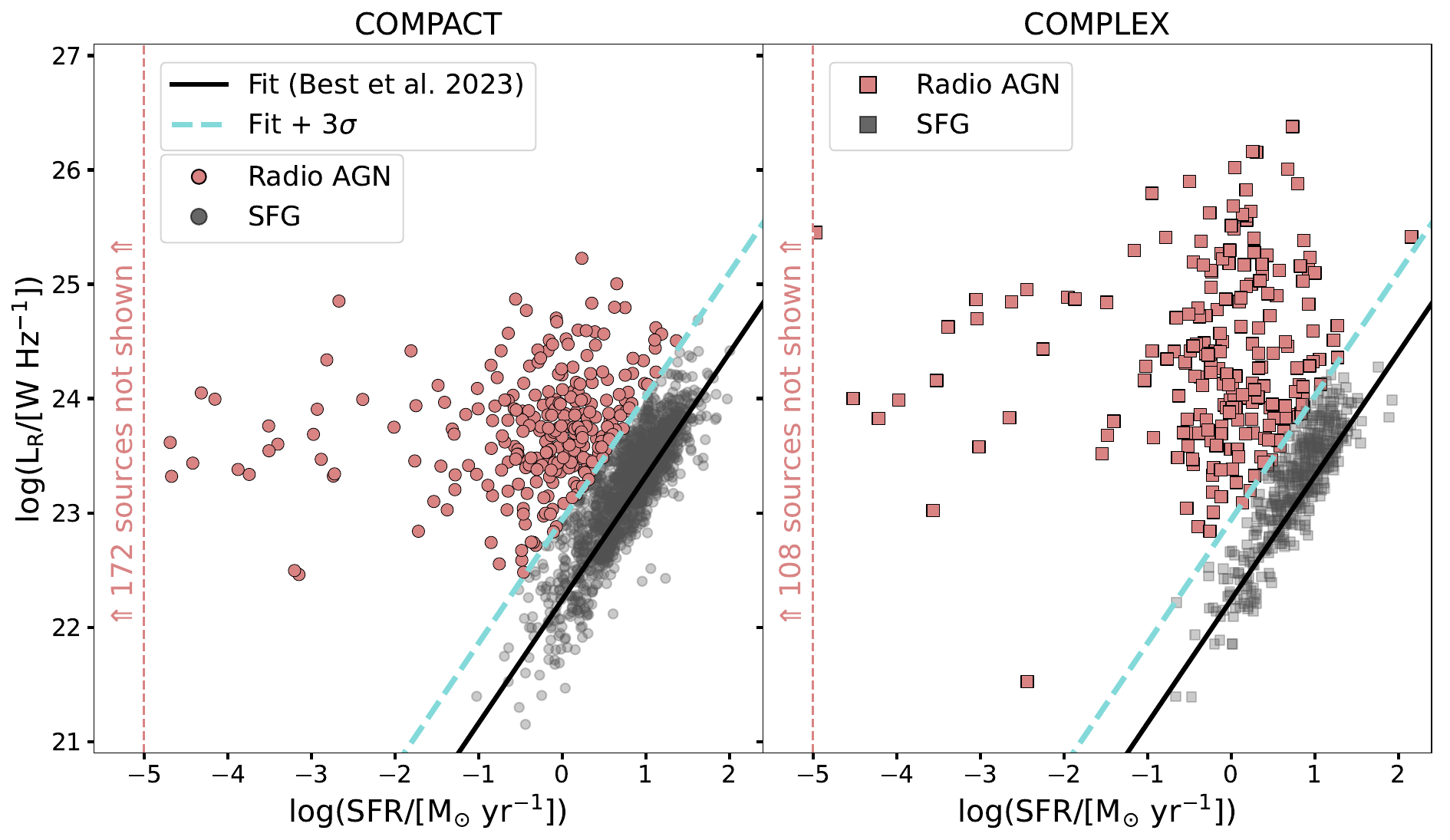}
\caption{SFR versus the radio luminosity of radio sources within the GAMA09 galaxy sample. The black solid line, taken from \citet{Best2023radioAGN}, describes the relation between SFR and $L_{\rm R}$ for star-forming galaxies hosting compact (left panel, grey circles) and complex (right panel, grey squares) radio sources. Sources lying $3\sigma$ above this relation (cyan, dashed line) form the sample of compact and complex radio AGN (light red circles, squares, respectively).}
\label{fig:radioAGNcut}
\end{figure*}

Overall, 24,613/36,631 (67\%) of the LOFAR sources are classified as `compact', and 12,018 as `complex' radio emitters. Fig.~\ref{fig:compact_complex_cutouts} shows two examples of a prototypical compact and complex source in our sample, where the LS9 one-band image is overlaid with radio contours spanning several factors of the local noise rms. 

Finally, all mass-complete sources used in the final incidence analysis are visually inspected to ensure the correct identification of the optical counterpart and of the radio morphology (marked by \texttt{vis\_inspected}$=$True). Among complex radio AGN, two common morphological classes are the FRI and FRII sources, which are powerful jetted AGN with core- and lobe-dominated emission, respectively \citep{FanaroffANDRiley1974}. During the visual inspection, radio AGN with FRII-like morphologies are identified, such that their incidence can be measured (see Sect. \ref{results:QLedd}). The identified secure and likely FRII sources in the GAMA09 field and their basic radio and optical properties (not guaranteed for completeness) are flagged as \texttt{FRII\_flag}$=$1 and $0.5$, respectively. Note that only those sources which can visually be resolved into edge-brightened, double lobed components at the 8\arcsec\ $\times$ 9\arcsec\ resolution of LOFAR are considered FRIIs here. Further details of the visual inspection process and results are presented in Appendix \ref{appendix:nwaylofarlegacy_visinspect}.

\subsubsection{Optical counterparts to the radio sources}
\label{sec:nwaylofarlegacy_radio}

This section summaries the procedure to find the optical counterparts to the LOFAR sources (see Appendix \ref{appendix:nwaylofarlegacy_visinspect} for details). 

Out of the total 36,631 LOFAR sources, 33,769 matched to an LS9 galaxy, with a maximum 8\arcsec\ search radius in NWAY. Magnitude priors on g, r, z and W1 were included to resolve counterpart ambiguity, since the optical counterparts of radio emitters tend to be found in redder galaxies \citep[e.g. ellipticals; see][]{Williams2019}.

Using the `optimal' \texttt{p\_any} $=0.06$ (see Fig.~\ref{fig:pany_threshold}) threshold to remove statistically unlikely matches, left 25,806 radio sources with reliable counterparts in LS9. Then, a cut of signal to noise SNR $>5$, defined as the ratio of the peak radio flux to the error in the peak flux, resulted in 22,754 matches between the LOFAR and LS9 catalogues. A Rayleigh distribution with $\sigma_{R}$ approximately equal to one validates the matching procedure (see Fig.~\ref{fig:rayleigh}). As with the X-ray catalogue, the LOFAR detections with an LS9 optical counterpart are matched to the GAMA09 galaxies using a simple 2\arcsec\ positional match. In addition to this, six large radio galaxies with a GAMA09 counterpart, are appended to the sample after a further visual inspection procedure (see Appendix \ref{appendix:nwaylofarlegacy_visinspect}). With this addition, there are in total 22,759 LOFAR detections with an LS9 optical counterpart and 2,619 radio sources among the GAMA09 galaxies (see Table \ref{tableOfsources}). From the latter subsample, three radio sources are classified as giant radio galaxies (GRGs; flagged with \texttt{GRG\_flag}) having largest linear sizes $>0.7$~Mpc \citep[e.g.][]{Saripalli2005}. Fig.~\ref{fig:src22763} shows an example of a GRG with a largest linear size $\sim$1.4~Mpc.

Of this overall total of 2,619 radio sources, 1,901 have compact morphology, 718 complex. In the following section, we further characterise these LOFAR-LS9-GAMA09 radio sources in terms of the origin of their radio emission and the properties of their host galaxies.




\begin{figure*}[ht!]
\centering
\includegraphics[width=1.07\linewidth]{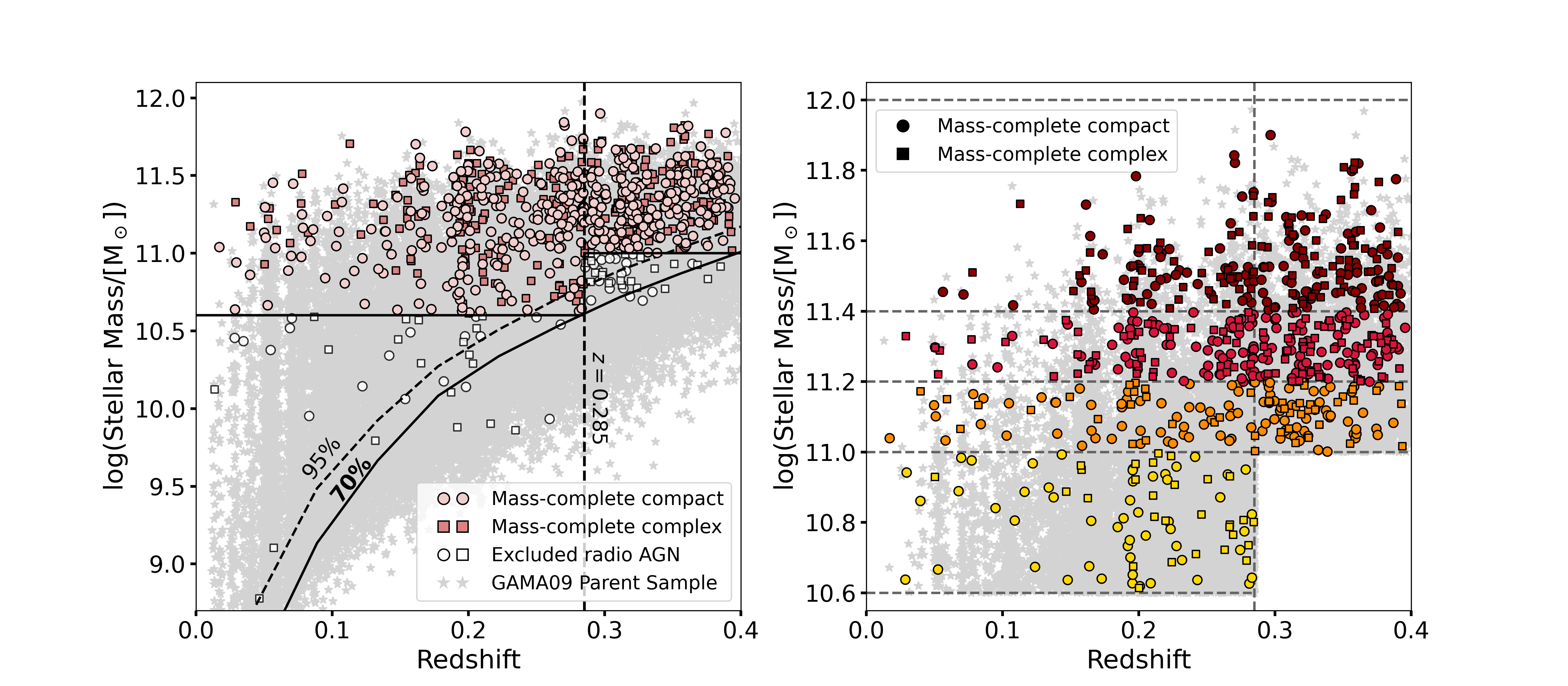}
\caption{Left: Stellar mass versus redshift distribution of the GAMA09 galaxies (grey points) and of the compact (light filled circles) and complex (dark filled squares) radio AGN. A vertical line divides the sample into two redshift bins. Completeness curves (70\%, 95\% with solid, dashed black lines, respectively) and horizontal thresholds are used to exclude sources incomplete in stellar mass (unfilled markers). Right: Zoom-in of the mass-complete sample, split into four stellar mass bins (yellow, orange, red, crimson).}
\label{fig:radio_zvsMstar}
\end{figure*}

\begin{figure}[h]
\centering
\includegraphics[width=0.95\linewidth]{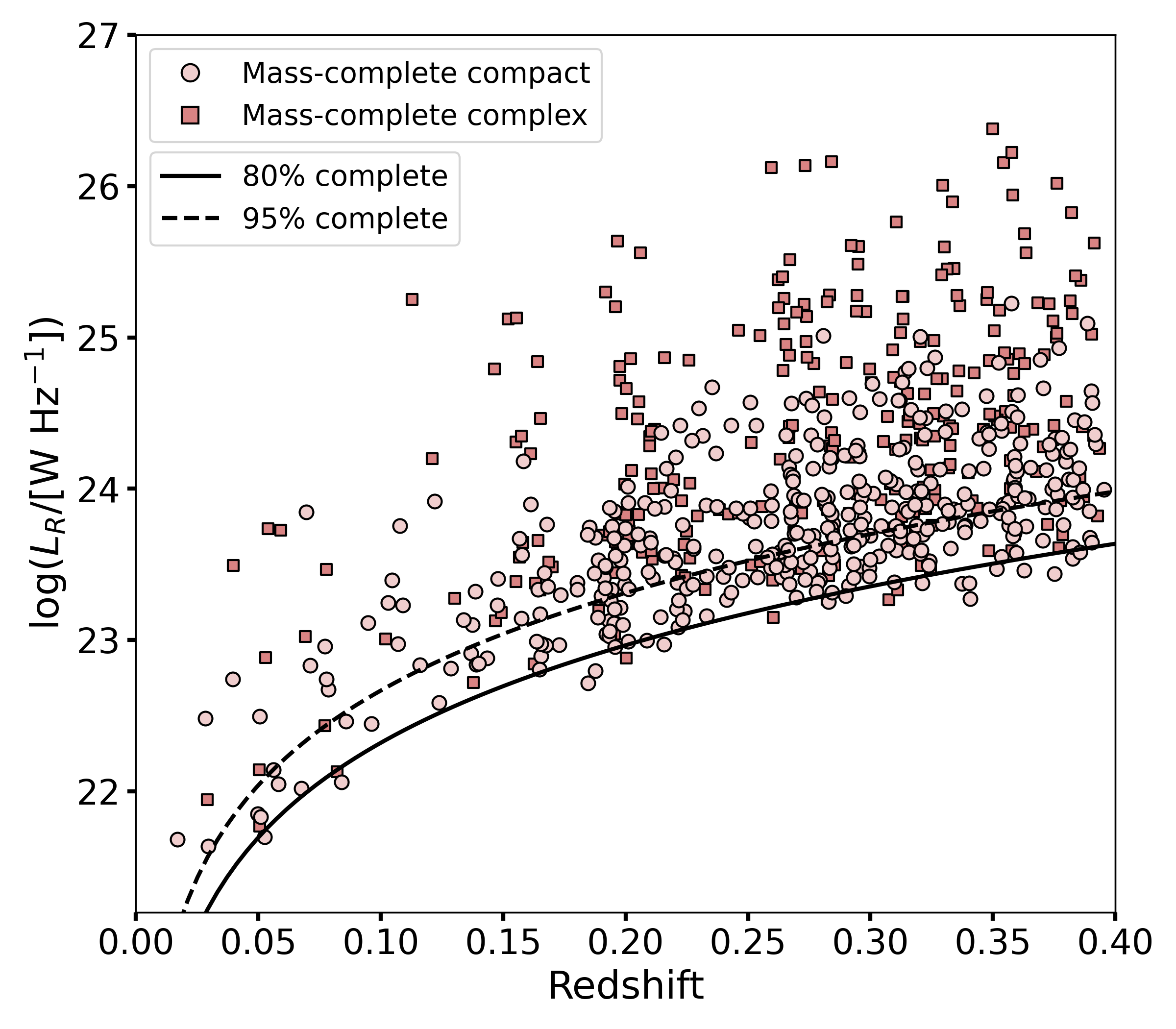}
\caption{Redshift versus radio luminosity distribution of the compact and complex G9 radio AGN in the two redshift bins (colours and symbols are as above). Black dashed and dotted curves show the 80\% and 95\% radio luminosity completeness limits, respectively.}
\label{fig:radio_zvsLr}
\end{figure}

\begin{figure}[h]
    \centering
    \includegraphics[width=0.95\linewidth]{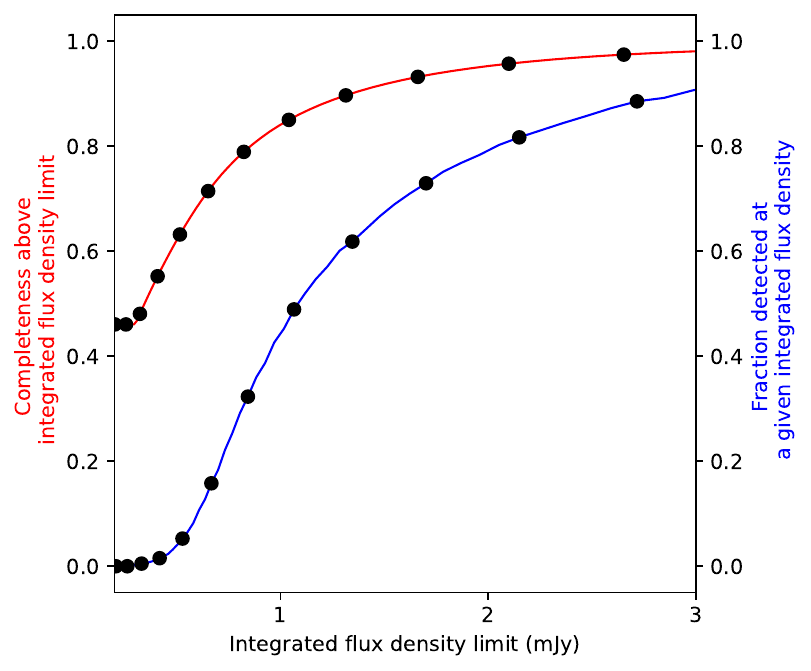}
    \caption{Point-source completeness functions for the LOFAR-eFEDS field, calculated by injecting simulated sources with a range of radio intensities onto the field's residual image. The red and blue lines show the cumulative completeness above and the fraction of detected sources at a given integrated flux density, respectively. The former is used to derive the luminosity completeness curves shown in orange on Fig.~\ref{fig:radio_zvsLr}.}
    \label{fig:lofarLRcompleteness}
\end{figure}

\subsubsection{Radio AGN vs. star-forming galaxies}
\label{sec:radioAGN}

Radio emission can have a variety of origins, including star formation, AGN radio jets, AGN wind interactions and coronal emission \citep{Panessa2019} and so it is vital for studies of radio AGN to be able to distinguish among these. 

Different methods to separate star-forming galaxies from radio AGN are widely discussed in the literature and have been refined significantly over the years with the advent of large surveys, for example radio SEDs and correlations with infrared parameters \citep{Gaby2017, Gurkan2018, Yun2001, Delvecchio2021}; brightness temperature \citep{Morabito2022}; using correlations between SFR (or proxies thereof, e.g. H$\alpha$) and radio emission to identify excess emission \citep{Smith2021, Best2005, Kauffmann2008}; emission line diagnostics, BPT diagrams \citep{Baldwin1981, Kewley2006}, or combinations of the above and other methods, as discussed in \citet{BestandHeckman2012, Sabater2019, Hardcastle2019}, and references therein.

The method used in this work takes advantage of the highly reliable FUV to FIR SED fitting of the GAMA sources (recall Sect. \ref{sec:GAMA}) to calculate the SFR of all GAMA09 galaxies, as well as the tight correlation between SFR and radio luminosity for star forming galaxies \citep{Condon1992, Smith2021, Best2023radioAGN, Heesen2023}. This relation is effectively able to trace recent star formation via synchrotron radiation emitted from massive stars ending their short lifetimes in supernovae explosions. Radio AGN can then be identified by measuring an excess with respect to the predicted SFR-related radio emission (`radio-excess AGN'). 

Fig.~\ref{fig:radioAGNcut} shows the SFR (in units of $M_{\odot}~{\rm yr}^{-1}$) plotted against the radio luminosity of the 1,901 compact (left panel) and 718 complex (right panel) radio sources detected by LOFAR and associated to a GAMA09 galaxy. Radio luminosity is calculated the standard way:
\begin{equation}
\label{eq_radiolum}
    L_{\rm 144MHz}~{\rm [W~Hz^{-1}]} = L_{\rm R}=4 \pi~d_L^2~F_{\rm Tot}~10^{-30}~(1+z)^{\alpha-1},
\end{equation}
where $d_L$ is the luminosity distance in cm, $F_{\rm Tot}$ is the total integrated flux\footnote{The following convention for the radio flux density as a function of frequency, $S_\nu$, is used here: $S_\nu \propto \nu^{-\alpha}$.} in units of Jansky (Jy) and $(1+z)^{\alpha-1}$ is the K-correction, with radio spectral index $\alpha=0.7$ \citep{Condon1992}. Note that $L_{\rm 144MHz}$ and $L_{\rm 150MHz}$ are used interchangeably. 

The black solid line is the best fit derived by \citet{Best2023radioAGN} using the LoTSS Deep Fields (accounting for non-detections such that the relation is not biased by radio imaging depth): 
\begin{equation}
    \label{eq_radioAGNcut}
    \log_{10}(L_{\rm R}/{\rm [W\, Hz^{-1}]}) = 22.24 + 1.08~\log_{10}({\rm SFR}/[M_{\odot}~{\rm yr^{-1}}]).
\end{equation}
This relation is fully consistent within 0.1~dex with other recent relations \citep[e.g.][]{Smith2021} and tracks well the star-forming cloud of objects shown in grey in Fig.~\ref{fig:radioAGNcut}. As the overall population of sources above and below the best fit line are asymmetric \citep[see Fig.~8 from][]{Best2023radioAGN}, the approximately Gaussian spread of the distribution of radio luminosities below Eq.~\ref{eq_radioAGNcut}, which has $\sigma=0.22$, is used to determine the cut for a source to be considered a radio AGN. All sources to the left of this cut, corresponding to 3$\sigma$ (0.7~dex) above the relation (dashed cyan line), are defined as radio AGN, as they have radio luminosities in excess of what is expected from pure star formation. The 172 compact and 108 complex sources with very low SFR ($\log({\rm SFR}/[M_{\odot}~{\rm yr^{-1}]\rm{)} < -5}$) are marked with text on the left hand side of each panel in Fig.~\ref{fig:radioAGNcut}.

Henceforth, only the radio AGN sample (light red points; marked in catalogue by \texttt{G9\_radioAGN}$=$True) will be considered in our subsequent analysis. From the total G9 radio sources, 445/1901 and 319/718 are compact and complex AGN, respectively. This sample has already been used in \citet{Popesso2023} to study the incidences of radio AGN in brightest cluster galaxies (BCGs). The fraction of compact/complex in the low (high) redshift bin is 221/139 (183/139), showing that there is no decreased detection in the fainter/extended complex sources, over the relatively small redshift range probed.

\begin{figure*}[ht!]
\centering
\includegraphics[width=0.85\linewidth]{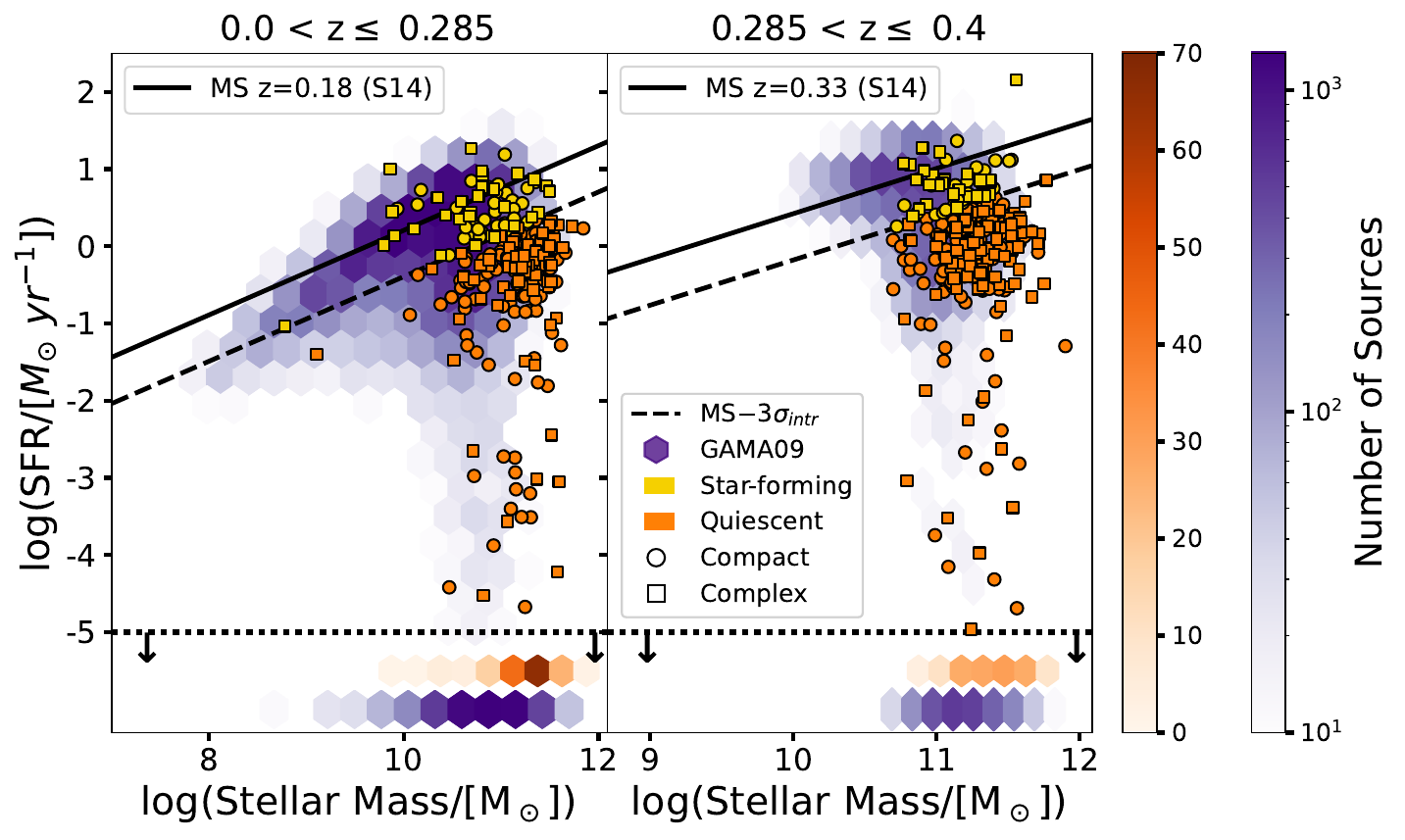}
\caption{Stellar mass versus SFR of the GAMA09 galaxy sample (purple hexbins) and the compact/complex (circles/squares) radio AGN. The solid black line in each redshift panel marks the star-forming galaxy main sequence from \citep[S14;][]{Speagle2014}. Sources $3\sigma$ below this line (black, dashed) are considered to be quiescent galaxies (orange), otherwise they are classed as star-forming (yellow). Quiescent sources below $\log \rm{(SFR)}=-5$ are marked as upper limits and their distributions are shown in the bottom of each panel.}
\label{fig:qvsSF_radio}
\end{figure*}

\subsubsection{Stellar mass and radio luminosity complete radio AGN samples}
\label{sec:masslumincomp_radio}

Fig.~\ref{fig:radio_zvsMstar} (left) shows the redshift versus stellar mass distribution of the GAMA09 parent sample (grey points) and of the radio AGN among them, defined in Sect. \ref{sec:radioAGN}, where the compact and complex radio emitters are marked with light circles and dark squares, respectively. The same stellar mass completeness curves, calculated as described in Sect. \ref{sec:mass_lumin_completeness} above, are shown in black, since the completeness is dictated by the underlying GAMA09 galaxy mass distribution. White-filled markers are the radio AGN which are excluded as a result of this cut. Overall, there are 682 radio AGN in the `mass-complete' (to 70\%) sample, of which 404 are compact and 278 are complex. This corresponds to a total fraction of GAMA09 galaxies hosting a radio AGN detected by LOFAR of about 3\% (682/21462).

Fig.~\ref{fig:radio_zvsLr} shows the radio luminosity distribution with respect to redshift of the mass-complete G9 radio AGN. The colours and symbols are as above. Black dashed and dotted lines show the 80\% and 95\% radio luminosity completeness thresholds, respectively. The generation of the completeness is similar to that described in \citet{Shimwell2019} (their Fig.~14 and Section 3.6). A residual image of the entire LOFAR-eFEDS field is generated using PyBDSF. Then, 45,000 sources with flux densities ranging from 0.1~mJy to 10~Jy are injected into the residual image (in the image, not u-v, plane). The injected sources are searched and counted with PyBDSF. The injection procedure is done 50 times to improve the injection/detection statistic. Fig.~\ref{fig:lofarLRcompleteness} shows that the point-source completeness depends on the integrated flux density of the injected sources. For instance, 50\% of the injected sources with flux densities above 0.34~mJy are detected. The completeness is 80\% for sources brighter than 0.85~mJy, and it increases to 95\% for sources with flux densities above 1.88~mJy. At the same completeness level, this LOFAR-eFEDS data requires sources to be a factor of $2-3$ brighter than those in the LoTSS-DR1 images of \citet{Shimwell2019} to be detected, mainly due to the higher noise level of this low declination field. 
In a similar way to the X-ray AGN, a weighting per bin is applied to account for radio luminosity dependent incompleteness (see details in Sect. \ref{sec:calc_incidences}).

The radio physical size is also calculated via Eq.~\ref{radiosizeEq}, to better classify the complex sample and comment on potential surface brightness limitations (see Sect. \ref{discussion}).
\begin{equation}
\label{radiosizeEq}
    R_{\mathrm {kpc}}= \theta * d_L/(1+z)^2,
\end{equation}
where $\theta$ is major axis in radians and $d_L$ is the luminosity distance in kpc.

\subsubsection{Host galaxy properties of radio AGN}
\label{sec:qvsSF}

In analogy with our analysis of the X-ray AGN sample, Fig.~\ref{fig:qvsSF_radio} shows the stellar mass versus SFR of the GAMA09 parent sample (purple hexbins), LOFAR radio AGN (yellow, orange; compact: circles, complex: squares) and the MS relation in black for the two redshift bins introduced in Sect. \ref{sec:mass_lumin_completeness}, along with a dashed line marking SFRs $3\sigma_{\rm intr}$ below the MS. The final numbers of sources are listed in Table \ref{tableOfsources}. All sources with $\log({\rm SFR})\leq-5$, lying below the black dotted line, are marked as upper limits. Their stellar mass distributions are plotted as orange (compact and complex radio AGN) and purple (GAMA09 galaxies) hexbins, shifted to an arbitrary low SFR value to avoid overlap. In the low (high) redshift bin, there are a total of 164 (116) and 5453 (2121) radio AGN and GAMA09 galaxies, respectively which have $\log({\rm SFR})\leq-5$. 

\begin{figure}[ht!]
    \centering
    \includegraphics[width=0.95\linewidth]{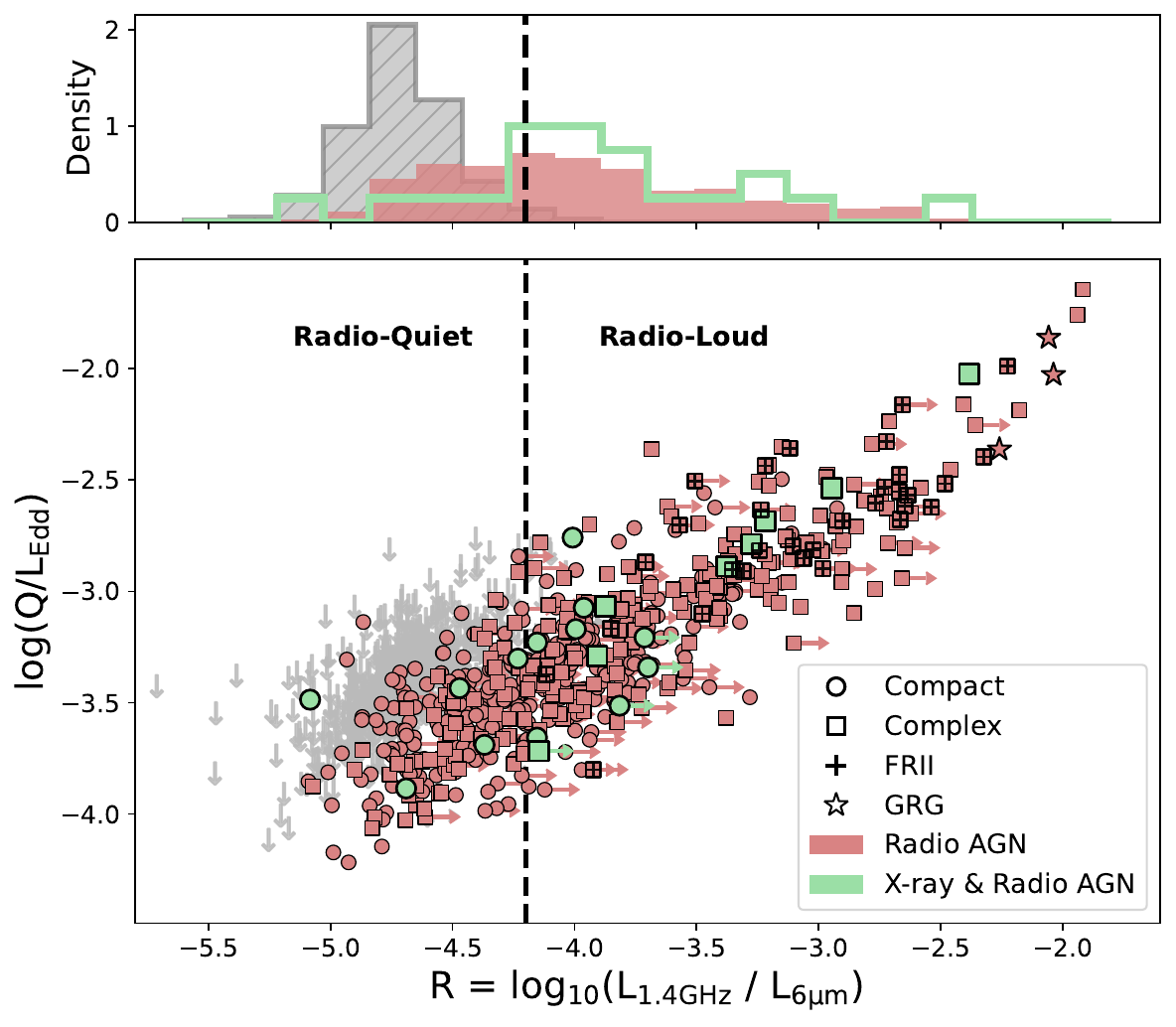}
    \caption{Radio-loudness, using 6$\mu$m luminosity as a proxy for accretion luminosity, plotted against $\lambda_{\rm{Jet}}$ for the mass-complete compact (circles) and complex (squares) radio AGN (light red), star-forming galaxies (grey upper limits), those radio AGN also detected in X-rays (green), and those which have secure FRII morphologies (black crosses) or are giant radio galaxies (stars). It can be seen that sources with  $\rm{log}(\lambda_{\rm{Jet}})\gtrsim -3.0$ are almost exclusively `radio-loud'.}
    \label{fig:radioloudness}
\end{figure}

\subsection{Combined X-ray and radio AGN sample characterization}
\label{sec:radioANDxray}

The LOFAR-LS9-GAMA09 (radio) and eFEDS-LS9-GAMA09 (X-ray) catalogues can be combined by matching the GAMA09 source IDs (or coordinates). This results in 121 sources emitting in both wavelength regimes (marked by \texttt{G9\_radioXray\_sources}$=$True), of which 74 and 32 are mass-complete X-ray and radio sources, respectively. However, only 24 are radio and X-ray AGN, by the criteria defined in the previous sections. 

For X-ray detected AGN, following \citet{Aird2012}, we adopt the following definition of the Eddington rate:
\begin{equation}
     \lambda_{\rm Edd}=\frac{L_{\rm Bol}}{L_{\rm Edd}}=\frac{25~L_{\rm X, Hard}}{1.26 \times 10^{38}~\rm{erg~s^{-1}}~(0.002~M_*)/M_{\odot}},
    \label{sBHAR_xray}
\end{equation}
where a simple bolometric correction factor of 25 is chosen to convert from hard ($2-10$~keV) X-ray luminosity to bolometric luminosity\footnote{Note that although bolometric correction factors depend on AGN luminosity, the chosen value of 25 agrees well with the range of 15-30 found in past studies on large samples for the luminosity range probed in this study \citep{Vasudevan2009}.} ($L_{\rm Bol}$). The Eddington luminosity ($L_{\rm Edd}$) is calculated using an estimate of the black hole mass as $M_{BH} \sim 0.002~M_*$, assuming the mass of the bulge is equal to $M_*$ \citep{Marconi2003}. The goal of $\lambda_{\rm Edd}$ is to serve as a mass-scaled scaled power indicator, not necessarily as the true `Eddington ratio' of the AGN, which is inherently difficult to constrain given the $\sim0.4$~dex systematic uncertainties on black hole mass measurements.

For radio AGN, the derivation of a mass-scaled jet power is more complicated. Two common methods to estimate jet power ($Q$) are either to calculate the work done by jets to inflate cavities in nearby cluster AGN using X-ray observations and combine with an estimate of the source age, or to infer it from correlations between narrow emission line luminosity and radio emission \citep{Willott1999, Hardcastle2007, MerloniHeinz2008, Cavagnolo2010, Daly2012, GodfreyShabala2013, HeckmanandBest2014, Ineson2017, Hardcastle2018}. Although there are several caveats that come with using such scaling relations (see Sect. \ref{sec:Qcaveats}), in this work we adopt the empirical relation from \citet{HeckmanandBest2014} (their Eq.~2) to define jet power, as it is based on a combination of the approaches mentioned above:
\begin{equation}
    Q = 2.8 \times 10^{37}~ \Bigg(\frac{L_{\rm 1.4GHz}}{10^{25}~{\rm W~Hz}^{-1}}\Bigg)^{0.68}~{\rm W},
    \label{Qeq}
\end{equation}
where the LOFAR 144MHz radio luminosity is converted to 1.4~GHz radio luminosity assuming a spectral index of $\alpha=0.7$. As can be seen, the equation is non-linear. Normalising by $L_{\rm{Edd}}$, we then define the specific black hole kinetic power:
\begin{equation}
    \lambda_{\rm Jet}=\frac{Q}{L_{\rm Edd}}.
    \label{eq_jetpower_Ledd}
\end{equation}

\begin{figure}[t!]
    \centering
    \includegraphics[width=0.95\linewidth]{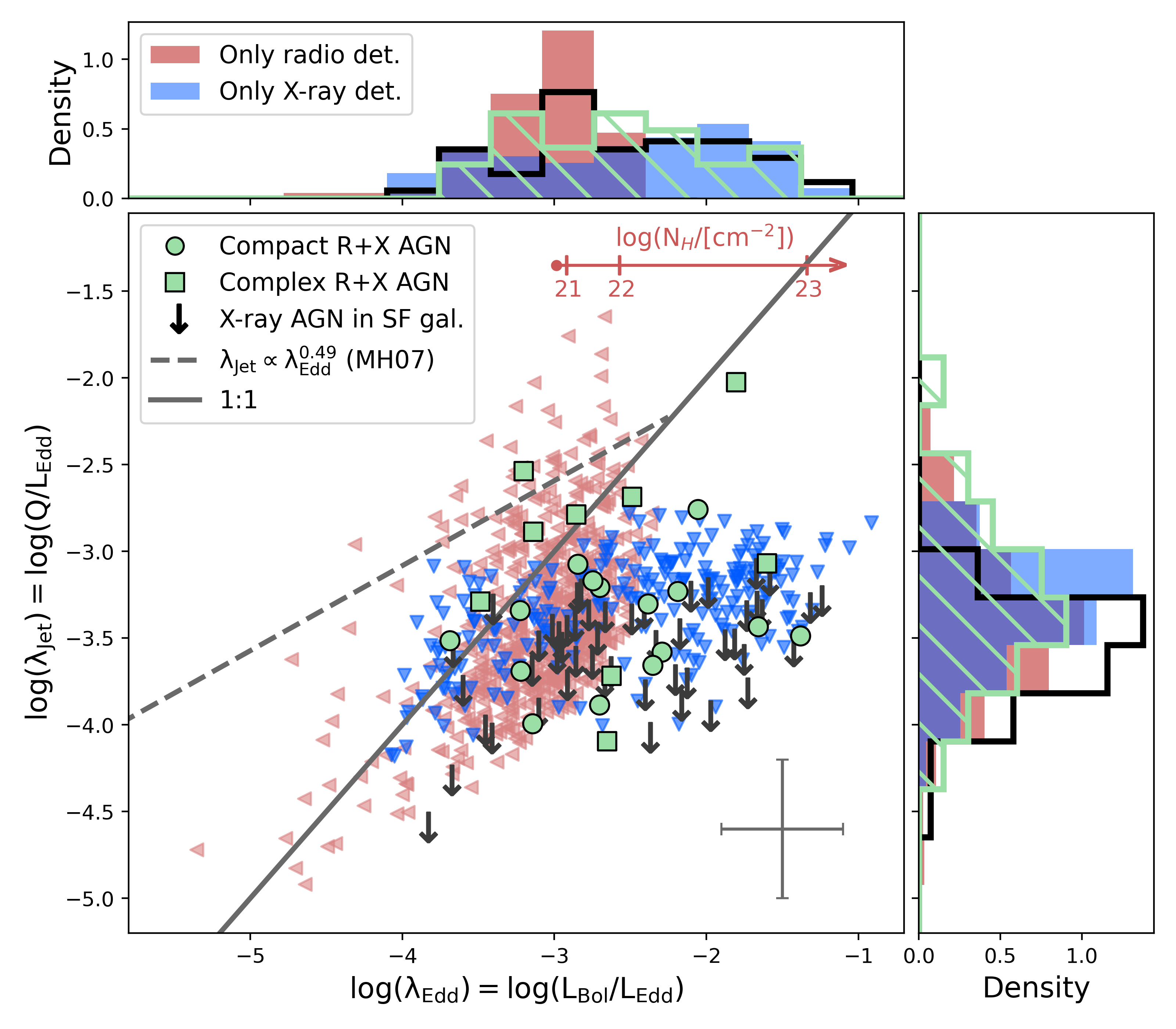}    
    \caption{The mass-complete distribution of $\lambda_{\rm Edd}$ versus $\lambda_{\rm Jet}$ for the radio and X-ray AGN (green), X-ray AGN in star-forming galaxies (black), only radio-detected AGN (light red) and only X-ray detected AGN (blue). Solid grey and dashed grey lines mark the 1:1 and $\lambda_{\rm Jet} \propto \lambda_{\rm{Edd}}^{0.49}$, respectively. Horizontal arrow marks the effect of obscuration on only radio-detected AGN (see text).}
    \label{fig:balanceOfpower}
\end{figure}

Fig.~\ref{fig:radioloudness} compares $\lambda_{\rm Jet}$ with the well-known measure of `radio-loudness' ($R$), a measure of the dominance of the radio emission over mid-IR emission, with 6$\mu$m luminosity ($L_{6\mu m}$), assumed to be a direct tracer of the reprocessed primary emission from accretion processes (in this case). $L_{6\mu m}$ is calculated via a log-linear interpolation (or extrapolation) of WISE fluxes and the threshold for a source to be considered `radio-loud' is $R>-4.2$ \citep{Klindt2019}. Only mass-complete sources and those passing the 80\% radio luminosity completeness curve (see Sect. \ref{sec:masslumincomp_radio}) are plotted on Fig.~\ref{fig:radioloudness} (552/682 radio AGN; 21/24 radio and X-ray AGN). Sources lacking good (any) WISE data are marked as lower limits in their radio-loudness, where $L_{6\mu m}$ is calculated from the WISE 5$\sigma$ point source sensitivities \citep{Wright2010}. Star-forming galaxies are marked as upper limits in $Q/L_{\rm Edd}$ as their possible AGN emission is indistinguishable from their star-formation emission. 
One can clearly see the relatively tight correlation, as expected for higher kinetic power objects to be more `radio-loud'. Partial correlation analysis reveals a strong positive correlation (Pearson coefficient, $r$, of 0.911) when controlling for stellar mass as a covariate, although the correlation becomes weaker ($r=0.428$) when controlling for radio luminosity, which make sense as it is a common variable in both axes. Overall, 60\% (47\%, 77\%) of the total (compact, complex) radio AGN sample are radio-loud (light red circles and squares), compared to only 2\% for the star-forming galaxies (grey upper limits). Around 76\% (61\%, 100\%) of the total (compact, complex) radio AGN also detected in X-rays (green circles and squares) are radio loud. FRII-like morphologies and giant radio galaxies, marked with black crosses and stars, respectively, tend to populate the high radio-loudness regime, in line with their expected powerful jets.

Fig.~\ref{fig:balanceOfpower} plots the radiative versus kinetic power for all the radio and X-ray AGN detected among GAMA galaxies. Different samples of sources are shown, namely radio and X-ray AGN (green), X-ray AGN in star-forming galaxies (black upper limits), X-ray AGN with no radio detections (blue), and radio AGN with no X-ray detection (light red). All samples are complete for stellar mass. For the radio-undetected sources, the 99\% flux limit (3.2~mJy) from Fig.~\ref{fig:lofarLRcompleteness} is adopted. For the X-ray-undetected sources the eFEDS survey-average $0.5-2$~keV flux 80\% completeness limit equal to $6.5\times 10^{-15}$~erg~cm$^{-2}$ from \citet{Brunner2022} is converted to a $2-10$~keV luminosity using a $\Gamma=2$ at the redshift of the source. A sample representative error of 0.4~dex in both $x-y$ variables is plotted in the bottom right corner (although the uncertainty on $Q$ may be larger, see Sect. \ref{sec:Qcaveats}). 
The grey solid line marks the 1:1 relation, whereas the grey dashed line is adapted from Eq.~3 of \citet{MerloniHeinz2007} and describes the radiatively inefficient ADAF mode where $\lambda_{\rm Jet}~\propto~\lambda_{\rm Edd}^{0.49}$ (with intrinsic scatter of 0.39), as observed for local radio galaxies in groups and clusters. It can be seen that the loci of all samples lie in the region where the two lines start to diverge, making any statements about the effect of different accretion modes on jet power in this work unfeasible. In fact, the bulk of the sources, especially the ones detected only in X-rays, do not populate the $\lambda_{\rm Jet} \propto \lambda_{\rm Edd}^{0.49}$ (radiatively inefficient, kinetically dominated) branch, where the fundamental plane is supposed to be valid. The eFEDS observations are not sensitive enough to probe the low power population, where an accretion mode transition would be more obvious; at the same time, the survey volume is too small to detect many high power sources.

In general, the radio-detected population scatters around $\lambda_{\rm Jet} \sim \lambda_{\rm {Edd}}$ line, whereas the X-ray detected one has $\lambda_{\rm Jet} \ll \lambda_{\rm {Edd}}$, as expected from kinetically and radiatively dominant accretion modes, respectively. There is, however, a hint that X-ray AGN that are radio-loud and radiatively efficient (upper right region of Fig.~\ref{fig:balanceOfpower}) may appear as a distinct population from their radio-quiet counterparts at the same $\lambda_{\rm Edd}$ (see also \citet{Ichikawa2023} for a discussion of the balance of power in higher redshift radio and X-ray detected AGN).  

The G9 radio AGN sample also contains some radio-detected sources with $\lambda_{\rm Jet} \gg \lambda_{\rm {Edd}}$ at high intrinsic jet kinetic power. However, the location of these sources could be affected by X-ray obscuration that has not been accounted for in the $\lambda_{\rm Edd}$ estimate. In Fig.~\ref{fig:balanceOfpower}, a horizontal arrow shows the effect that different levels of $\log(N_H/[{\rm cm}^{-2}])=21, 22, 23$ have on a source with flux equal to the eFEDS 80\% limit, an average redshift of 0.24 and an average $\log(M_*/M_{\odot})=11$. Obscured AGN (undetected by eROSITA) are intrinsically more luminous; high levels of obscuration would shift radio-detected sources toward the 1:1 line. 

\section{Calculating the incidence of AGN among GAMA09 galaxies}
\label{sec:calc_incidences}

In this section we present the methodology adopted to calculate AGN incidence as a function of stellar mass and specific black hole kinetic (from radio) or radiative (from X-rays) power, along with extra corrections accounting for completeness.

\subsection{Stellar mass -- redshift binning}
\label{sec:mstarzbinning}

Firstly, as shown on Fig.~\ref{fig:radio_zvsMstar} (left), the radio data is split into two redshift bins, to limit evolutionary or redshift dependent completeness effects on the analysis. They are determined by dividing the sample of sources equally in two: (i) $0<z\leq0.285$; (ii) $0.285<z\leq0.4$. The same two redshift bins are then adopted for the X-ray analysis as shown on Fig.~\ref{fig:xray_zvsMstar} (left), for consistency, and are represented, where relevant, with light grey and black colours throughout the paper.  

Secondly, four stellar mass bins are introduced, as shown on the right panels of Figs.~\ref{fig:xray_zvsMstar} and \ref{fig:radio_zvsMstar}, in ranges of $\log(M_*/M_{\odot})$: \textit{(i)} $10.6-11.0$; \textit{(ii)} $11.0-11.2$; \textit{(iii)} $11.2-11.4$; \textit{(iv)} $11.4-12.0$. These are chosen to achieve an optimal splitting of the parameter space whilst keeping the bin sizes larger than the average error on stellar masses calculated by GAMA (around 0.1~dex).

\begin{figure}[t!]
\centering
\includegraphics[width=1.1\linewidth]{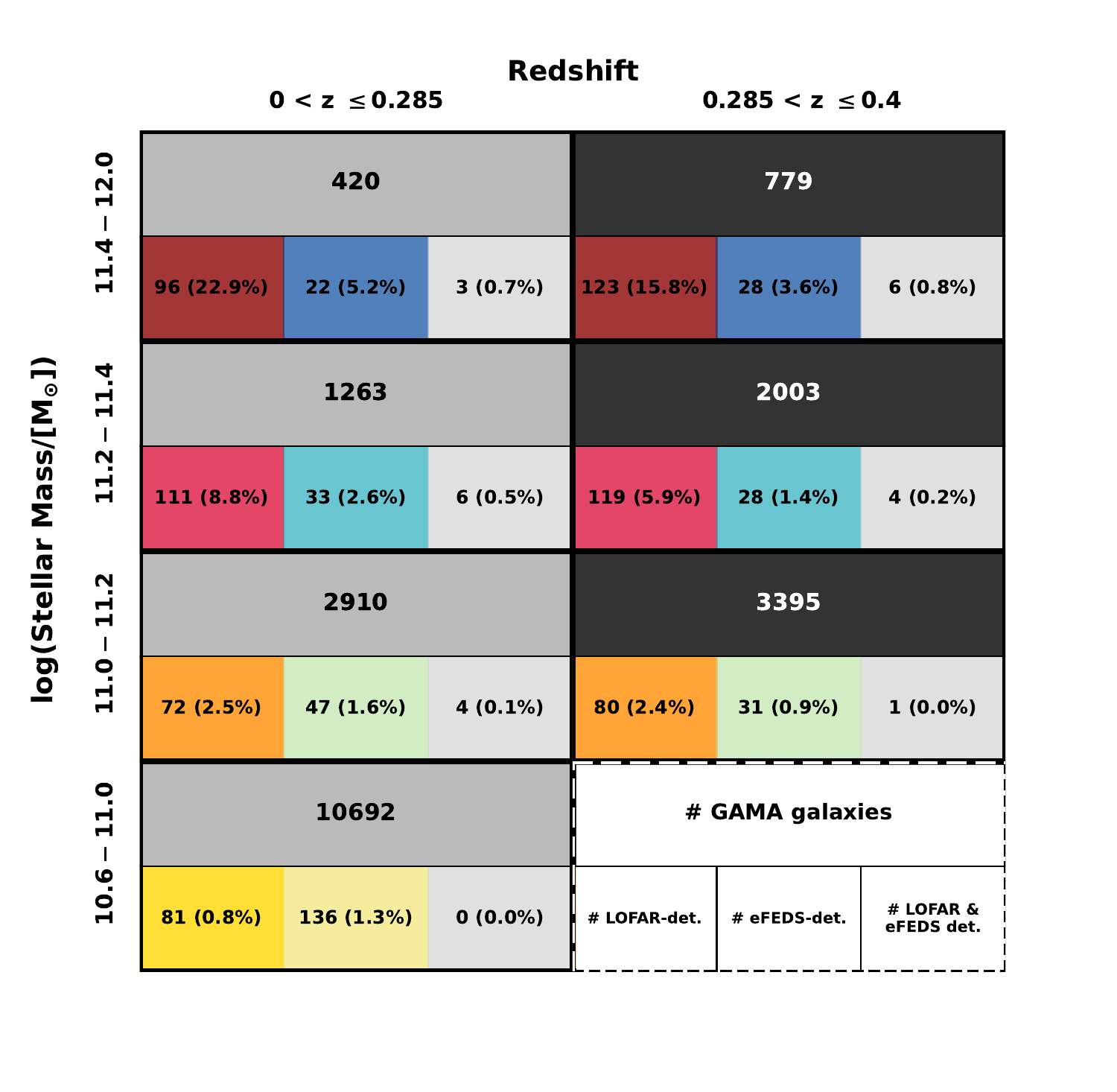}
\vspace{-1cm}
\caption{Infographic showing the incidence of radio AGN, X-ray AGN and AGN detected at both wavelengths, in different stellar mass--redshift bins (colours are as defined above). The legend is in the bottom right (e.g. there are 10692 GAMA09 galaxies in the lowest stellar mass bin, of which 81, 0.8\%, are LOFAR detected, etc.). Note that the AGN detected in both radio and X-ray are a subset of the individual pure radio and pure X-ray detected numbers.}
\label{fig:venn}
\end{figure}

These redshift and stellar mass bins are used throughout to combine both the mass-complete G9 radio and X-ray sources, as well as the GAMA09 galaxies themselves (serving as the parent sample). A summary infographic is shown in Fig.~\ref{fig:venn}, showing the numbers and incidences of X-ray AGN, radio AGN and AGN detected at both wavelengths, in the different $M_* - z$ bins. 

\subsection{Measuring AGN incidences as a function of mass-scaled power indicators}

In Section~\ref{sec:results} we will present a detailed analysis of AGN (both X-ray and radio selected) for the different stellar mass and redshift bins introduced above, as a function of mass-scaled AGN power indicators, which are derived from X-ray or radio luminosity over stellar mass ($L_{\rm X}/M_*$, $L_{\rm R}/M_*$). These measurable quantities can further be used as proxies of the fundamental dimensionless power rates: the specific black hole radiative power ($\lambda_{\rm Edd}$) and the specific black hole kinetic power ($\lambda_{\rm Jet}$). The analysis of the incidence of radio AGN as a function of mass-scaled jet power indicators is presented here for the first time.
   
Normalising the luminosity by stellar mass is important as it unmasks correlations with respect to the underlying radiative and kinetic power output distribution. For example, \citet{Aird2012, Bongiorno2012,Georgakakis2017, Birchall2022} and others show that the increasing fraction of X-ray detected AGN with stellar mass is just a selection effect of magnitude-limited surveys being able to detect objects down to lower accretion rates at higher mass for the same luminosity. In other words, looking at the incidence of AGN as a function of just luminosity is degenerate to the high accretion rate, small mass black holes and low accretion rate, large mass black holes.

The method to calculate the incidence of AGN (valid for all target samples: radio-only, X-ray only, or radio and X-ray samples) is to estimate the confidence intervals on binomial population proportions using Beta distributions \citep{Cameron2011}. In essence, this returns a measure of the fraction of target objects compared to GAMA09 parent galaxies in each given bin. The errors on these values are denoted by the 16th and 84th percentiles (1$\sigma$) of the distribution. This method is favoured over others, including for example that of \citet{Gehrels1986}, as the Poisson error on population proportions is systematically underestimated for small samples or large samples with extreme population proportions (either very low or very high detection fractions). As seen in Table \ref{tableOfsources} and Fig.~\ref{fig:venn}, the GAMA09 sample is large, yet the X-ray and radio detections, especially when split into different bins, are sometimes orders of magnitude less, necessitating the use of \citet{Cameron2011} confidence intervals. Moreover, the effects that the SF properties of the host galaxy can have on the incidences are examined by splitting up the sample (see Sect. \ref{sec:results_incRadio}), thus reducing further the statistics. This is also done for the compact and complex radio morphologies. 

Once the fraction of galaxies hosting the given target sample of AGN as a function of the different (mass-scaled) parameters have been calculated, a power-law is fit to the data in the form of $y=A \times (10^{x-x_0})^B$, using UltraNest\footnote{\url{https://johannesbuchner.github.io/UltraNest/example-line.html}} \citep{UltraNest2021}. A power-law slope ($B$) and normalisation ($A$) at a given (log) x-axis value ($x_0$) can then be obtained and compared across the different stellar mass and redshift bins in order to extract trends.

\begin{figure*}[ht!]
\centering
\includegraphics[width=0.85\linewidth]{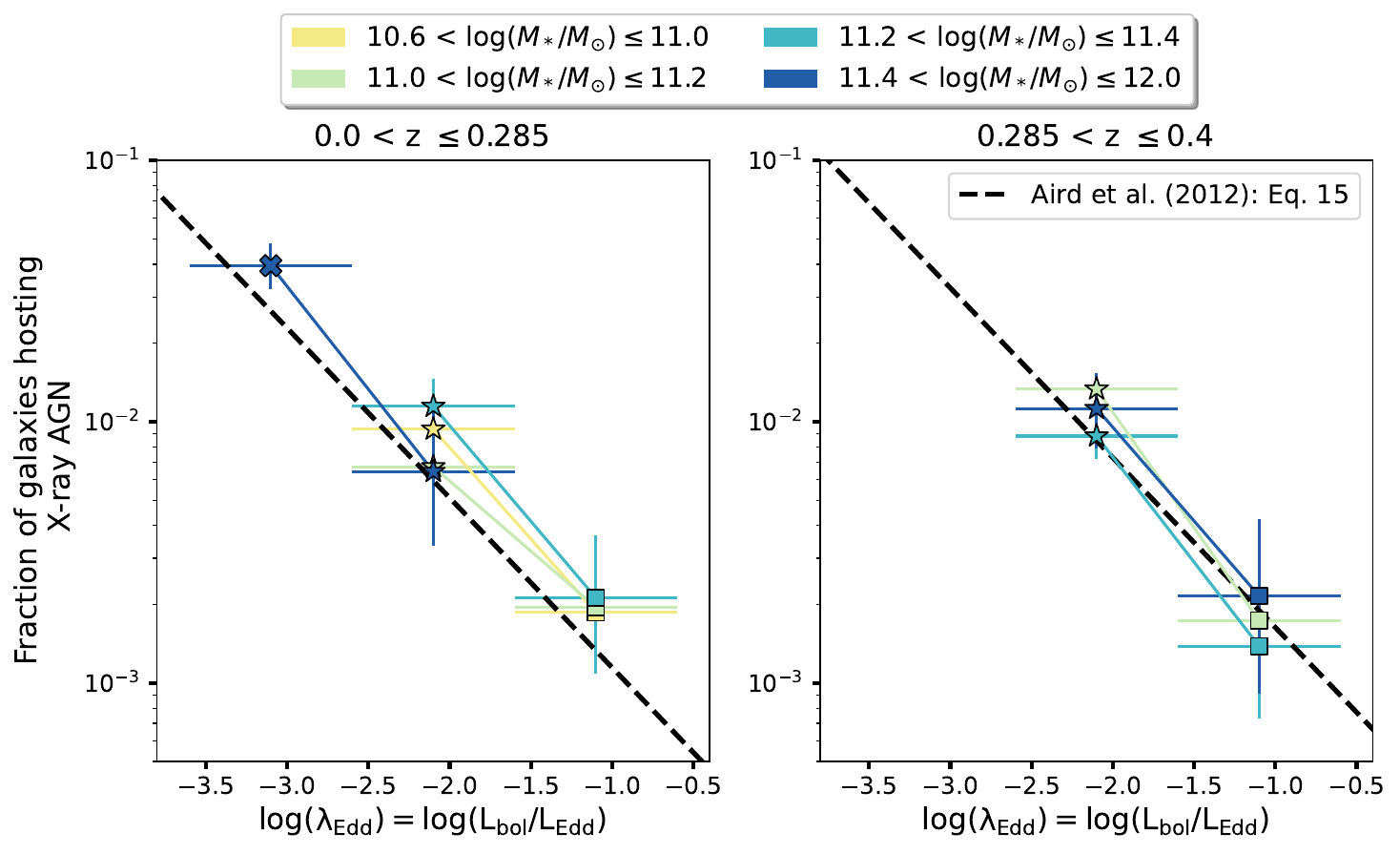}
\caption{Fraction of GAMA09 galaxies hosting eROSITA-eFEDS detected X-ray AGN as a function of $\lambda_{\rm{Edd}}$ in different stellar mass (yellow, green, teal, blue) and redshift (two panels) bins. The results agree well with those of \citet{Aird2012} (black dashed lines), corroborating the idea that there is a mass-invariant triggering and fuelling mechanism at play in X-ray AGN.} 
\label{fig:xrayfracs}
\end{figure*}

\begin{figure*}[ht!]
\centering
\includegraphics[width=0.9\linewidth]{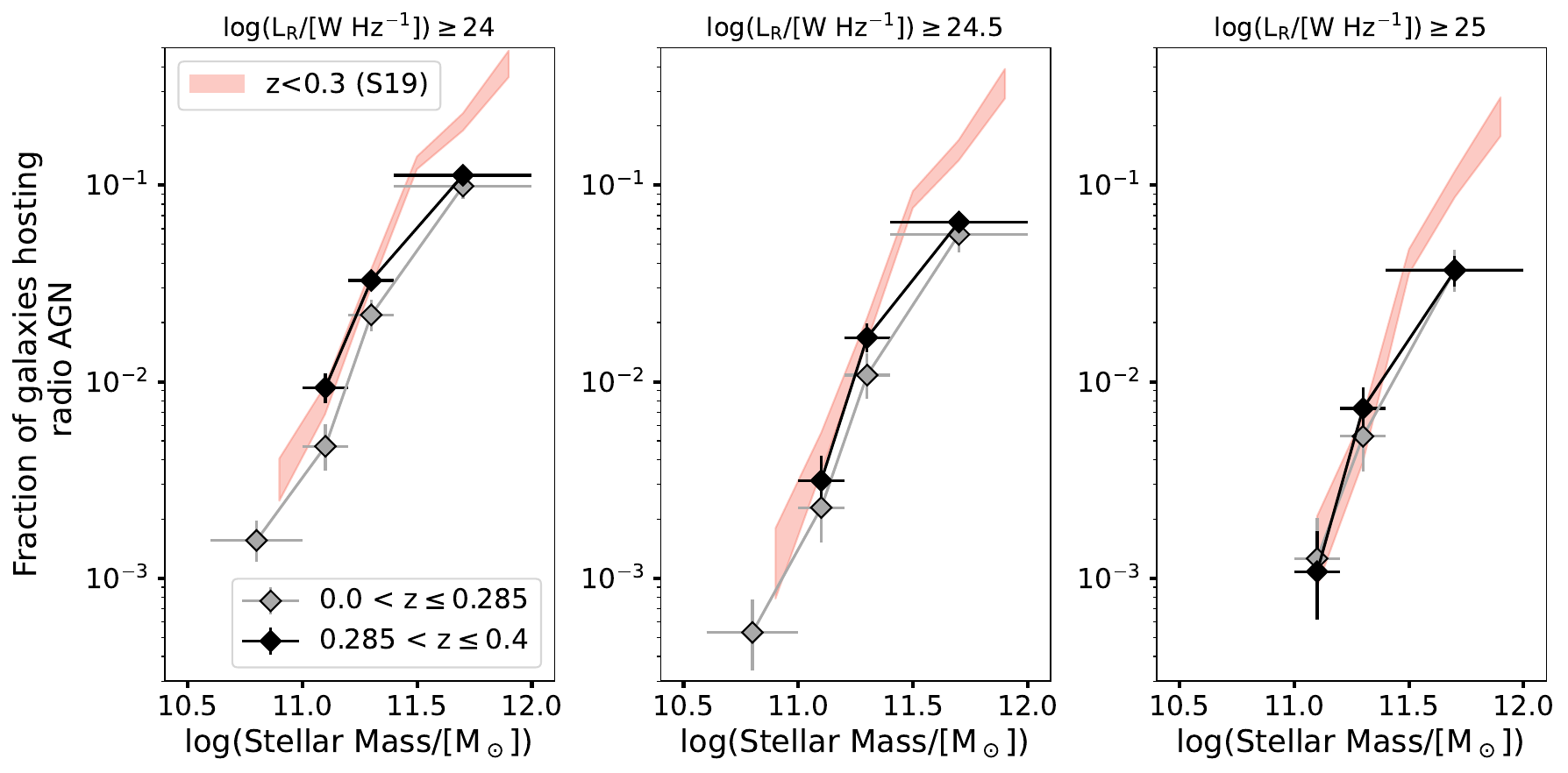}
\caption{Fraction of GAMA09 galaxies hosting both complex and compact radio AGN as a function of stellar mass, in different redshift (purple and green) and luminosity (panels) bins. A strong increase in the fraction of detected radio AGN with increasing stellar mass is observed. Orange shaded curves mark the results from Fig.~5 of \citet{Sabater2019} in the redshift range $0<z<0.3$.}
\label{fig:sabaterfracs}
\end{figure*}

\subsection{Accounting for radio and X-ray luminosity incompleteness}

Using the information regarding the flux sensitivity of the eROSITA and LOFAR instruments observing the GAMA09 field (see Figs.~\ref{fig:xray_zvsLx} and \ref{fig:radio_zvsLr}), it is possible to apply a correction to the incidence in the bins which are not fully complete in luminosity. Note that a weighting per bin, instead of per source, is the appropriate method here as the incompleteness is a result of the survey limitations, not of the sources themselves. 

For example, for a given $L_{\rm R}-z$ bin, one can calculate the median luminosity, convert it back to an observed radio total integrated flux and interpolate to find the survey sensitivity at that flux level. The incidence in that bin would then be weighted by a factor of 1/sensitivity (for any sensitivity greater than 50\%, otherwise it is considered incomplete and removed). Similarly, for every $L_{X}-z$ bin, the sensitivity can be directly interpolated from Fig.~\ref{fig:xray_zvsLx}, given the median luminosity and redshift. Note that these corrections mainly affect the low $\lambda_{\rm{Jet, Edd}}$ and the low $M_*$ bins at higher redshift. 

Appendix \ref{missedhighSFRradioAGN} discusses the correction applied to account for the potential missed radio AGN in highly star-forming galaxies, resulting from the 3$\sigma$ cut in Fig.~\ref{fig:radioAGNcut} preferentially removing higher mass galaxies. Note that this correction only affects the lowest $\lambda_{\rm Jet}$ sources (crosses on Fig. \ref{fig:epsilon_graph} below). Faint X-ray AGN in highly star-forming galaxies may equally be missed (recall the selection in Fig.~\ref{fig:xrbhotgas}). However, as shown in \citet{Merloni2016}, the X-ray emission for a typical $10^8~M_{\odot}$ AGN in $10^{10.5}~M_{\odot}$ main sequence star-forming host dominates over star-formation for $\lambda_{\rm Edd}>10^{-5}$. The G9 X-ray AGN do not extend to such low $\lambda_{\rm Edd}$ and so any incompleteness from this effect would be negligible in the context of this work.

\section{Results}
\label{sec:results}

The overall AGN sample statistics given in Fig.~\ref{fig:venn} show that 3\% and 1.5\% of GAMA09 galaxies are detected as radio and X-ray AGN, respectively. Taking only the mass-complete samples, 7\% of X-ray AGN are also radio AGN, in line with the commonly expected population of `radio loud' QSOs. Yet only 4\% of LOFAR-detected radio AGN are X-ray detected.

As host galaxy stellar mass increases, it gets increasingly likely to host both radio and X-ray AGN \citep[as found in, e.g.][]{Best2005, Smolcic2009, Brusa2009}. However, at $11.4<\log(M_*/[M_\odot])\leq12$, there is a factor $\sim$4 higher probability to host a radio AGN compared to an X-ray AGN, in both the low and high redshift bins. 

At face value, Figures \ref{fig:qvsSF_xray} and \ref{fig:qvsSF_radio} show that radio AGN tend to lie mostly (87\%) in quiescent galaxies, in contrast to X-ray AGN which are found in star-forming galaxies 62\% of the time. However, in Sect. \ref{sec:results_incRadio} we will show, albeit with limited statistics, that quiescent and star-forming radio AGN, once completeness has been accounted for, have a similar incidence as a function of mass-normalised jet power. 

\begin{figure*}[t!]
\centering
\includegraphics[width=0.85\linewidth]{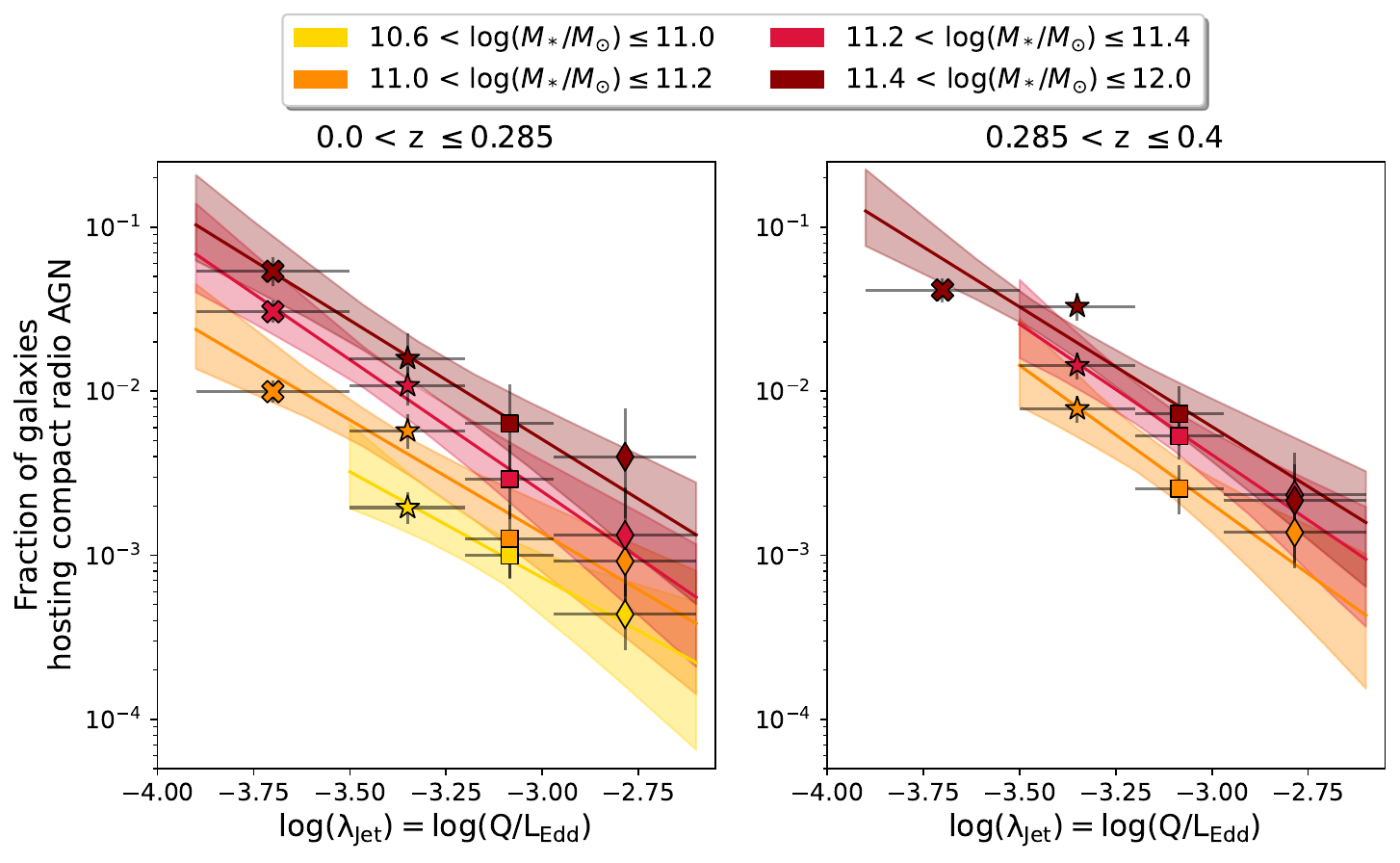}
\caption{Fraction of GAMA09 galaxies hosting compact radio AGN as a function of the specific black hole kinetic power, $\lambda_{\rm{Jet}}$, in different stellar mass and redshift bins. The power-law fit slope and normalisation values are shown in Fig.~\ref{fig:powresults} and Table \ref{table:allpowfits}.}
\label{fig:epsilon_graph}
\end{figure*}

\begin{figure}[ht!]
\centering
\includegraphics[width=0.95\linewidth]{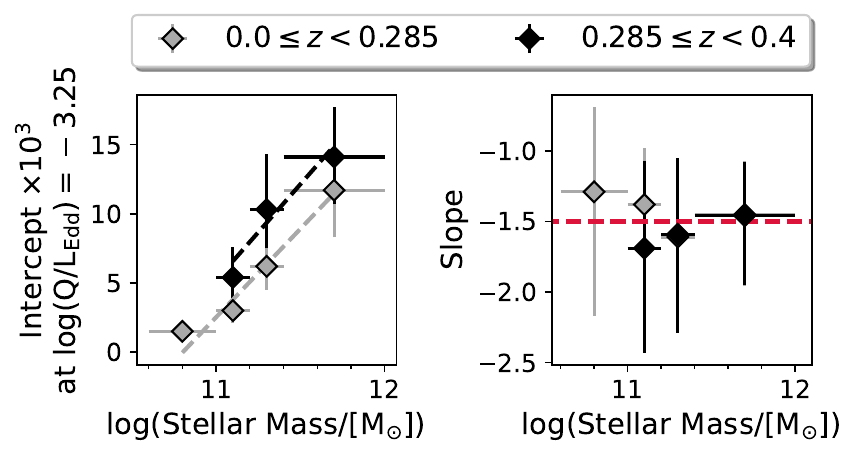}
\caption{Results of a power-law fit, $y=A \times (10^{x-x_0})^B$, to all the different mass, redshift bins present in Fig.~\ref{fig:epsilon_graph}. The left panel plots the normalisation ($A*10^{3}$) as the y-intercept at $x=x_0=-3.25$; the right panel plots the slope ($B$). The slope is consistently around $-1.5$ for all $M_*$ values (red dashed line). The normalisations show a slight mass dependence of the incidence, with some redshift evolution. Light grey and black dashed lines show the result from a linear fit (parameters listed in Table \ref{table:allpowfits}).}
\label{fig:powresults}
\end{figure}

\subsection{Incidence of eFEDS X-ray AGN}
\label{results:xray}

Fig.~\ref{fig:xrayfracs} shows the fraction of GAMA09 galaxies hosting eROSITA-eFEDS detected X-ray AGN as a function of $\lambda_{\rm{Edd}}$ in different stellar mass and redshift bins. Note that the stringent X-ray luminosity completeness limits leave too few sources to split the X-ray incidences into AGN residing in quiescent versus star-forming galaxies, therefore, they are combined.

As seen in past studies \citep[e.g.][]{Aird2012,Bongiorno2012}, the X-ray incidences across the wide range of $\log(M_*/M_{\odot})$, from 10.6 to 12.0, are remarkably similar. Regardless of the stellar mass, the X-ray incidence depends on the value of $\lambda_{\rm{Edd}}$, with higher accretion rate AGN having a lower incidence (rarer) than those at lower accretion rates. Specifically, around 0.1\% and 1\% of galaxies host an X-ray accreting at $\lambda_{\rm Edd} \sim 0.1$ and $\sim 0.01$, respectively. This has been attributed to a universal, stellar mass-invariant (and thereby black hole mass-invariant?) fuelling and triggering mechanism present in X-ray detected AGN. It is associated with a universal underlying $\lambda_{\rm{Edd}}$ distribution with power-law slope $-0.65$, independent of host galaxy stellar mass, that evolves to higher normalisations with increasing redshift, as shown in Fig.~\ref{fig:xrayfracs}.

Our results shown in Fig.~\ref{fig:xrayfracs} serve as a validation of the methods described in this work and a proof of concept that the soft response of eROSITA (with thorough consideration for completeness) is able to recover past results obtained mainly with harder X-ray instruments, less susceptible to absorption (e.g. {\it XMM-Newton, Chandra}), at least for the low-redshift samples probed here. Note that using only the unobscured AGN selection, the effects of absorption leading to incompleteness, become present for the lowest $\lambda_{\rm Edd}$ sources (see Appendix \ref{appendix:softsel}).

\subsection{Incidence of radio AGN}
\label{sec:results_incRadio}

\subsubsection{As a function of stellar mass}
\label{result:radio_stellarmass_inc}

Fig.~\ref{fig:sabaterfracs} shows that the fraction of GAMA09 galaxies hosting (detectable) radio AGN is a strongly increasing function of stellar mass, in different redshift and cumulative luminosity bins. For $\log(L_R/[\rm{W~Hz^{-1}}])\geq24$, around 10\% of galaxies host radio AGN at the highest masses $\log(M_{*}/M_{\odot})>11.5$, whilst at $\log(M_{*}/M_{\odot})=11.1$ the prevalence is only around 1\% (up to $z<0.4$). These results agree well with \citet{Sabater2019}, as shown by the orange shaded regions over-plotted onto Fig.~\ref{fig:sabaterfracs}, taken from their Fig.~5 (left panel). Of course, as for the case of X-ray selected AGN, this strongly increasing radio AGN incidence as a function of stellar mass is again a selection effect resulting from the underlying $\lambda_{\rm{Jet}}$ distribution and our survey flux limits, which is why is it essential to probe quantities normalised by stellar mass (see next section).

Unfortunately, it is difficult with present data to comment on the redshift evolution of radio AGN incidence as a function of radio luminosity \citep[e.g.][]{Smolcic2017}, yet a weak increasing trend in normalisation with redshift is apparent, as expected from past studies.

\subsubsection{As a function of $\lambda_{\rm{Jet}}$ for compact radio morphologies}
\label{results:QLedd}

As motivated in Sections \ref{sec:intro} and \ref{sec:methods} above, it is important to examine the AGN incidence as a function of mass-normalised power indicators, which for the radio regime are not as straightforward as the X-ray one, where $\lambda_{\rm{Edd}} \propto L_{\rm X}/M_*$. For the radio AGN incidence as a function of the simple observable $L_{\rm R}/M_*$, refer to Fig.~\ref{fig:sbhar_graph} in the Appendix \ref{appendix:LrMstar}, but note that this parameter is an indirect (and complex) tracer of the underlying jet power.

To examine the physical nature of jet powering, Fig.~\ref{fig:epsilon_graph} shows the fraction of GAMA09 galaxies hosting compact radio AGN as a function of the specific black hole kinetic power, $\lambda_{\rm{Jet}}$ in different redshift and stellar mass bins. Power-law slopes and normalisation of the fit to the data points are shown in Fig.~\ref{fig:powresults} and summarised in Table \ref{table:allpowfits}. A linear parameterisation of the power-law normalisation as a function of stellar mass can be given by ${A=10^{-3}~[m \times (\log M_*-11.4)+c]}$, where ($m,c$)=($12.8^{+4.2}_{-4.7}$, $7.63^{+1.3}_{-1.4}$) and ($13.9^{+8.2}_{-6.8}$, $10.7^{+1.9}_{-2.0}$) for the low and high redshift bins, respectively.

Similarly to the X-ray AGN incidence, the radio AGN incidence decreases as $\lambda_{\rm{Jet}}$ increases, because higher radio power objects become less common at all masses in the sample. On the other hand, there is a non-zero mass dependence, shown by the increasing power-law normalisations with stellar mass, that is not present in Fig.~\ref{fig:xrayfracs}. In fact, at $\log\lambda_{\rm Jet}=-3.25$, the highest mass galaxies are 7.8 and 2.6 times more likely to host radio AGN compared to the lowest mass bins in the low and high redshift bin, respectively. Possible reasons why the incidence of radio AGN shows this mass dependence, along with the caveats in the calculation of $Q$ are discussed in Sect. \ref{discussion}.

An important takeaway from Fig.~\ref{fig:epsilon_graph} is that the slopes of the observed power-law distributions for all stellar mass ranges probed are the same, with a value equal to about $-1.5$. This shows clearly that it is not only the massive galaxies that host powerful jetted AGN, nor do only the low mass galaxies host low-power jets. 

There is also a slight tendency for increased detection fractions with increasing redshift (see increasing intercept values in Fig.~\ref{fig:powresults}), possibly relating to an increased characteristic $\lambda_{\rm{Jet}}$ distribution at different epochs. Nevertheless, a larger redshift range would be needed to probe any redshift dependence further.

\begin{figure}[ht!]
\centering
\includegraphics[width=0.95\linewidth]{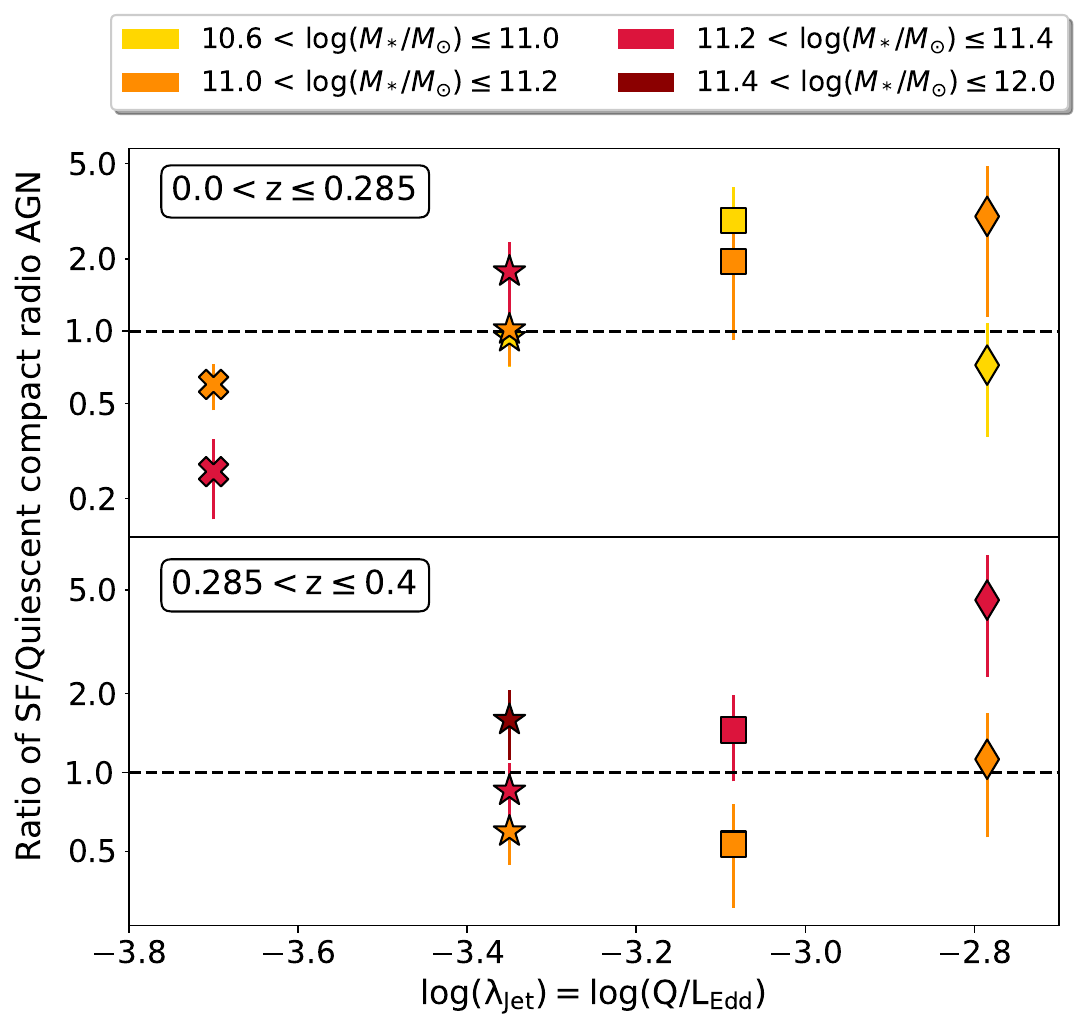}
\caption{Ratio of the measured incidence of radio AGN in star-forming to quiescent galaxies (colours and symbols as above), showing that the fraction of quiescent galaxies hosting radio AGN is similar to that of star-forming ones.}
\label{fig:results_qvsSF}
\end{figure}

We also study the incidence of radio AGN in quiescent versus star-forming galaxies. Indeed, as mentioned in the introduction, the differences between quiescent and star-forming hosts, such as temperature and fraction of gas, could have direct effects on the powering of jets \citep[see e.g.][]{Kondapally2022}. At each $\lambda_{\rm Jet}$ value, we sample 1000 points in the range of the 1$\sigma$ uncertainty on the incidence of quiescent and star-forming radio AGN separately. We then find the average ratio between the two, with the standard deviation on the mean giving the 1$\sigma$ error. Fig.~\ref{fig:results_qvsSF} shows the ratio of the measured incidence of compact\footnote{Note that the sample of complex morphology star-forming hosts is too small to robustly compare to its quiescent equal.} radio AGN in star-forming versus quiescent galaxies in the same redshift, stellar mass and $\lambda_{\rm Jet}$ bins as above. It can be seen that the fraction of quiescent galaxies hosting radio AGN is similar to that of star-forming galaxies. In general, there is no evidence of a suppressed radio AGN incidence in star-forming galaxies (with the exception of the lowest $\lambda_{\rm Jet}$ sources in the low redshift bin). Importantly, this indicates that, contrary to older findings \citep[e.g.][]{Matthews1964, Dunlop2003, Best2005, Hickox2009}, radio AGN are not predominantly hosted by `red and dead' giant elliptical galaxies, when the incidences are properly computed from complete samples. Indeed, the LOFAR survey and availability of ample multi-wavelength data is thus finally enabling the field of radio astronomy to probe radio AGN in even the most star forming galaxies, by allowing a better understanding of the origin of the radio emission.

However, due to the still limited sample size, it is not possible within the scope of this investigation to further probe the differences in jet powering resulting from the host galaxy properties \citep[see e.g.][for work on this topic in the radio and X-ray regimes]{Kondapally2022, Aird2019, Birchall2023}. Therefore, the two samples are combined, as already done for Fig.~\ref{fig:epsilon_graph}, in order to increase sample statistics. 


{\renewcommand{\arraystretch}{1.5}%
\begin{table}[ht!]
\centering
\begin{tabular}{c|cc|cc}
\hline
\hline
       &
       \multicolumn{2}{c}{$\rm{0<z\leq0.285}$}  & 
       \multicolumn{2}{c}{$\rm{0.285<z\leq0.4}$} \\
\hline
~~~$\rm{log(M_*/M_{\odot})}$~~~ & $\vardiamondsuit$    & $\diamondsuit$     &$\vardiamondsuit$     & $\diamondsuit$     \\
\hline
$10.6-11.0$ & $11.7 ^ {+10.0}_{-6.6}$ & 
$21.9^{+9.7}_{-8.0}$ & - & -                        \\
 $11.0-11.2$ & $31.4 ^{+21.0}_{-16.7}$ & $16.2^{+13.2}_{-9.0}$ & $23.6 ^{+13.7}_{-10.7}$ & $21.2^{+10.7}_{-8.5}$ \\
$11.2-11.4$ & $32.1 ^{+16.9}_{-14.2}$ & $32.6^{+12.9}_{-11.3}$ & $13.1 ^{+8.5}_{-6.2}$ & $22.7^{+8.3}_{-7.0}$  \\
$11.4-12.0$ & $25.6^{+12.4}_{-10.2}$   & $30.9^{+11.3}_{-9.9}$ & $25.5 ^{+10.9}_{-9.1}$ & $26.1^{+8.6}_{-7.5}$ \\

  \hline
  \hline
\end{tabular}
\caption{Percentage of total (compact and complex) radio AGN that have FRII-like morphologies. The statistics are only shown for the incidence of the filled diamond $\lambda_{\rm Jet}$ sources ($\vardiamondsuit$ on Fig.~\ref{fig:compact_complex_graph} where the boosted incidence is present) and for the range $-3<\log \lambda_{\rm Jet}\leq-1.5$ (unfilled diamond, $\diamondsuit$) in each stellar mass and redshift bin.}
\label{table:fr2fracs}
\end{table}}

\subsubsection{As a function of $\lambda_{\rm{Jet}}$ for both compact and complex radio morphologies}
\label{results:QLedd_complex}

\begin{figure*}[ht!]
\centering
\includegraphics[width=0.85\linewidth]{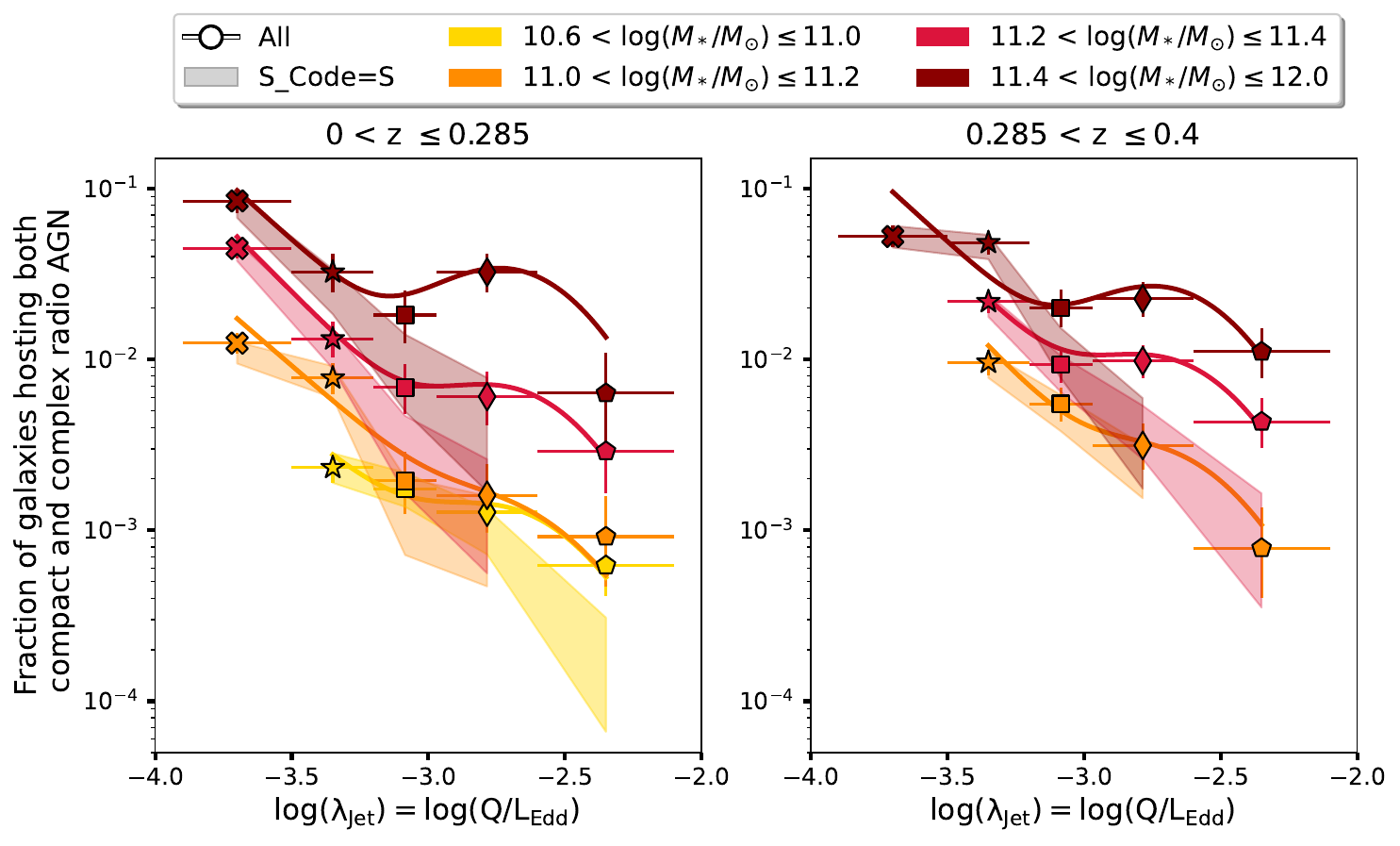}
\caption{Fraction of GAMA09 galaxies hosting both compact and complex radio AGN as a function of the specific black hole kinetic power, $\lambda_{\rm{Jet}}$, in different stellar mass and redshift bins. The markers plot all the sources, while the shaded region shows the incidences of just the \texttt{S\_code}$=$S sources. In contrast to the compact-only case, a larger mass dependence is present, as well as a boosted incidence at high $\lambda_{\rm{Jet}}$ (diamonds) values, indicating a jet power dependence. A power-law plus Gaussian fit is used to approximate the data (solid curves), for which fit parameters are presented in Table \ref{table:allpowfits}.}
\label{fig:compact_complex_graph}
\end{figure*}

Until now we have only studied the incidence of compact radio sources, as they provide the largest statistics. Fig.~\ref{fig:compact_complex_graph} shows instead the incidence of radio AGN with {\it both} compact and complex radio morphologies, as a function of $\lambda_{\rm{Jet}}$. An additional high $\lambda_{\rm{Jet}}$ bin (pentagon) is added as complex sources reach higher radio luminosities than the compact sample (recall Fig.~\ref{fig:radioAGNcut}). Focusing first on the plotted markers (all compact and complex), the main difference, compared to Fig.~\ref{fig:epsilon_graph}, is that the incidences are boosted at high $\lambda_{\rm{Jet}}$ values (diamonds), especially for the higher stellar mass bins where our sample contains a larger number of complex radio AGN. There is also an increased mass dependence of the incidence compared to the compact-only case. This implies that, although both compact and complex radio AGN are in general more frequently detected in more massive galaxies, it is the complex sample preferentially driving this mass dependence.

\begin{table*}[t!]
\centering
\begin{tabular}{l|cccc|cccc}
\hline
\hline
                          & \multicolumn{4}{c|}{Compact-only (pow)}                                                          & \multicolumn{4}{c}{Compact and Complex (pow+Gauss)}                                           \\
                          
                          & \multicolumn{2}{c}{Pow norm.}               & \multicolumn{2}{c|}{Pow slope}                     & \multicolumn{2}{c}{Pow norm.}               & \multicolumn{2}{c}{Gauss norm.}                 \\
                          \hline
$\rm{log(M_*/M_{\odot})}$ & z1                   & z2                   & z1                      & z2                      & z1                   & z2                   & z1                     & z2                     \\
\hline
$10.6-11.0$               & $1.5^{+0.5}_{-0.6}$  & $-$                  & $-1.29^{+0.88}_{-0.6}$  & $-$                     & $2.0^{+0.5}_{-0.6}$  & $-$                  & $1.0^{+0.6}_{-0.7}$    & $-$                    \\
$11.0-11.2$               & $3.0^{+0.9}_{-0.9}$  & $5.4^{+1.5}_{-2.2}$  & $-1.38^{+0.52}_{-0.4}$  & $-1.69^{+0.74}_{-0.62}$ & $4.2^{+1.1}_{-0.9}$  & $8.1^{+2.3}_{-2.4}$  & $0.8^{+1.2}_{-1.5}$    & $2.0^{+1.7}_{-2.0}$    \\
$11.2-11.4$               & $6.2^{+1.7}_{-2.1}$  & $10.3^{+2.8}_{-4.0}$ & $-1.61^{+0.51}_{-0.39}$ & $-1.59^{+0.7}_{-0.54}$  & $9.8^{+2.9}_{-3.2}$  & $15.4^{+4.3}_{-5.4}$ & $5.6^{+3.3}_{-4.0}$    & $8.2^{+4.2}_{-5.0}$    \\
$11.4-12.0$               & $11.7^{+3.4}_{-4.0}$ & $14.1^{+3.4}_{-3.6}$ & $-1.45^{+0.51}_{-0.38}$ & $-1.46^{+0.49}_{-0.38}$ & $21.7^{+5.8}_{-6.0}$ & $21.1^{+5.2}_{-4.5}$ & $30.5^{+10.9}_{-19.6}$ & $23.2^{+10.7}_{-16.4}$  \\
\hline
\hline
\end{tabular}
\caption{Fit results for the power-law and power-law plus Gaussian trends fit to the incidence of radio AGN in different stellar mass and redshift bins ($z1$ and $z2$ for the low and high redshift bins, respectively). The power-law norm (pow norm.) is the value of the intercept $\times~ 10^{3}$ at $\log \lambda_{\rm Jet}=-3.25$. Gauss norm. represents the normalisation of the Gaussian function, with an additional multiplicative factor of 1000. These values are used to plot the trends in Figs.~\ref{fig:epsilon_graph} and \ref{fig:compact_complex_graph}.}
\label{table:allpowfits}
\end{table*}

Interestingly, when plotting the subset of compact and complex sources satisfying the condition that their radio emission is modelled only by a single Gaussian (\texttt{S\_code}$=$S), the upturn at high jet powers disappears, resembling the simple power-law distributions seen in compact radio AGN (Fig.~\ref{fig:epsilon_graph}). Note that a significant fraction, 52\% (144/278), of complex (mass-complete) radio AGN have a single Gaussian component. These are radio sources which are morphologically simple, but too large for being classified as compact. 

We approximate the complex and compact radio AGN incidence with a power-law plus a Gaussian, marked by solid curves on Fig.~\ref{fig:compact_complex_graph}. The power-law slopes are fixed to the individual best-fit values from the compact-only incidence (Fig.~\ref{fig:powresults}) and the Gaussian is centred at $\mu=\log \lambda_{\rm Jet}=-2.71$ with fixed width $\sigma=0.19$. Both power-law and Gaussian normalisations are left free in the fit. All fit parameters are presented in Table \ref{table:allpowfits}.

Fig.~\ref{fig:scodehist} shows the histogram of the different subsets of compact (light blue filled), complex with \texttt{S\_code}$=$S (black, unfilled), complex with \texttt{S\_code}$=$M (magenta, hatched) and complex \texttt{S\_code}$=$M sources with physical sizes $R_{\rm {kpc}}>60$~kpc (dark blue filled). Complex (i.e. non-compact) sources show a clear bimodal distribution in both radio luminosity (top right) and $\lambda_{\rm{Jet}}$ (bottom right). Thus we conclude that the sources modelled with multiple Gaussians and sources with large physical sizes drive the upturn in the incidence seen at higher jet powers. Possible explanations for this are discussed in Sect. \ref{sec:Qcaveats}.

Lastly, since all mass-complete sources were visually inspected, it is possible to also give an estimate of the prominence of FRIIs in the sample (flagged as \texttt{FRII\_flag}$=$1). Table \ref{table:fr2fracs} shows the percentage of total (compact and complex) radio AGN with FRII-like morphologies in different stellar mass and redshift bins. Percentages are only shown for the incidence of the filled diamond $\lambda_{\rm Jet}$ sources ($\vardiamondsuit$ on Fig.~\ref{fig:compact_complex_graph} where the boosted incidence is present) and for the extended range of $-3<\log \lambda_{\rm Jet} \leq-1.5$, which includes all of the most powerful FRIIs of the sample. It can be seen that FRII objects are in fact not so `rare' and make up around $10-30$\% of the high jet power sources at $z<0.4$. There does not seem to be a distinct trend with host galaxy stellar mass, although a larger sample with higher resolution would be needed to test this.

\subsection{Incidences of both X-ray and radio AGN}

The incidence as a function of $\lambda_{\rm{Edd}}$
is shown in Fig.~\ref{fig:XandRfracs}. This combines the limited number of radio and
X-ray detected AGN (same as Fig.~\ref{fig:xrayfracs}). The mass bins are now combined into one, keeping all mass completeness requirements fulfilled. Only the 18/24 X-ray and radio AGN which have radio luminosity in excess of the 95\% completeness threshold (dashed line on Fig.~\ref{fig:radio_zvsLr}) are included. Again it is seen that the pure X-ray detected sources follow well the \citet{Aird2012} power-law. The incidence of radio and X-ray AGN is around 10\% less than that of the pure X-ray AGN, in line with the generally accepted fraction of `radio-loud' AGN in X-ray surveys  \citep[see e.g.][]{Kellermann2016}.

\section{Discussion}
\label{discussion}

In Sect. \ref{sec:results} we have shown that a universal AGN triggering and fuelling mechanism is present for X-ray detected AGN (as also found by several previous studies), yet this phenomenon does not seem to translate into a fully mass-invariant jet power mechanism, as seen by the incidence of radio AGN as a function of $\lambda_{\rm{Jet}}$ in Fig.~\ref{fig:epsilon_graph}. It even shows an additional jet power dependence when including complex radio emitters (Fig.~\ref{fig:compact_complex_graph}). This section discusses the possible reasons for these differences, given the caveats and limitations of the surveys, and asks whether there even should be a universal jet powering mechanism present in AGN. Firstly though, the GAMA survey completeness to AGN and its reliability to estimate stellar masses is vetted.

\begin{figure}[t!]
\centering
\includegraphics[width=\linewidth]{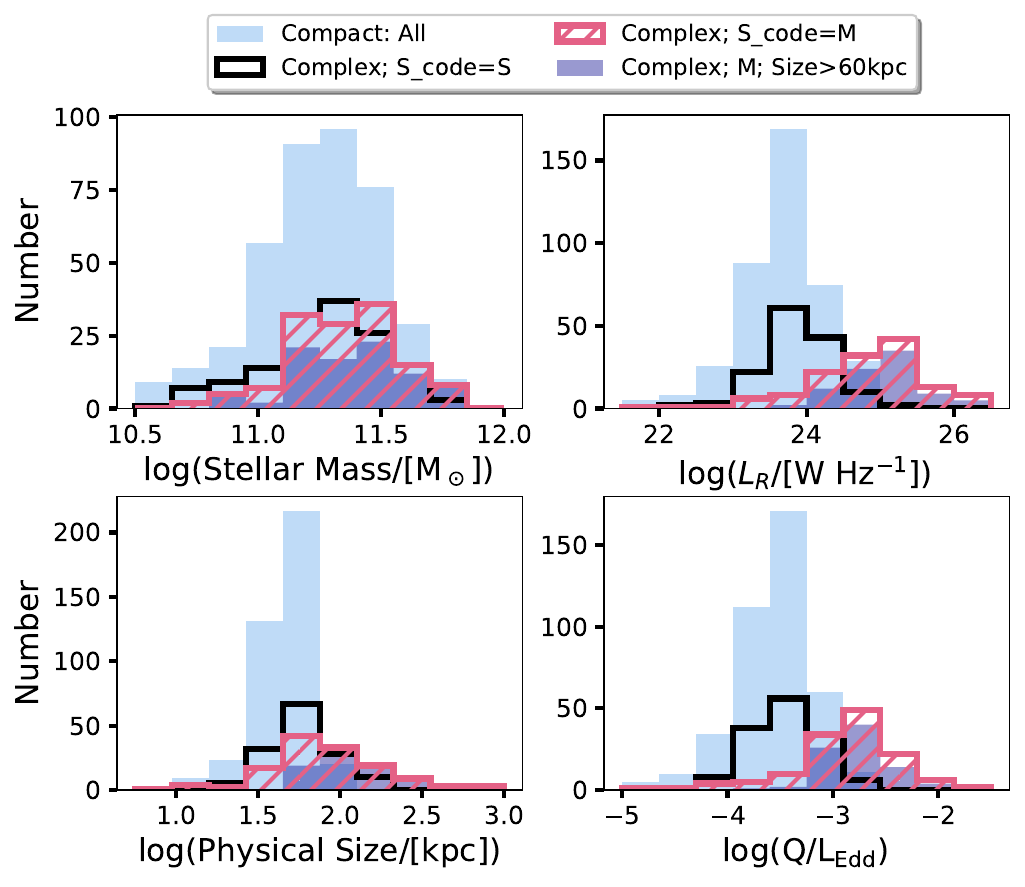}
\caption{Histograms showing the stellar mass, radio luminosity, physical size and $\lambda_{\rm{Jet}}$ distributions of different subsets of radio AGN: compact (light blue filled), complex \texttt{S\_Code}$=$S (black, unfilled), complex \texttt{S\_Code}$=$M (magenta, hatched) and complex \texttt{S\_Code}$=$M sources with physical sizes $>60$~kpc (dark blue filled). This shows that the sources responsible for the boosted radio AGN incidences at high jet powers are the ones best modelled by multiple Gaussian components and large physical sizes.}
\label{fig:scodehist}
\end{figure}

\subsection{Incompleteness due to the lack of bright quasars in GAMA}

GAMA is inherently a galaxy survey and bright Type I QSOs may have been missed \citep[or de-selected; see Section 3.1 of][]{Baldry2010}, therefore it is important to consider the implications this might have on our measures of AGN incidences. \citet{Comparat2022} estimate the fraction of missed eROSITA eFEDS X-ray extra-galactic point sources (AGN) in GAMA09, in the redshift range $0.05<z<0.3$, by firstly taking their Legacy Survey DR8 (LS8) counterparts, with spectroscopic or good quality photometric redshifts \citep{MaraCTPeFEDS}, within the GAMA09 footprint. Then, after matching to the GAMA09 galaxy catalogue, they find that 88.8\% of the X-ray sources with an LS8 r-band magnitude $<19.8$ have a GAMA09 counterpart, meaning that around 10\% of X-ray bright QSOs may be missed. This is in line with the known 10\% incompleteness mentioned in Sect. \ref{sec:GAMA} and accounted for in Sect. \ref{sec:mass_lumin_completeness}. A much smaller fraction of missed radio AGN is expected, considering their dominant kinetic, rather than radiative emission. 

Meanwhile, for the brighter objects that are present in the GAMA sample, it could be questioned whether the stellar mass estimates are robust and not overestimated due to the AGN contamination (there is no AGN component in the SED fitting for GAMA galaxies). However, we show in Fig.~\ref{fig:mstarcomparison} that this is not the case for our samples. We compare the GAMA derived stellar masses of the G9 X-ray AGN to the ones from Hyper Supreme-Cam (HSC) obtained for a subset of the eFEDS X-ray sources \citep[which have better AGN and host galaxy decomposition,][]{Aihara2018,Li2023}. Using the $\sim$300 matched objects, the stellar masses agree well, as shown in Fig.~\ref{fig:mstarcomparison} (bottom), with a slight systematic offset of around 0.1~dex (constant with mass) for the GAMA derived measurements, well within the stellar mass bins used in this work (note that both HSC and GAMA use the \citet{Chabrier2003} IMF). Thus, this is not a dominant uncertainty, especially not for the radio results as those objects are not expected to be hosted by bright quasars \citep[corroborated by none of the G9 radio AGN lying within the `WISE' wedge of luminous AGN,][]{Mateos2013}. 

\citet{Thorne2022} confirm that the addition of an AGN component to the SED fitting has no significant impact on the GAMA derived stellar masses. Fig.~\ref{fig:mstarcomparison} (top) shows the excellent agreement between $M_*$ measurements from \citet{Thorne2022} and the ones used in this work for the G9 radio AGN. However, \citet{Thorne2022} find that stellar properties such as SFRs may be overestimated in cases where an AGN component is significant but not considered. This could potentially lead to an underestimate of radio-excess AGN, although it is not deemed a large impact for the same reasons as outlined above.


\begin{figure}[t!]
\centering
\includegraphics[width=1.05\linewidth]{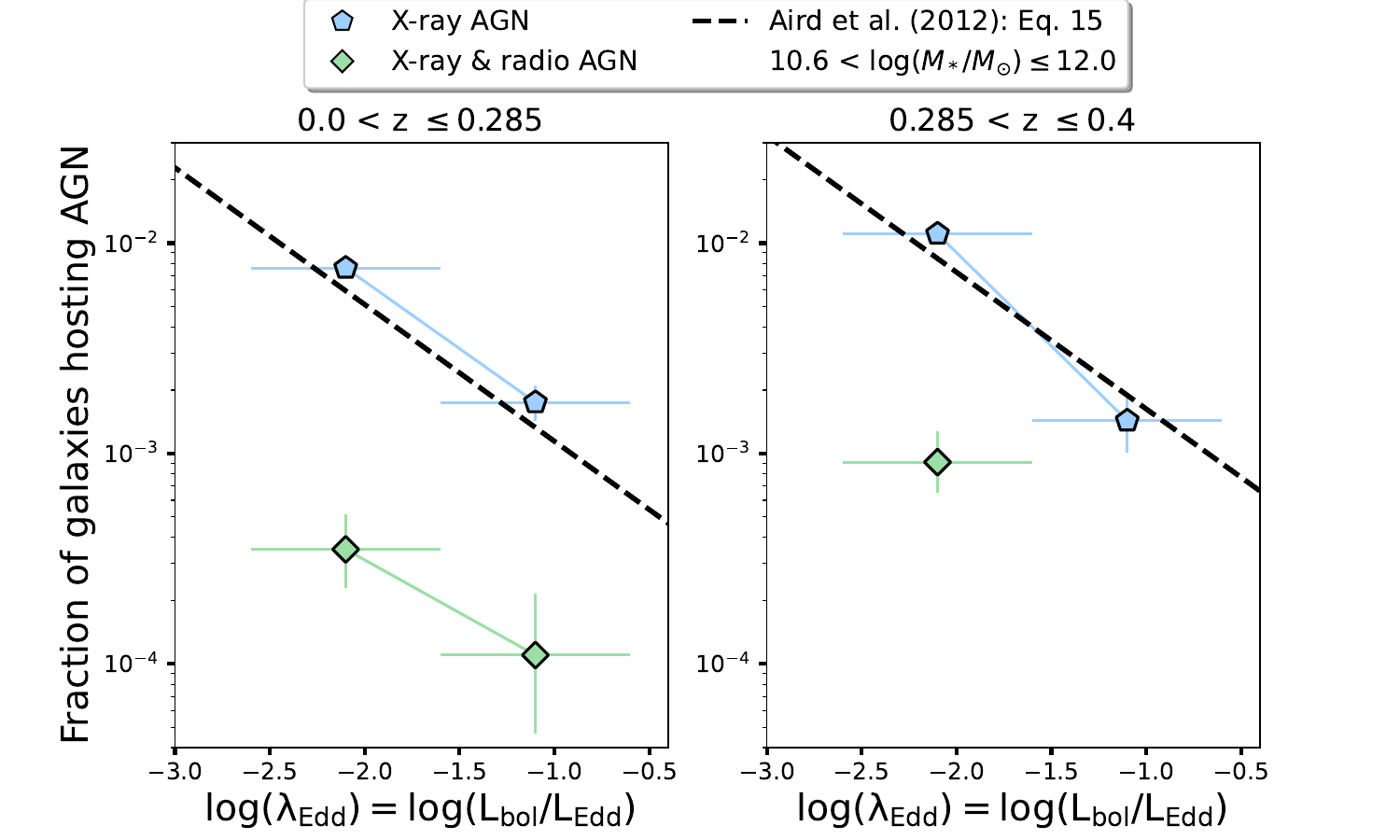}
\caption{Fraction of GAMA09 galaxies hosting purely X-ray-detected AGN (blue) and both X-ray and radio AGN (green) as a function of $\lambda_{\rm{Edd}}$. The stellar mass, radio luminosity and X-ray luminosity are ensured for completeness. The pure X-ray fractions follow well the \citet{Aird2012} trend, whilst the radio and X-ray AGN lie an order of magnitude below it.}
\label{fig:XandRfracs}
\end{figure}

\subsection{Caveats in the inference of jet power}
\label{sec:Qcaveats}

Jet power is a notoriously difficult quantity to estimate accurately. Even though the most widely used relation is applied in this work, there are numerous caveats that must be discussed. 

\subsubsection{The $Q-L_R$ relation}

Firstly, as stated in \citep{Hardcastle2019}, a single conversion from radio luminosity to jet power is inherently flawed as, although jet power could be constant in time, radio luminosity by default cannot be: it must start from zero (and is predicted to decay at late times). Use of such relations requires an assumption to be made on the type of radio source it has been calibrated to, usually large sources in relatively rich environments.

Secondly, there exist a large number of radio luminosity to jet power conversions in the literature, not only with (slightly) different normalisations but also with vastly varying power-law slopes, ranging from 0.4 \citep{Birzan2004, Best2006, Best2007} all the way to 1 \citep[][see their Fig.~12 for a visual comparison with respect to the simulated jet powers]{Hardcastle2018}. The slope of the $Q-L_R$ relation has a direct impact on the mass dependence of the incidence, whereby the more sub-linear the slope is, the more mass-invariant the incidence becomes.

\begin{figure}[t!]
\centering
\includegraphics[width=1\linewidth]{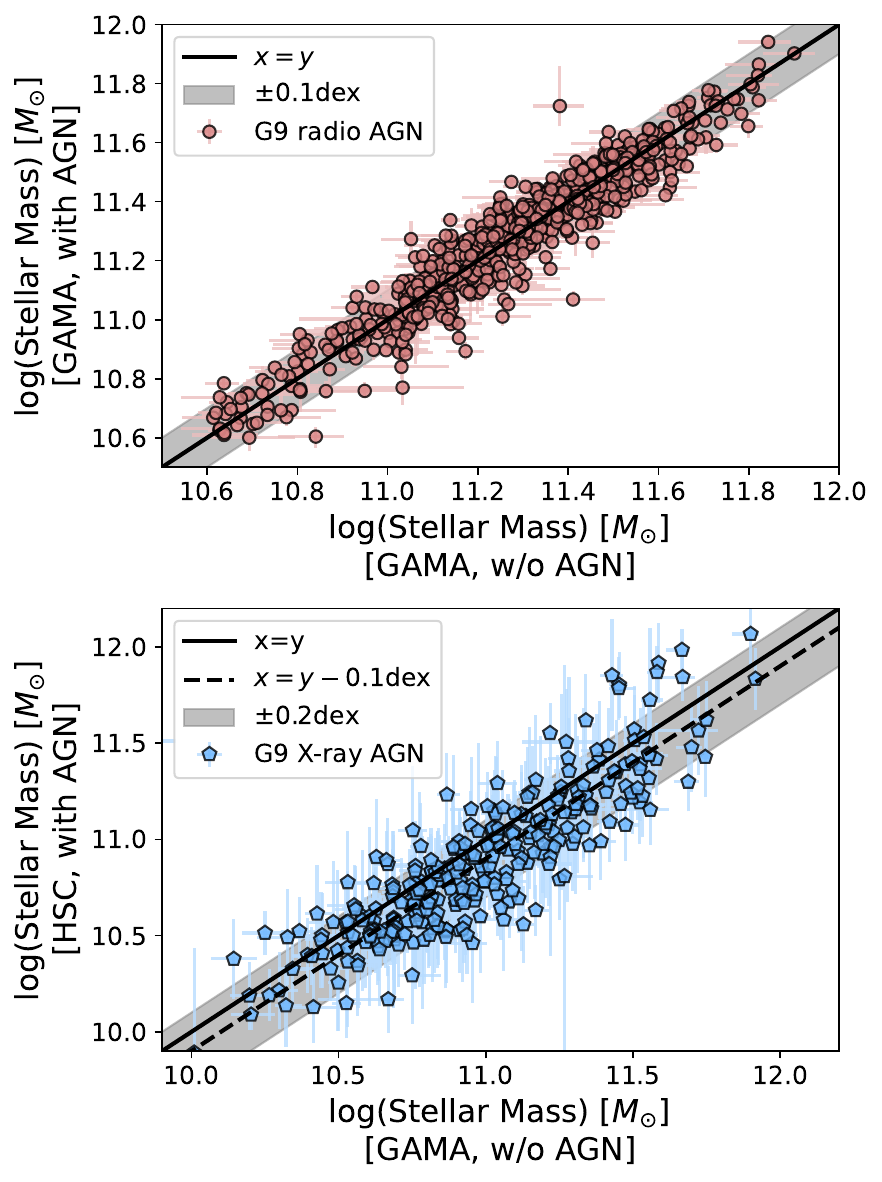}
\caption{Top: Comparison of GAMA $M_*$ measurements with and without AGN component in the SED fitting for the G9 radio AGN sample (with 1$\sigma$ errorbars). Bottom: Comparison of $M_*$ derived from GAMA without AGN component versus those derived from HSC with AGN-host decomposition for a subset of G9 X-ray AGN. }
\label{fig:mstarcomparison}
\end{figure}

Thirdly, some works predict jet power conversions to depend on additional parameters such as radio spectral index \citep{GodfreyShabala2013}, proton content \citep{Croston2018} or spin (see below). \citet{HeinzandSunyaev2003} and \citet{Merloni2003} derive, for a self-similar freely-expanding jet coupled to a radiatively inefficient accretion flow, $L_{\rm Kin} \propto L_{R}^{12/(17+8\alpha)}$, where $L_{\rm Kin}$ is the kinetic luminosity, equivalent to $Q$ here. Thus, jet power scaling depends on the observed radio spectral index and this acts to flatten the $Q-L_R$ relation for steeper $\alpha$ values. This may be important as a constant $\alpha=0.7$ is used here, which better describes larger, more diffuse sources (typically in the `complex sample). Meanwhile, compact sources with synchrotron self-absorbed cores have much flatter $\alpha \sim 0$, causing a potentially underestimated jet power. However, not all the G9 radio AGN sources in our sample can be associated with a jet coupled to a low accretion rate, radiatively inefficient mode of accretion (see Fig.~\ref{fig:balanceOfpower}), so the applicability of the \citet{HeinzandSunyaev2003} scaling for the full sample is questionable.

Moreover, the scaling we adopt from Eq.~\ref{Qeq} \citep{HeckmanandBest2014} is a cavity-based relation, making use of the minimum energy condition to find the energy needed to inflate the radio lobes to their observed volume, calibrated with pressure of the surrounding medium estimated from X-ray observations. These are observations which have been conducted mainly on the relatively few nearby, large cluster AGN with deep X-ray and radio coverage. Therefore, they may not be accurate in describing the jet power of the large majority of low-luminosity ($\log(L_R/[\rm{W~Hz^{-1}}])<25$), compact sources. In fact, Eq.~\ref{Qeq} may underestimate the jet power for these sources if there are diffuse, large scale lobes that are too faint to detect. Such surface brightness limitations in LOFAR are a known observational bias \citep[see e.g.][]{Hardcastle2019} but we do not attempt to correct for this here. \citet{Croston2019} also report that FRI cavity scaling relations may overestimate FRII jet powers by up to an order of magnitude.

The jet composition (e.g. proton content, electron energy distribution) and collimation/interaction with environment could further impact the observed relation. For example, it has been found that low-luminosity, larger physical size radio AGN jets have a higher proton content, consistent with entrainment arguments \citep[][Croston et al., in prep.]{CrostonHardcastle2014, Hardcastle2018, Hardcastle2019, Croston2018}. Although this effect is taken into account in the pressure calculation within the cavity-based method, there may be a systematic underestimation for low-power sources with respect to Eq.~\ref{Qeq} (i.e. the assumption of a universal proton content for all types of radio AGN is invalid).

These arguments could equally explain the upturn in the incidence of high jet power sources (diamond points) shown in Fig.~\ref{fig:compact_complex_graph}. This is because the systematically underestimated jet power for lower luminosity radio-detected objects, once corrected for, would shift these sources to higher jet powers, thus populating a power-law distribution at the level of the current diamond points.

The mass dependence of the incidence could also inherently be due to an unaccounted for correlation of jet power with stellar mass itself. However, \citet{TurnerShabala2015} find no such correlation using their modelled FRI and FRII AGN ($z<0.1$) in realistic galaxy environments. Note that if future black hole masses estimates become available for such large samples, it could be educational to look for direct correlations of the radio incidence with black hole mass, instead of using stellar mass as their proxy.

Lastly, spin ($a$) and magnetic flux density/configuration are also important parameters in the context of driving jets in AGN \citep{BlandfordZnajek1977, SikoraBegelman2013}. In fact, the Blandford-Znajek mechanism \citep{BlandfordZnajek1977}, predicts the spin to directly impact jet power as $Q \propto a^2$ \citep{Meier2002, Amarantidis2019}. However, this is only one specific model and there are too few objects (with high selection bias) for which spin measurements exist \citep{Reynolds2021} to be able to qualitatively test the effect of spin on jet power, and its dependence on stellar mass or accretion rate. 


\subsubsection{Observed radio morphology}

The striking difference between the purely compact radio AGN incidence as a function of $Q/L_{\rm Edd}$ and the one including the compact and complex sources, opens up the discussion about whether the jet powering mechanisms may vary with observed (LOFAR) radio morphology \citep[see also e.g.][]{Mingo2022}. 

Firstly, a given source radio morphology determination may be plagued with orientation effects or be dependent on the observing frequency. For example, the uncertainty in observing pure-core or core and lobe emission in projection, may hinder the ability to draw conclusions on the fundamental nature of jet powering, due to contaminating factors such as environment. Also, the increase in sensitivity towards older electron populations at lower frequencies \citep{Condon1992}, could be problematic for jet power estimates of sources no longer possessing an active core. This is complicated by the fact that jet power is a lifetime average quantity, in contrast to the X-ray emission that can respond relatively fast to accretion events. Therefore, measuring `instantaneous' jet power is almost impossible.

To try to understand why there is such a strong increase in the fraction of galaxies hosting radio AGN at high jet powers, the stellar mass, radio luminosity, physical size and $\lambda_{\rm{Jet}}$ distributions of radio AGN exhibiting different morphological properties are shown in Fig.~\ref{fig:scodehist}. The stellar mass distributions of all subsets are similar, but a dichotomy is present in the radio luminosity between the compact and the complex single component sources versus the complex multiple component and also large sources. Fig.~\ref{fig:radioAGNcut} showed hints of the relatively numerous complex radio AGN population at $\log(L_R/[\rm{W~Hz^{-1}}])>25$. This bi-modality is then translated to the $\lambda_{\rm{Jet}}$ distribution, resulting in the upturn in the incidence being caused by these large, complex (multiple components) radio AGN, more specifically the subset with physical sizes $>60$kpc. Note that the lack of upturn in the lowest stellar mass bins in Fig.~\ref{fig:compact_complex_graph}, is due to the lack of large complex sources in those mass bins as shown in Fig.~\ref{fig:scodehist}.

However, there are conflicting results in the literature about how morphology affects jet power estimates. For example, \citet{GodfreyShabala2013} find good agreement between the jet power relations derived for both FRIs versus FRIIs and low versus high power sources. They find that the supposed higher $L_{\rm R}/Q$ ratios in FRII sources, due to lower fraction of the energy in non-radiating particles, are counteracted by the effects of lower density environments, spectral ageing and strong shocks, bringing FRII $Q-L_{\rm R}$ relations into agreement with that of FRIs. 
On the other hand, \citet{TurnerShabala2015} find that larger physical size sources tend have higher jet powers compared to smaller sources with equivalent morphology and luminosity, and FRII-like objects have a factor of two higher jet powers than FRI-like sources (see their Fig.~9).


\begin{figure}[t!]
\centering
\includegraphics[width=\linewidth]{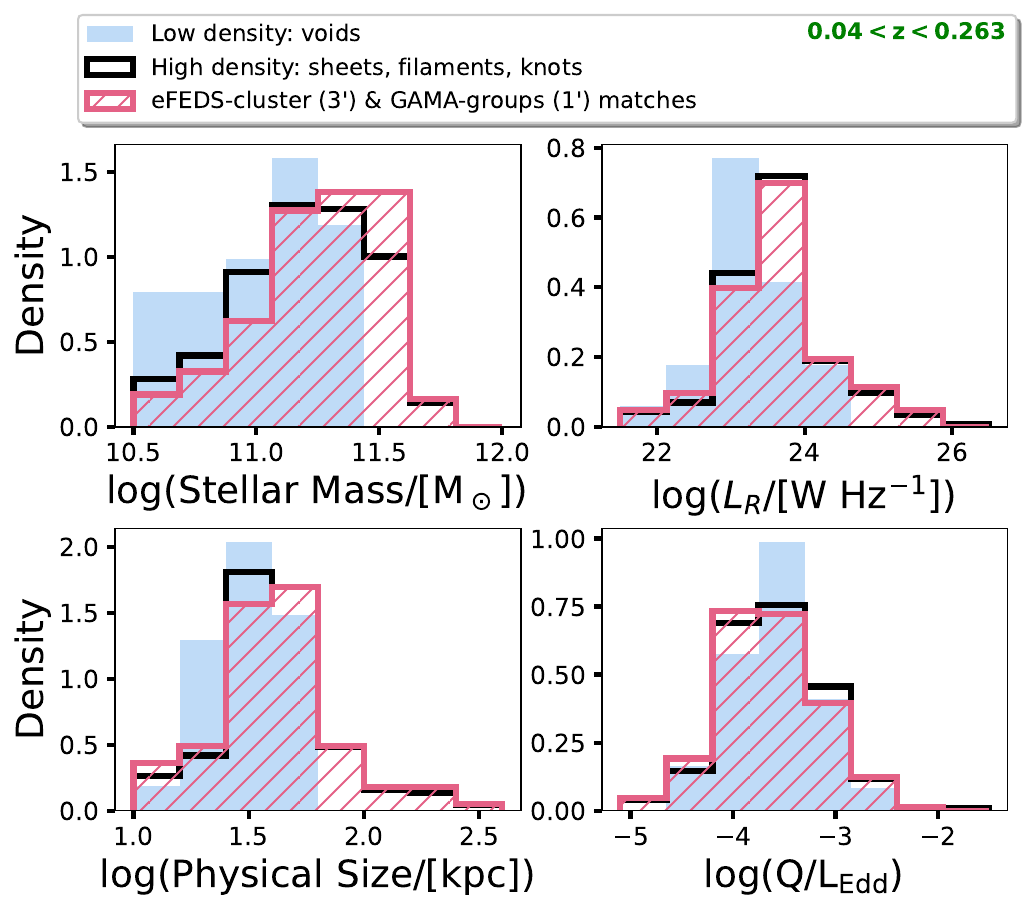}
\caption{Histograms showing the parameter distributions of the G9 radio AGN split into different environments. Radio AGN in dense environments (black unfilled, magenta hatched) are not the cause for the upturn in the measured incidence seen in Fig.~\ref{fig:compact_complex_graph}, as they have the same $\lambda_{\rm{Jet}}$ distribution as the ones in low density (light blue filled) environments. Note that the GAMA cosmic web environment catalogue is limited to $0.04<z<0.263$, so this is applied to all datasets shown here, along with the appropriate mass completeness cuts.}
\label{fig:filamentvoid}
\end{figure}

\subsubsection{Radio AGN environment}

It is known that radio emission can be boosted by denser (e.g. group- or cluster-like) environments, due to the reduced adiabatic expansion losses and more synchrotron shocks as the multiple component, large jets interact with the surrounding interstellar and galactic medium \citep{BarthelArnaud1996, McNamara2006, HardcastleKrause2013}. Therefore, a straightforward hypothesis for the increased incidence at high $\lambda_{\rm{Jet}}$ compared to the compact-only case, would be that complex sources are more often located in dense environments. This is because the incidence of complex sources, which become increasingly dominant at increasing $\lambda_{\rm{Jet}}$, would then be counted in preferentially higher $\lambda_{\rm{Jet}}$ bins due to the boost in radio luminosity.

To test this, we match the G9 radio AGN to the publicly released GAMA DR2 Geometric Environment catalogue\footnote{Available at: \url{http://www.gama-survey.org/dr4/schema/dmu.php?id=96}} \citep{Eardley2015}, in which galaxies have been rigorously sorted into four Cosmic Web environments: knots, filaments, sheets and voids (in decreasing order of density). This is only valid for $0.04<z<0.263$, so all samples and completeness limits have been adjusted accordingly. In total, there are 27 radio AGN in low-density voids and 228 in sheets, filaments and knots.

In addition, the radio AGN are also matched to the eFEDS X-ray cluster catalogue \citep{Liu2022eFEDSclusters} using a 3~arcmin match radius and to the GAMA09 groups catalogue\footnote{Available at: \url{http://www.gama-survey.org/dr4/schema/dmu.php?id=115}} \citep{Robotham2011groups} within 1~arcmin, as a further probe of dense environments. Out of the 264 mass-complete radio AGN in $0.04<z<0.263$, 171 and 26 match to the GAMA groups and eFEDS cluster catalogue, respectively. 

Fig.~\ref{fig:filamentvoid} shows that, as expected, the radio AGN in higher density environments reside preferentially in larger stellar mass galaxies, and do have increased radio emission, as well as larger physical sizes. However, they are not preferentially hosting radio AGN with higher $\lambda_{\rm{Jet}}$ compared to those in low density voids.

This is interesting as numerous past work finds environment to be an important factor in the driving and propagation of jets \citep[e.g.][]{HardcastleKrause2013, English2016, Croston2019, Morris2022,  Mingo2022}. The incidence of radio AGN is also reported to be higher in dense, cluster environments \citep{Best2004, Best2005, Best2007, Sabater2013, BestandHeckman2012, HeckmanandBest2014}. Yet, not many studies have probed differences in environment using also mass-scaled parameters and samples well controlled for mass and luminosity completeness.

We therefore conclude that environment is {\it not} the primary cause of the observed increased incidence of powerful, complex, large scale radio AGN at high stellar masses.



\subsubsection{Radio size and lifetimes}

An alternative explanation for the boosted incidence of high jet power sources in the complex radio morphology sample would be the idea put forward by \citet{Hardcastle2018, Hardcastle2019}. They state that large, powerful radio sources are just the older, longer-lived population of normal radio AGN, rather than a marker of different host galaxy or larger scale environmental properties. \citet{Hardcastle2019} show that simulated sample of small, low power radio AGN (possibly akin to the `compact' sample here) are better modelled with log-uniformly distributed lifetimes (more short lived than long lived), whereas large, high power AGN (similar to the `complex' sample), fit better to uniformly distributed lifetimes.

A consequence of the longer-lived nature of the complex radio AGN could manifest through the boosted detection fractions of this population, compared to the compact sources, potentially explaining the elevated incidence measured at high $\lambda_{\rm{Jet}}$  (Fig.~\ref{fig:compact_complex_graph}). Recall here that there is a $\sim10-30\%$ FRII fraction (see Table \ref{table:fr2fracs}), objects known to have large physical sizes and powerful jets, among the high $\lambda_{\rm{Jet}}$ sources at all masses. This would also explain the similar stellar mass distributions between the different subsets in Fig.~\ref{fig:scodehist}, but clear dichotomy in the radio luminosity and therefore $\lambda_{\rm{Jet}}$ distribution. Furthermore, this could be the reason why the mass dependence of the incidence increases with respect to the compact-only case: at higher jet powers for all masses, there will be more larger, longer lived radio AGN; in fact this is shown by the diverging incidences between the single (shaded region) and multi-component (markers) sources on Fig.~\ref{fig:compact_complex_graph}. Just considering the single component sources, the incidence as a function of stellar mass is approximately similar to the compact only case.

\subsection{Towards probing disk-jet coupling through Radio and X-ray Incidences}
\label{sec:discussion_radioXray}

The incidence analysis presented here is a powerful probe of the bulk behaviour of a population, in this case the disk-jet coupling in radio and X-ray AGN. So far, we have shown that there is a non-zero mass dependence of the incidence of compact radio AGN as a function of $\lambda_{\rm Jet}$ and that an additional jet power dependence becomes apparent when we include complex radio AGN. However, the origin of these differences and whether they can be related to intrinsic differences in accretion modes is unclear.

Drawing a firm connection between empirical incidences and accretion models may be hampered by the fact that the radio and X-ray AGN in our sample populate a mixture of different accretion modes (see Fig.~\ref{fig:balanceOfpower}). Therefore, there may be no dominant population driving clear trends in the total incidence of radio AGN. This is complicated by the fact that it is very difficult to get a reliable measurement of the fundamental accretion mode without having simultaneous measurements of accretion and jet-driven power for many of our objects, with the majority of the AGN being detected either in radio or in X-rays, but not in both.
Indeed, the accretion rate regime in which the \citet{Merloni2003} LK fundamental plane branch holds ($\log \lambda_{\rm Edd}\ll 3-4$) is not only too faint for the eFEDS X-ray sensitivity limit, but also lower than that of the radio AGN sample presented here. On the other hand, for the `high radiative' (HR; radio-quiet) branch \citep{MerloniHeinz2008}, mostly populated by the X-ray detected AGN, even the origin of the radio emission is highly debated \citep[e.g. wind, jet, or coronal origin; see][]{Panessa2019}, which is important as it would require different physical models than the one of a scale-invariant, freely expanding jet model presented in \citet{HeinzandSunyaev2003}.

Past studies of (bright) radio AGN have made attempts to separate the radiatively efficient and inefficient accreting sources, by classifying them into `high' and `low' excitation radio galaxies (HERGs/LERGs), depending on whether strong optical emission lines are present or not \citep{BestandHeckman2012}. Interestingly,  \citet{Smolcic2009} find that LERGs preferentially live in higher mass galaxies compared to HERGs, in line with their expected lower accretion rates, but both populations can span similar ranges in radio luminosity, although HERGs tend to be skewed towards the more luminous regime \citep{BestandHeckman2012}. These properties put HERGs exactly in the region of mass and intrinsic jet power where we observed a marked increase in the incidence for `complex' radio sources (Fig.~\ref{fig:compact_complex_graph}). Indeed, Fig.~\ref{fig:radioloudness} shows that the majority (96\%) of sources above $\rm{log}(\lambda_{\rm Jet})\gtrsim-3.0$, the value above which there is a boosted incidence, are complex. This means that the majority of the sources lying above the $\lambda_{\rm Jet} \propto \lambda_{\rm Edd}$ line in Fig.~\ref{fig:balanceOfpower} (i.e. `jet dominated' AGN) are also complex in their observed (LOFAR) morphology. 
Considering that the incidences of compact versus complex radio AGN showed such striking differences in their mass and jet power dependencies, and that it is the complex morphologies which dominate at these high jet (and radiative) powers, it could be postulated that morphology (and/or radio physical sizes, as shown on Fig.~\ref{fig:scodehist}), could be a signpost of jet domination, and a tracer of a specific accretion mode in this sample. This may be important in qualitatively understanding the observed radio AGN incidence results, which we elaborate upon below.

Let us assume the intrinsic distribution of accretion rate, $P_{\dot{m}}=d\,P(\dot{m}, M)/d\log \dot{m}$, as a function of dimensionless accretion rate ($\dot{m}=\eta \dot M c^2/L_{\rm Edd}$, with $\eta$ the accretion efficiency) and black hole mass ($M=M_{\rm BH}/M_{\odot}$), follows a power-law distribution described by $P_{\dot{m}} \propto \dot{m}^{-\alpha}~M^{\beta}$. Observationally (Fig.~\ref{fig:xrayfracs}), we find that the observed incidence follows a power-law $P_{\lambda_{\rm Edd}}=$ d\,$P(\lambda_{\rm Edd}, M)/\rm{d}\log \lambda_{\rm Edd} \propto \lambda_{\rm Edd}^{-\gamma} M^{\delta}$,  where $\gamma\sim0.65$ and $\delta\sim0$. Using a scaling of $\lambda_{\rm Edd}$ with $\dot{m}$ in different accretion modes, it is possible to determine the value of $\alpha$ (assuming $\delta=0$). Adopting a two-modes relation \citep{MerloniHeinz2008}:
\begin{equation}
\label{lambdascaling}
  \lambda_{\rm Edd} \propto \begin{cases}
    \dot{m}, & \dot{m}>\dot{m}_{\rm crit}. \\
    \dot{m^2}, & \dot{m}<\dot{m}_{\rm crit},
  \end{cases}
\end{equation}
where $\dot m_{\rm crit}$ is the critical dimensionless accretion rate at which a putative state transition takes place, one finds $\alpha=\gamma$ for $\dot{m}>\dot{m}_{\rm crit}$ and  $\alpha=2\gamma$ for $\dot{m}<\dot{m}_{\rm crit}$.

Therefore, under this scenario the incidence of X-ray AGN as a function of $\lambda_{\rm Edd}$ is expected to flatten at low accretion rates below the critical value. This flattening, and possible turnover, has recently been found observationally by \citet{Aird2017, Torbaniuk2024}, who probe the X-ray AGN incidence at $\rm{log}(\lambda_{\rm Edd})\ll-3$ using larger samples and deeper X-ray data, as well as by \citet{Volonteri2016} using cosmological hydrodynamical simulations. On the other hand, for high accretion rates, the incidence of X-ray AGN has been shown to drop off dramatically as the sources reached their Eddington limits \citep{Bongiorno2012, Georgakakis2017}. 


The same parameterisation can be made for the observed radio AGN incidence, $P_{\lambda_{\rm Jet}}=d\,P(\lambda_{\rm Jet}, M)/d\log \lambda_{\rm Jet} \propto \lambda_{\rm Jet}^{-x}~ M^{y}$, where $x \sim 1.5$ (Fig.~\ref{fig:powresults}) and $y$ is poorly constrained, but likely $>0$. Next, we make a first assumption that the same intrinsic mass accretion distribution is responsible for the energy provided to power the jet, meaning that the jet power at the base is governed by the internal energy of the system \citep{HeinzandSunyaev2003} and so $P_{\lambda_{\rm Jet}} \propto P_{\dot{m}}$. Although the mass dependence is found to be small but non-zero, we start by examining the simplest $y=0$ case. Given a generalised formulation of Eq.~\ref{lambdascaling} of $\lambda_{\rm Edd} \propto \dot{m}^A$, we can write $\lambda_{\rm Jet}^{-x} \propto \dot{m}^{-A*\alpha}$. Then, $\lambda_{\rm Jet}$ scales linearly with $\dot{m}$ in the LK mode, where $A=2$ \citep{HeinzandSunyaev2003, MerloniHeinz2007}. Therefore, we obtain $x=2\alpha=1.3$, which is consistent, within errors, with the steep compact radio AGN incidence slope of $\sim -1.5$, and with the prediction above that compact sources trace the relatively radiatively inefficient branch.

For the complex morphologies, we cannot make concrete statements as the observed incidence does not follow a power-law distribution and the assumption of $y=0$ is not appropriate.
A larger sample of high-power sources, possibly exploring larger volumes than what is accessible in the GAMA09 field will be needed to properly characterise the functional form of their incidence, and we defer this analysis to future work.




\section{Conclusions}

In this work, the incidence of radio and X-ray AGN as functions of several mass-normalised power indicators are presented, aiming to test if the mass-invariant AGN triggering and fuelling mechanisms seen in the X-ray selected AGN samples also translate to a mass-invariant jet powering mechanism for radio AGN. 

Firstly, even with the soft response of eROSITA, we are able to recover mass-invariance of the X-ray AGN incidence as a function of $\lambda_{\rm{Edd}}$ found with X-ray AGN samples selected from \textit{Chandra} and \textit{XMM-Newton} (Fig.~\ref{fig:xrayfracs}). This is possible after carefully controlling for incompleteness due to the loss of obscured AGN, and is enabled by the good spectroscopic capabilities of the eROSITA CCDs \citep{Meidinger2020}.

The novelty of this work is the consideration of the incidence of radio AGN as a function of specific black hole kinetic power, $\lambda_{\rm{Jet}}$. To this end, a new fully characterised sample of radio AGN is defined based on LOFAR observations of the eFEDS field, complemented with LS9 optical counterparts. Using radio morphology, we distinguish compact and complex radio sources.
For a sub-sample of radio sources matched to the GAMA09 galaxy spectroscopic sample, we categorise their host galaxies being either quiescent or star-forming, we measure their stellar masses and define as radio AGN those with excess radio emission with respect to the level expected from pure star formation in a given galaxy. These GAMA09 galaxies provide a well-controlled parent sample of known completeness as a function of stellar mass, within which we then compute the incidence of radio AGN. For the GAMA09 galaxies, we also compile a complete sample of 34 sources hosting radio AGN with secure FRII morphologies, three of which classifying as giant radio galaxies.
\\
The main results of our work are summarised below: 

\begin{itemize}
    \item The strongly increasing incidence of radio AGN detected above the LOFAR flux limit as a function of stellar mass is recovered (Fig.~\ref{fig:sabaterfracs}), as in past studies \citep[e.g.][]{Sabater2019}, and hints of increasing normalisation with increasing redshift are present.
    
    \item The fraction of quiescent versus star-forming GAMA09 galaxies hosting compact radio AGN are similar (Fig.~\ref{fig:results_qvsSF}), suggesting that radio AGN are \textit{not} only found in `red and dead' galaxies.

    \item The fraction of GAMA09 galaxies hosting compact radio AGN as a function of $\lambda_{\rm{Jet}}$ shows approximately constant power-law slope of $-1.5$, but increasing normalisation with increasing stellar mass, and redshift (Figs.~\ref{fig:epsilon_graph} and \ref{fig:powresults}). The strong observed increase of the incidence as a function of stellar mass (see first point), is a {\it selection bias} favouring radio AGN with lower specific radio luminosity to be detected in higher mass galaxies as a result of flux limited surveys.
    
    \item The constant slope of the incidence for different stellar masses highlights that it is not only the most massive radio AGN that host the most powerful radio AGN and the low mass radio AGN that host the low power radio AGN. This slope is also steeper than the power-law describing the X-ray AGN incidence as a function of $\lambda_{\rm Edd}$, $\sim-0.65$, which may be understood in the context of reprocessing of accretion energy into coronal X-ray and radio (jetted) emission and suggests that compact radio AGN trace a relatively radiatively inefficient mode of accretion. Incidence analysis is thus useful to gain insight on average properties of a given accretion mode.

    \item Including also the complex radio morphology sources reveals a striking boosted incidence at high $\lambda_{\rm{Jet}}$, due to large physical size ($>60$~kpc) and multiple component radio AGN, in the (high) stellar mass bins where such sources are dominant (Fig.~\ref{fig:compact_complex_graph}). We find that this enhanced incidence is {\it not} due to environmental effects (i.e. to powerful complex radio AGN residing in more massive haloes), as probed by the GAMA sample. A different mass dependence of the incidence, in comparison to the pure compact case, is also clear at all $\lambda_{\rm Edd}$. 
    
    \end{itemize}
    
We discuss the numerous caveats associated with calculating jet powers for different radio luminosities, sizes and morphologies, which complicate the interpretation of the physical mechanisms driving jets. Finally, we explore the disk-jet connection in various accretion modes by postulating different accretion rate and black hole mass scaling of the radio and X-ray AGN incidence.

Overall, larger sample sizes probing larger volumes and depths, along with improved jet power estimation, taking into account the numerous influential factors, are the key to better understand the disk-jet connection through AGN incidences. Future machine learning based algorithms will also help in finding and characterising X-ray and radio AGN \citep{Alegre2022, Mostert2022, Barkus2022}, such that incidence analysis can be completed for the large current/upcoming wide and deep surveys, for example WEAVE/LOFAR \citep{Smith2016}, eRASS \citep{Merloni2024}, ASKAP \citep{McConnell2020} and VLASS \citep{Lacy2020}. 

\vspace{+1cm}

\begin{acknowledgements}
    The authors thank the anonymous referee for their careful reading of the paper and their constructive comments. This work is based on data from eROSITA, the soft X-ray instrument aboard SRG, a joint Russian-German science mission supported by the Russian Space Agency (Roskosmos), in the interests of the Russian Academy of Sciences represented by its Space Research Institute (IKI), and the Deutsches Zentrum für Luft- und Raumfahrt (DLR). The SRG spacecraft was built by Lavochkin Association (NPOL) and its subcontractors, and is operated by NPOL with support from the Max Planck Institute for Extraterrestrial Physics (MPE). The development and construction of the eROSITA X-ray instrument was led by MPE, with contributions from the Dr. Karl Remeis Observatory Bamberg \& ECAP (FAU Erlangen-Nuernberg), the University of Hamburg Observatory, the Leibniz Institute for Astrophysics Potsdam (AIP), and the Institute for Astronomy and Astrophysics of the University of Tübingen, with the support of DLR and the Max Planck Society. The Argelander Institute for Astronomy of the University of Bonn and the Ludwig Maximilians Universität Munich also participated in the science preparation for eROSITA. The eROSITA data shown here were processed using the eSASS software system developed by the German eROSITA consortium.

    \\
    ZI acknowledges the support by the Excellence Cluster ORIGINS which is funded by the Deutsche Forschungsgemeinschaft (DFG, German Research Foundation) under Germany´s Excellence Strategy – EXC-2094 – 390783311.
    \\
    Part of this work was supported by the German \emph{Deut\-sche For\-schungs\-ge\-mein\-schaft, DFG\/} project number Ts~17/2--1. 
    \\
    FdG acknowledge the support of the ERC Consolidator Grant ULU 101086378
    \\
    LOFAR data products were provided by the LOFAR Surveys Key Science project (LSKSP; https://lofar-surveys.org/) and were derived from observations with the International LOFAR Telescope (ILT). LOFAR (van Haarlem et al. 2013) is the Low Frequency Array designed and constructed by ASTRON. It has observing, data processing, and data storage facilities in several countries, that are owned by various parties (each with their own funding sources), and that are collectively operated by the ILT foundation under a joint scientific policy. The efforts of the LSKSP have benefited from funding from the European Research Council, NOVA, NWO, CNRS-INSU, the SURF Co-operative, the UK Science and Technology Funding Council and the J\"ulich Supercomputing Centre.

    \\
    Funding for the Sloan Digital Sky Survey V has been provided by the Alfred P. Sloan Foundation, the Heising-Simons Foundation, the National Science Foundation, and the Participating Institutions. SDSS acknowledges support and resources from the Center for High-Performance Computing at the University of Utah. The SDSS web site is www.sdss.org.
    
    SDSS is managed by the Astrophysical Research Consortium for the Participating Institutions of the SDSS Collaboration, including the Carnegie Institution for Science, Chilean National Time Allocation Committee (CNTAC) ratified researchers, the Gotham Participation Group, Harvard University, Heidelberg University, The Johns Hopkins University, L’Ecole polytechnique fédérale de Lausanne (EPFL), Leibniz-Institut für Astrophysik Potsdam (AIP), Max-Planck-Institut für Astronomie (MPIA Heidelberg), Max-Planck-Institut für Extraterrestrische Physik (MPE), Nanjing University, National Astronomical Observatories of China (NAOC), New Mexico State University, The Ohio State University, Pennsylvania State University, Smithsonian Astrophysical Observatory, Space Telescope Science Institute (STScI), the Stellar Astrophysics Participation Group, Universidad Nacional Autónoma de México, University of Arizona, University of Colorado Boulder, University of Illinois at Urbana-Champaign, University of Toronto, University of Utah, University of Virginia, and Yale University.
\end{acknowledgements}

%
%


\bibpunct{(}{)}{;}{a}{}{,} 

\bibliographystyle{aa} 
\bibliography{radioxrayinc.bib} 

\begin{appendix}

\section{Catalogues}
\label{appendix:catalogues}


The eFEDS X-ray catalogue, optical host-galaxy counterparts, X-ray spectroscopy results are available on the eROSITA Early Data Release website\footnote{ \url{https://erosita.mpe.mpg.de/edr/eROSITAObservations/Catalogues/}}. 

The full field 144MHz LOFAR-eFEDS source catalogue containing 45,207 entries (light red region in Fig.~\ref{fig:catskyplot}), produced using PyBDSF at a resolution of 8\arcsec\ $\times$ 9\arcsec\ is made available on the LOFAR Surveys Data Releases website\footnote{\url{https://lofar-surveys.org/efeds.html}}. The columns are the same as in the LoTSS DR1 from \citet{Shimwell2019}, but more details can be found in the PyBDSF documentation\footnote{\url{https://pybdsf.readthedocs.io/en/latest/write_catalog.html}}. 

The LOFAR-eFEDS value-added catalogue (VAC), which underpins this work (available on the LOFAR Surveys Data Releases website) is provided for the 36,631 sources in the eFEDS region where the X-ray exposure time exceeds $500$s (blue shaded region on Fig.~\ref{fig:catskyplot}). Included in this VAC is the radio morphology classification, optical host galaxy identification using Legacy DR9 data and selection of radio AGN; all of the details can be found in Section \ref{methods:radio}. Table \ref{table:catalogue_cols} gives a complete description of all columns in the VAC. A boolean value of 1 (0) is equivalent to True (False) in the columns describing flagged samples. Note that only the mass-complete G9 radio AGN were visually inspected in this work; other sources, especially those with complex radio morphologies, may need further validation to ascertain the correctness of their optical counterparts as radio centring may not be trivial for such objects (as discussed in Appendix \ref{appendix:nwaylofarlegacy_visinspect}). This also means that the \texttt{FRII\_flag} entries are only valid for the subset which was visually inspected.

Twelve LOFAR sources with different handling in their optical counterpart identification and/or radio property calculation are marked with \texttt{special\_flag}=1. Full details are given in Appendix \ref{appendix:nwaylofarlegacy_visinspect} but a summary is presented here. Four sources (LOFAR Source id: 7310, 10975,
25001, 29295) have been manually rematched to their correct LS9 optical counterpart during the visual inspection process. Two sources (LOFAR Source id: 22763, 26644), with correctly identified optical counterparts, were moved from the compact to the complex radio morphology sample and had their radio properties updated manually, including summing associated component emission. Six sources (LOFAR Source id: 8153, 14599, 23634, 29440, 29781, 32863), from the additional visual inspection process to find large FRIIs, also had their optical counterparts and radio properties updated manually. For sources with manually assigned counterparts, the \texttt{p\_any}, \texttt{p\_i} entries are NULL. For sources with summed components, the \texttt{LOFAR\_Total\_flux} and \texttt{LOFAR\_E\_Total\_flux} are summed/ propagated accordingly from the individual component entries; the \texttt{LOFAR\_Peak\_flux} and \texttt{LOFAR\_E\_Peak\_flux} entries are NULL; the \texttt{LOFAR\_Maj} entry corresponds to the projected largest linear size in degrees; and \texttt{CTP\_Separation} is taken as the distance from the given LOFAR Source id and the matched LS9 counterpart.


\begin{table*}[]
    \centering
    \caption{Column descriptions for the LOFAR-eFEDS value-added catalogue.}
    \begin{tabular}{l|l}
    \hline
    \hline
        \textbf{Column Name} & \textbf{Description} \\
        \hline
        \textbf{LOFAR} \\
        \texttt{LOFAR\_Source\_id} & Unique number that identifies the source. \\
        \texttt{LOFAR\_RA} & Right ascension of the source (for the equinox of the image), in degrees. \\
        \texttt{LOFAR\_DEC} & Declination of the source (for the equinox of the image), in degrees. \\
        \texttt{LOFAR\_Total\_Flux} & Total, integrated Stokes I flux density of the source at the reference frequency, in Jy. \\
        \texttt{LOFAR\_E\_Total\_Flux} & 1$\sigma$ error on the total flux density of the source, in Jy. \\
        \texttt{LOFAR\_Peak\_Flux} & Peak Stokes I flux density per beam of the source, in Jy/beam. \\
        \texttt{LOFAR\_E\_Peak\_Flux} & 1$\sigma$ error on the peak flux density per beam of the source, in Jy/beam. \\
        \texttt{LOFAR\_Maj} & FWHM of the major axis of the source, in degrees. \\
        \texttt{LOFAR\_pos\_err} & Positional error calculated using Eq.~\ref{poserr_eq}. \\
        \\

        \textbf{LOFAR: Morphology Flags} \\
        \texttt{LOFAR\_scodeS\_flag} & True if \texttt{LOFAR\_S\_code=S}, i.e. fit with only a single Gaussian by PyBDSF. \\
        \texttt{LOFAR\_fluxratio\_flag} & True if \texttt{LOFAR\_Total\_Flux/LOFAR\_Peak\_Flux} $<3.6$. \\
        \texttt{LOFAR\_maj\_flag} & True if \texttt{LOFAR\_Maj} $<19.1$\arcsec. \\
        \texttt{LOFAR\_isolated\_flag} & True if no nearest neighbours within 45\arcsec. \\
        \texttt{LOFAR\_compact\_flag} & True: compact source; all four flags above are True (see Sect. \ref{sec:compactvscomplex}). False: complex source.   \\
        \\
        \textbf{Legacy Survey DR9} \\
        \texttt{LS9\_UNIQUE\_OBJID} & Unique source identifier: \texttt{BRICKID\_OBJID} (\texttt{RELEASE}=9010 for sources present here). \\
        \texttt{LS9\_TYPE} & Morphological model. \\
        \texttt{LS9\_RA} & Right ascension at equinox J2000. \\
        \texttt{LS9\_DEC} & Declination at equinox J2000. \\
        \texttt{LS9\_pos\_err} & Positional error. \\
        \texttt{LS9\_mag\_g\_dered} & De-reddened g-band magnitude. \\
        \texttt{LS9\_mag\_r\_dered} & De-reddened r-band magnitude. \\
        \texttt{LS9\_mag\_z\_dered} & De-reddened z-band magnitude. \\
        \texttt{LS9\_mag\_W1\_dered} & De-reddened W1-band magnitude. \\
        \texttt{LS9\_mag\_W2\_dered} & De-reddened W2-band magnitude. \\
        \texttt{LS9\_mag\_W3\_dered} & De-reddened W3-band magnitude. \\
        \texttt{LS9\_mag\_W4\_dered} & De-reddened W4-band magnitude. \\
        \\
        \textbf{NWAY Match} \\
        \texttt{CTP\_Separation} & Separation, in arcsec, between LOFAR source and best-match LS9 counterpart. \\
        \texttt{p\_i} & Probability for the counterpart to be the correct one. \\
        \texttt{p\_any} & Probability for a source to have any counterpart in the search region. Optimal: \texttt{p\_any} $>0.06$. \\
        \\
        \textbf{GAMA09} \\
        \texttt{uberID} & Unique GAMA ID of object. \\
        \texttt{CATAID} & Unique numeric GAMA object identifier. \\
        \texttt{RAcen} & Right Ascension of flux-weighted centre (ICRS). \\
        \texttt{Deccen} & Declination of flux-weighted centre (ICRS). \\
        \texttt{Z} & Spectroscopic redshift. \\
        \texttt{StellarMass\_50} & Median stellar mass from MCMC chain. \\
        \texttt{StellarMass\_16} & 16th percentile stellar mass from MCMC chain. \\
        \texttt{StellarMass\_84} & 84th percentile stellar mass from MCMC chain. \\
        \texttt{SFR\_50} & Median SFR from MCMC chain. \\
        \texttt{SFR\_16} & 16th percentile SFR from MCMC chain. \\
        \texttt{SFR\_84} & 84th percentile SFR from MCMC chain. \\
        \texttt{SC} & Science sample class. \texttt{SC} $\geq6$ is used here. \\
        \\
        \textbf{Additional Flags}
        \\
        \texttt{radioAGN\_flag} & True if LOFAR source (with SNR$>$5, \texttt{p\_any}$>$0.06) fulfils radio-excess criterion defined by Eq.~\ref{eq_radioAGNcut}. \\
        \texttt{vis\_inspected} & True if source has been visually inspected (see Appendix \ref{appendix:nwaylofarlegacy_visinspect}). \\
        \texttt{G9\_radioAGN} & Final sample of mass-complete G9 radio AGN, with visual inspection results applied (see Table \ref{tableOfsources}).\\
        \texttt{FRII\_flag} & \texttt{FRII\_flag} $=1, 0.5, 0$: secure, likely, unlikely FRII-morphology, respectively. \\
        \texttt{GRG\_flag} & Giant radio galaxy flag (largest linear size $>0.7$~Mpc). \\
        \texttt{G9\_radioXray\_sources} & True if the LOFAR source has an X-ray match in eROSITA eFEDS (see Table \ref{tableOfsources}). \\
        \texttt{special\_flag} & True if source required special cross-matching and/or property estimation (see Appendix A). \\
    
        \hline
        \hline
    \end{tabular}
    \label{table:catalogue_cols}
\end{table*}

\section{Unobscured X-ray AGN Incidences}
\label{appendix:softsel}

\begin{figure}[h!]
\centering
\includegraphics[width=0.95\linewidth]{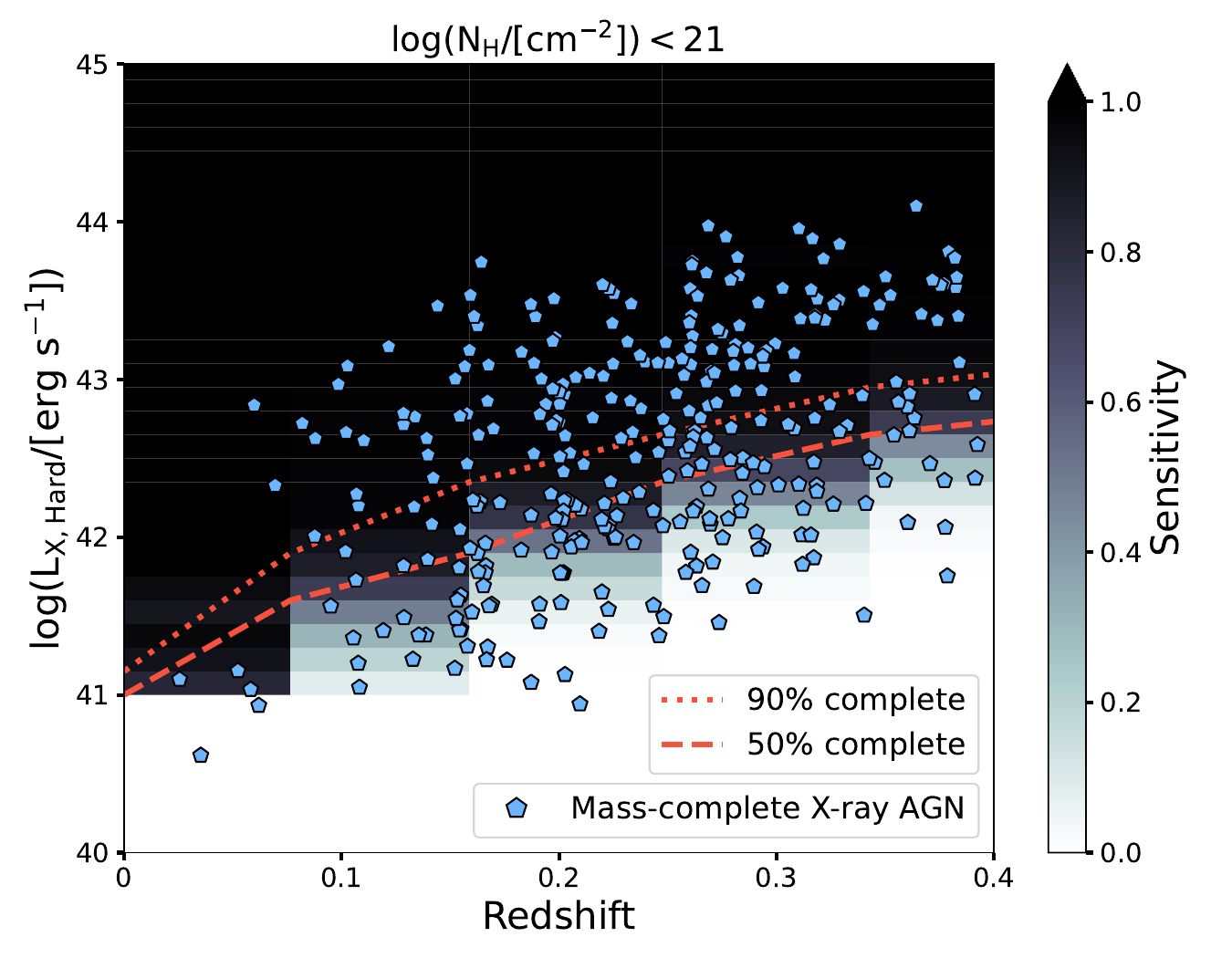}
\caption{Redshift versus intrinsic hard ($2-10$~keV) X-ray luminosity of the G9 X-ray AGN (pentagons). A sensitivity grid for only unobscured ($\log N_H <21$~cm$^{-2}$) sources is plotted in the background from simulations done by \citet{TengSimulation2022} and used to compute the 50\% and 90\% X-ray luminosity completeness limits, respectively.}
\label{fig:unobs_selection}
\end{figure}

\begin{figure}[h!]
\centering
\includegraphics[width=0.95\linewidth]{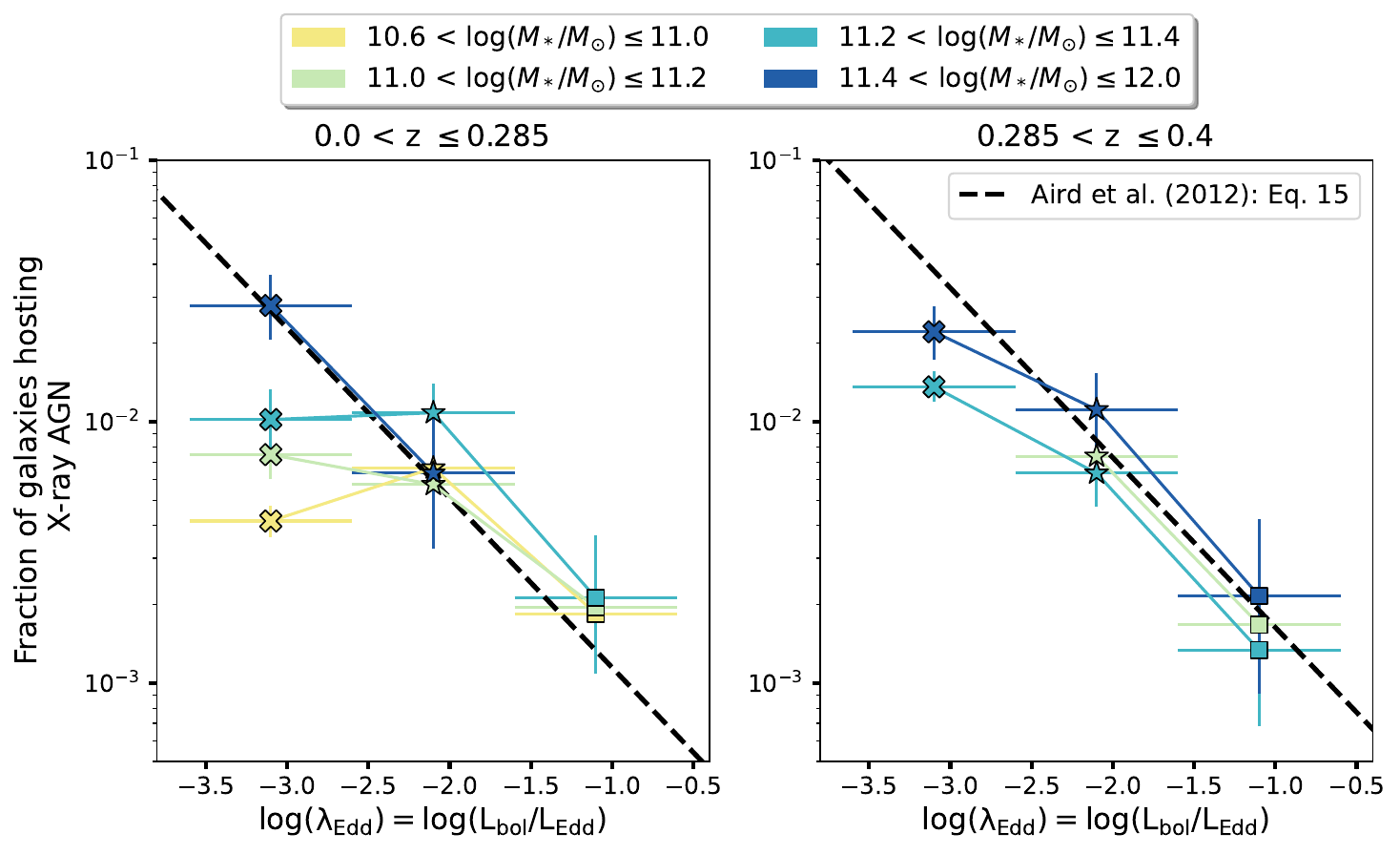}
\caption{Fraction of GAMA09 galaxies hosting X-ray AGN as a function of $\lambda_{\rm Edd}$ in different stellar mass and redshift bins, using only a soft selection (meaning that the X-ray luminosity is corrected to be complete for $\log N_H <21$~cm$^{-2}$). As expected, the lowest $\lambda_{\rm Edd}$ bins are incomplete, compared to the black dashed line, due to obscuration.}
\label{fig:unobsXrayfracs}
\end{figure}

Fig.~\ref{fig:unobs_selection} shows the unobscured (i.e. assuming $\log N_H <21$~cm$^{-2}$) sensitivity corrections (orange curves), over-plotted on the G9 X-ray AGN sample. It is seen that many more X-ray AGN are now $>90\%$ complete, as would be expected from eROSITA's soft selection. However, when considering the fraction of galaxies hosting such X-ray AGN as a function of $\lambda_{\rm Edd}$ (Fig.~\ref{fig:unobsXrayfracs}) there is a clear lack of detections at the lower accretion rate end. This is due to the effects of obscuration affecting the lower luminosity population more, making them drop out of the sample. Fig.~\ref{fig:unobsXrayfracs} is interesting to show the levels of $\lambda_{\rm Edd}$ which start to become significantly affected by obscuration.

\section{Optical Counterparts to the Radio Sources and Visual Inspection Results}
\label{appendix:nwaylofarlegacy_visinspect}

This section provides more details on the steps taken to find the optical counterparts to the new set of radio sources characterised in this work and describes the results from the visual inspection.

Firstly, as noted in \citet{Williams2019}, the positional errors in the LOFAR catalogues, as outputted automatically by PyBDSF, are often underestimated. This is because the pipeline returns only the error on the FWHM of the major and minor axes of the Gaussian fits, without considering additional correlated noise known empirically to be present. Following \citet{Williams2019}, a $\sqrt{2}$ factor is applied to the catalogue positional error, and an additional astrometric uncertainty of $\sigma_{\rm ast}=0.6$ is added in quadrature. The latter is common in radio to optical studies \citep{Williams2019}, arising from the directional asymmetries in mapping from major/minor axes to RA/Dec space in the optical, especially considering the low declination of the LOFAR-eFEDS field (meaning the beam is elongated). This step is necessary for NWAY to be able to accurately identify counterparts, especially for the obvious bright radio sources with unrealistically small positional errors. Overall, the positional error ($\sigma_{\rm{pos}}$; catalogue column \texttt{LOFAR\_pos\_err}) on the radio sources in this work is calculated using the following equation:
\begin{equation}
    \label{poserr_eq}
    \sigma_{\mathrm{pos}}=\sqrt{2*\Big( \sigma_{\mathrm{RA}}^2+\sigma_{\mathrm{Dec}}^2 \Big) + \sigma_{\mathrm{ast}}^2}
\end{equation}
where $\sigma_{\rm{RA,Dec}}$ are the traditional errors on the RA, Dec position. 

Drawing a cumulative histogram of the $\sigma_{\rm{pos}}$ values of the LOFAR sources revealed that only 10\% of sources have $\sigma_{\mathrm{pos}}>1.6$\arcsec. Therefore, the maximum search radius to be used in NWAY to find optical counterparts for the radio sources is taken to be 8\arcsec\ (five times this value). The LS9 positional uncertainties of the sources in Table \ref{tableOfsources} are much smaller, having an average of 0.1\arcsec, which is taken as the constant $\sigma_{\rm{pos}}$ for all sources in the NWAY procedure.

Adding magnitude and/or colour priors significantly improves the accuracy of matching radio sources to their optical counterparts as radio emitters tend to be found in redder galaxies \citep[e.g. ellipticals; see][]{Williams2019}. Therefore, absorption corrected {\it g, r, z} and {\it W1} magnitudes from LR9 \citep[using the extinction map of][]{Schlegel1998dustextinctionmap} are added as internal priors in the NWAY match, using the `auto' feature, where NWAY learns to differentiate the magnitude or colour (or other source parameter) distributions between target and field sources `on the fly' \citep[for more details see Section B6.1 in][]{maraNWAY2018}. 
Lastly, the appropriate sky densities of each catalogue are calculated and a prior completeness fraction of 70\% is assumed for the NWAY matching process \citep[see e.g.][]{Williams2019, Smolcic2008}. The latter can be justified given that the radio sources in the LOFAR Deep Field \citet{KondapallyDeepFields2021} with total integrated flux $>1$~mJy, having optical counterparts above the LS9 r-band limit of 23.54 mag, is 68\%. The non-detected sources are also likely high redshift ($z \sim 3-4$) obscured radio quiet quasars and so their absence would not impact the results of this work, based on a local sample of radio AGN \citep[see e.g. Section 6.1 in][]{KondapallyDeepFields2021}.

\begin{figure}
\centering
\includegraphics[width=0.95\linewidth]{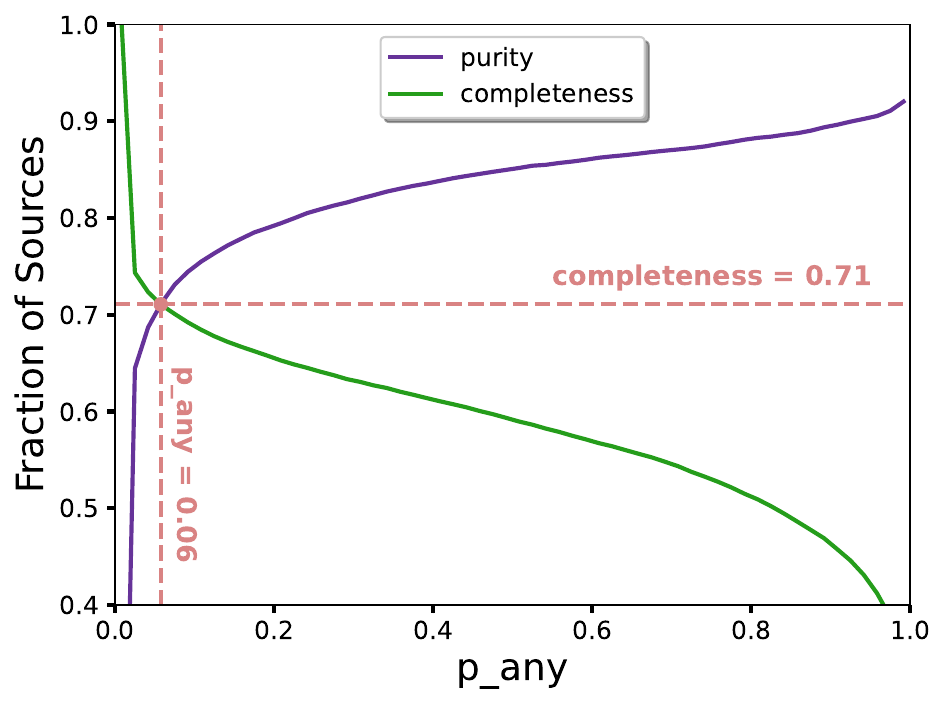}
\caption{Purity (purple) and completeness (green) curves, found by creating a `fake' catalogue of sources with shifted sky coordinates, to calculate the fractions of wrongly assigned counterparts as a function of \texttt{p\_any} (see text for details). The optimal \texttt{p\_any} threshold is equal to 0.06.}
\label{fig:pany_threshold}
\end{figure}

In this way, 33,769/36,631 LOFAR sources are matched to LS9 optical sources. To cut those matches which are statistically unlikely to be real matches, whilst keeping as many matches as possible (i.e. finding the balance between purity and completeness), an `optimal' \texttt{p\_any} cut is defined. This is done by creating a `fake' match catalogue where the Dec coordinates of the radio sources by are shifted by 60\arcsec\ \citep{MaraCTPeFEDS}. The real radio sources within 8\arcsec\ of such `fake' sources are removed and then the `fake' catalogue is again matched in the same way to LS9. A reverse cumulative ratio, effectively the `completeness', is then calculated between the \texttt{p\_any} distributions of the `fake' to real matches. The purity is defined as one minus this ratio. Fig.~\ref{fig:pany_threshold} depicts the trade-off between purity (purple curve) and completeness (green curve), and the optimal \texttt{p\_any}, located at the intersection, is $0.06$. Applying this cut on the real sample, results in 25,806/36,631 matches, or in other words 70\% match fraction, in agreement with \citet{Williams2019}. 

\begin{figure}
\centering
\includegraphics[width=0.95\linewidth]{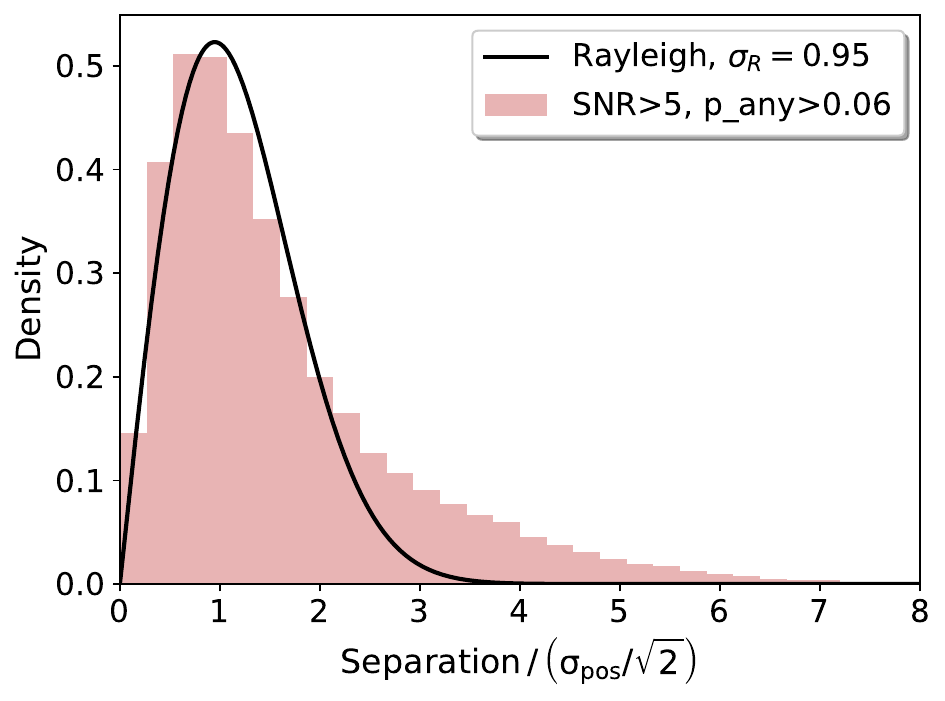}
\caption{Rayleigh curve fit to the histogram of the ratio of counterpart separation to the 1D positional error for the sample of 22,754 LOFAR-LS9 sources. Visually, a good agreement can be seen with $\sigma_{R} \sim 1$, as expected, although there is a remnant tail (see text for discussion).}
\label{fig:rayleigh}
\end{figure}
To further filter LOFAR detections which may be spurious, a cut of signal to noise SNR $>5$, defined as the ratio of the peak radio flux to the error in the peak flux is applied. This resulted in 22,754 matches between the LOFAR and LS9 catalogues.

\begin{figure*}[h!]
    \centering
    \includegraphics[width=0.27\linewidth, height=0.245\linewidth]{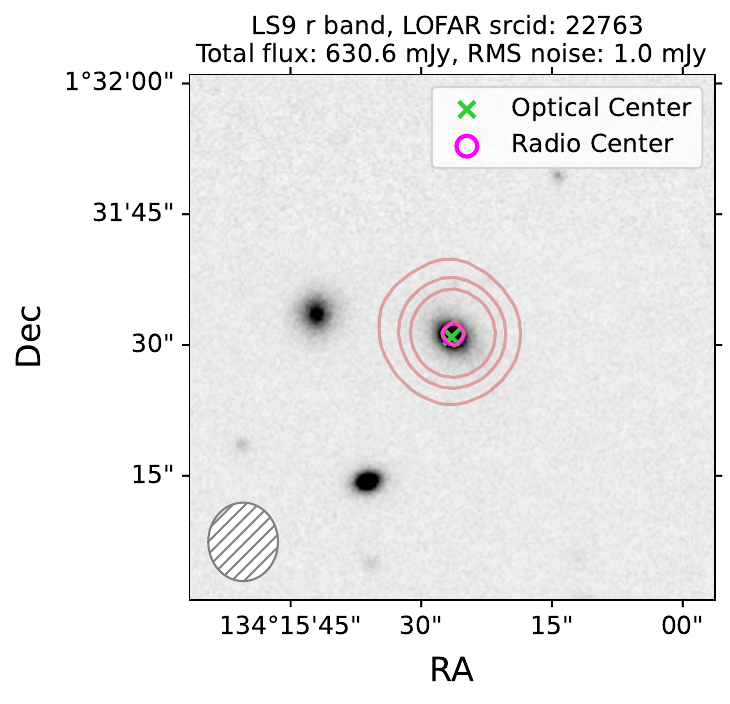}
    \includegraphics[width=0.3\linewidth]{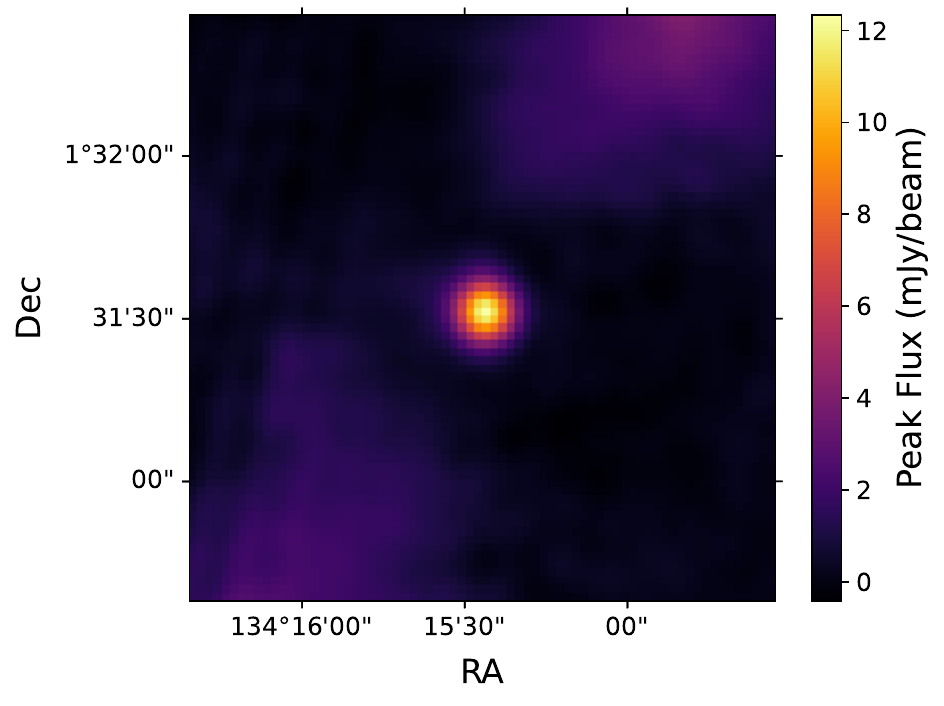}
    \includegraphics[width=0.3\linewidth]{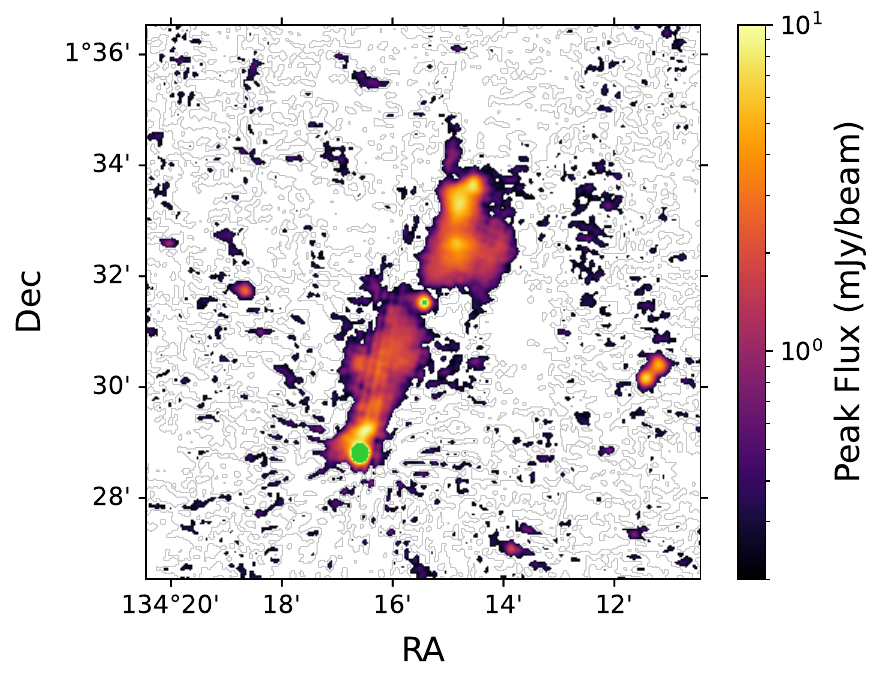}
    \caption{The three radio and optical cutouts of LOFAR Source id 22763, used for visual inspection and for classifying this source as an FRII (see text for details). Note that the green colour on the right-most panel indicates regions with peak flux greater than 10~mJy/beam, chosen to help better visualise the full dynamic range in the radio image.}
    \label{fig:src22763}
\end{figure*}

For positional (Gaussian) matching experiments, it is a common validation test to plot the distribution of the separation between counterparts divided by $\sigma_{\rm{pos}}/\sqrt{2}$ \citep{MaraCTPeFEDS, Pineau2017}, for which the best-fit should follow a Rayleigh distribution with $\sigma_{R}=1$ (note: the division by $\sqrt{2}$ is to plot the one-dimensional positional error). Fig.~\ref{fig:rayleigh} shows the radio-to-optical distribution for SNR $>5$, \texttt{p\_any} $>0.06$ sources in light red. Even though the best fit Rayleigh distribution (black curve) is at $\sigma_{R}=0.95$ and there is an excess of sources in the tail of the distribution, this is normal for `real' distributions \citep{Pineau2017}. In particular, radio sources often do not have symmetric and Gaussian RA/Dec errors, which is one of the assumptions made in having a Rayleigh distribution with $\sigma_{R}=1$, and the radio centring of complex morphology radio sources is not always trivial. 


Next, as described in Sect. \ref{sec:nwaylofarlegacy_radio}, we matched to the GAMA catalogue and appended six large FRII radio galaxies to the sample after an additional visual inspection process (described at the end of this section), giving a total of 2,619 radio sources among the GAMA09 galaxies. The final sample of mass-complete compact and complex G9 radio AGN (recall Sect. \ref{sec:masslumincomp_radio}) was then visually inspected by three of the authors of this work. Visual inspection was done to make sure that the NWAY identified counterpart was correct and also to characterise the radio morphology, in particular to identify FRII-like sources (see below).

Three cutouts per source were created to be visually inspected: 
\begin{enumerate}
    \item (1\arcmin$\times$1\arcmin)  LS9 one-band image centred on the optical coordinates, with radio contours overlaid, as in Fig.~\ref{fig:compact_complex_cutouts}.
    \item (1.8\arcmin$\times$1.8\arcmin) radio intensity image.
    \item (10\arcmin$\times$10\arcmin) radio intensity image (adjusted for visualising large dynamic ranges) to see the surroundings of the radio source in case of larger scale associated emission.
\end{enumerate}

As a result of the visual inspection, two sources, which had catalogued SNR$>$5, were removed as they appeared to be associated to noise in the radio images, usually due to calibration errors in the spokes around bright sources (LOFAR Source id: 10347, 27051). Then four sources had to be rematched to different optical counterparts as the radio centring from asymmetric or complex jetted morphologies was not associating the core of the emission properly (LOFAR Source id: 7310, 10975, 25001, 29295). There was one compact case (LOFAR Source id: 22763) where the wider area radio intensity cutout revealed four additional associated radio components (diffuse lobes and compact hot spots) in a large-scale FRII source, classifying as a giant radio galaxy. In this case, the emission from the five components was summed to give the total radio flux (similarly, the two components associated to LOFAR Source id: 29295 were also summed). Fig.~\ref{fig:src22763} shows the three cutouts created for this source, as described above, underlining the need to look at the larger scale environment. Likewise, a head-tail radio source (LOFAR Source id: 26644), catalogued as `compact' was moved to the complex sample. Otherwise, NWAY was almost 100\% successful at assigning the correct optical counterparts (to the authors' best knowledge), when the radio centring and calibration was accurate.

Moreover, during the visual inspection process, three visual inspectors would assign a value of 1, 0.5 or 0 depending on if a source showed secure, likely or unlikely FRII-like morphology. If the average of three verdicts was $>0.5, =0.5$ or $<0.5$, the source was classed as a `secure', `likely' or `unlikely' FRII, respectively, and flagged accordingly in the VAC (\texttt{FRII\_flag}). We find 28 FRIIs this way, which are combined with six further larger FRIIs, described below.

Lastly, the optical host identification method adopted in this paper is not tailored for finding large and powerful radio galaxies or sources with lobe components catalogued as separate IDs, thus these objects may be missed. To ascertain the completeness of the complex radio AGN sample, a test was made to visually inspect (with 10\arcmin\ $\times$ 10\arcmin\ cutouts) all LOFAR-eFEDS sources in the GAMA09 area with $F_{\rm Tot}>$10~mJy and \texttt{Maj}>19.1\arcsec\ and \texttt{LOFAR\_scodeS\_flag}$=$False, matching to at least one other nearest neighbour within 2\arcmin\ (528 catalogue entries in total). A total of 78/528 catalogue entries were flagged as containing possible large, disconnected radio components. Each entry was then matched to GAMA09 within 5\arcmin\ to visually search for host galaxy counterparts. Nine sources were identified to match to a host galaxy detected in GAMA09, three of which were already present in the G9 radio AGN sample. This brings the total identified secure FRIIs in this investigation to 34 sources. Three out of these 34 sources are classified as giant radio galaxies, marked in the VAC with \texttt{GRG\_flag}$=$True \citep[largest linear size $>0.7$~Mpc, e.g.][]{Saripalli2005}. The LOFAR components are summed to get the total flux and the largest linear projected size is used to calculate the extent of the source in kpc.


\section{Accounting for missed radio AGN in highly star-forming galaxies}
\label{missedhighSFRradioAGN}

In relation to Fig.~\ref{fig:radioAGNcut}, there is a further step of incompleteness that has to be addressed, as the radio AGN cut can introduce a selection effect of preferentially removing higher mass galaxies. This is because higher stellar mass corresponds to a higher SFR, a consequence of the main sequence of star-forming galaxies \citep[e.g.][see Sect. \ref{sec:qvsSF}]{Speagle2014}. Thus, as one moves to higher SFRs on Fig.~\ref{fig:radioAGNcut}, the sources have to have higher and higher radio luminosities, meaning stronger jets, to systematically dominate over the stellar emission. However, as indicated by the radio luminosity function for radio AGN \citep{Smolcic2008, Sabater2019, Kondapally2022}, higher power jetted systems are less common in the universe and therefore, it gets harder to pass the cut for higher SFRs. 


To mitigate this, the incompleteness is accounted for by determining the fraction of sources which could be missed using a combination of the main sequence presented in \citet{Speagle2014} and the radio AGN cut from \citet{Best2023radioAGN}. 

Firstly, the $\log M_*$ and $\log Q/L_{\rm{Edd}}$ bin limits are used to calculate a parallelogram in $\log M_*-\log L_{\rm{R}}$ space covered by those limits (e.g. top right corner of this parallelogram would have an $\log L_{\rm{R}}$ value calculated by combining the maximum bin limits of $\log M_*$ and $\log Q/L_{\rm{Edd}}$). Secondly, a linear relation in $\log M_*-\log L_{\rm{R}}$ space is computed, combining the radio AGN $3\sigma$ cut (relating $\log L_{\rm R}$ and $\log \rm{(SFR)}$) and the MS equation (relating $\log \rm{(SFR)}$ to $\log M_*$, within a given redshift bin). Anything with a radio luminosity below this linear relation is incomplete as it bridges into the star-forming MS and means that radio AGN below that line are missed. Therefore, the fraction of sources missed within a given $\log M_*$ and $\log Q/L_{\rm{Edd}}$ bin is simply the geometric ratio of the parallelogram area below vs. above the linear relation. The weighting, applied by multiplying the incidence in a given bin, is then the inverse of this missed fraction. Note that this correction only affects the lowest $L_{\rm R}/M_*$ or $\lambda_{\rm Jet}$ bins (crosses on Figures \ref{fig:epsilon_graph} and \ref{fig:sbhar_graph}), in some cases being <50\% complete and therefore removed.

\section{Incidence of radio AGN as a function of the "specific" radio luminosity $L_{\rm R}/M_*$}
\label{appendix:LrMstar}

\begin{figure}[h!]
\centering
\includegraphics[width=\linewidth]{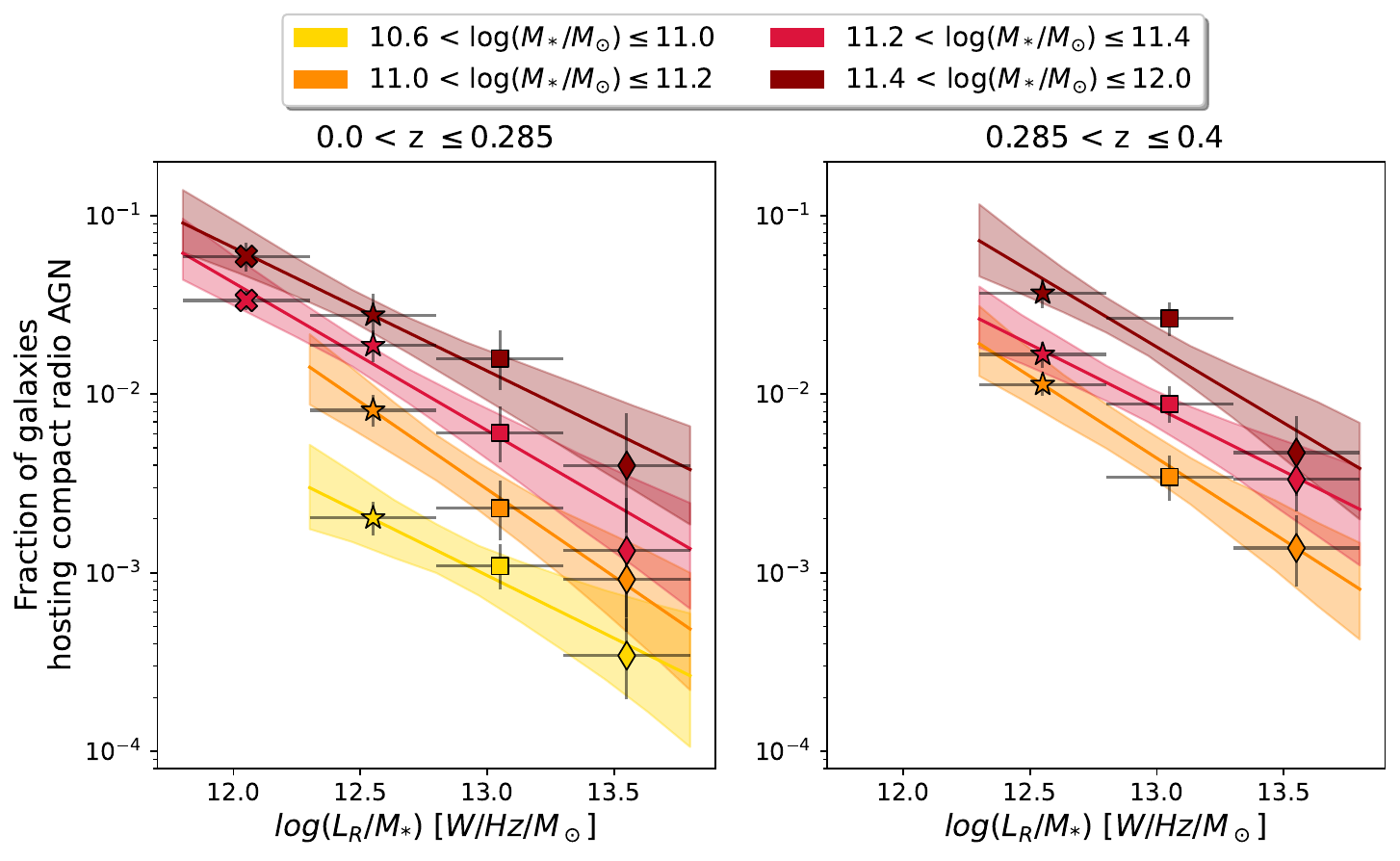}
\caption{Fraction of GAMA09 galaxies hosting compact radio AGN as a function of radio luminosity normalised by stellar mass ($L_{\rm R}/M_*$), in different redshift (panels) and stellar mass bins (colours). Power-laws, and associated errors, are fit to each stellar mass bin and are plotted with the corresponding colour.}
\label{fig:sbhar_graph}
\end{figure}

Fig.~\ref{fig:sbhar_graph} shows the fraction of combined quiescent and star forming GAMA09 galaxies hosting only compact radio AGN as a function of specific radio luminosity ($L_{\rm R}/M_*$), in different redshift and stellar mass bins. The average power-law slopes are constant around $-0.8$. 
A mass dependence is clearly seen by the increasing power-law normalisations with stellar mass. Quantitatively, at $\log(L_{\rm R}/M_{*})=13$, the highest mass galaxies are 13.5 and 4.2 times more likely to host radio AGN than lowest mass galaxies, across all values of $L_{\rm R}/M_*$, in the low and high redshift bins, respectively.

\end{appendix}

\end{document}